\documentclass[]{interact}

\usepackage{epstopdf}

\usepackage{comment}
\usepackage{hyperref}

\usepackage{apacite}

\usepackage{algorithm}
\usepackage{algpseudocode}
\usepackage{url}
\usepackage{graphicx}
\usepackage{float}
\usepackage{newfloat}
\usepackage{subcaption}
\usepackage{lipsum}
\usepackage{xcolor}
\usepackage{pdflscape}  
\usepackage[normalem]{ulem}

\usepackage{booktabs,siunitx,tabularx,threeparttable,makecell}
\usepackage{soul}
\usepackage{afterpage}
\usepackage{placeins}
\usepackage{caption}
\theoremstyle{plain}

\theoremstyle{definition}

\theoremstyle{remark}

\newif\ifanonymous
\anonymousfalse  

\ifanonymous
  \newcommand{\rev}[1]{\textcolor{red}{#1}}
\else
  \newcommand{\rev}[1]{#1}
\fi

\begin{document}


\title{Computationally Efficient Estimation of Localized Treatment Effects for Multi-Level, Multi-Component Interventions to Address the Opioid Crisis}

\ifanonymous
  \author{Anonymous Authors}
\else
\author{
\name{Abdulrahman A. Ahmed\textsuperscript{a} 
, M. Amin Rahimian *\textsuperscript{,a}
\thanks{$^*$Correspondence to: rahimian@pitt.edu}, 
Qiushi Chen \textsuperscript{b}, and Praveen Kumar\textsuperscript{c}}
\affil{\textsuperscript{a} Department of Industrial Engineering, University of Pittsburgh, Pittsburgh, USA; \\
\textsuperscript{b} Harold and Inge Marcus Department of Industrial and Manufacturing Engineering, The Pennsylvania State University, University Park, USA; \\
\textsuperscript{c} Department of Health Policy and Management, University of Pittsburgh, Pittsburgh, USA}
}
\fi

\maketitle

\begin{abstract}

The opioid epidemic remains a major public health challenge in the United States, requiring a \textit{multi-pronged intervention}  approach to mitigate harms to communities. Given the heterogeneity of the epidemic, it is crucial for policymakers to understand \textit{localized treatment effects} of different intervention components and utilize limited resources efficiently. 
While locally calibrated simulation models can project epidemic outcomes for any given intervention policy, collecting simulation results for all intervention combinations to estimate localized treatment effects for each community is impractical because the number of combinations grows exponentially with the number of interventions and the levels at which they are applied.
To tackle this, we develop a \emph{two-stage} metamodel framework with a \emph{two-step sequential design} for efficient sampling. The metamodel consists of a response function linking health outcomes to each intervention component’s treatment effect, and a Gaussian process regression (GPR) to learn spatial and socio-economic structures of the treatment effects based on locally-contextualized covariates. 
With two-step sequential sampling, we leverage spatial correlations and posterior uncertainty to sequentially sample the most informative counties and treatment conditions. 
We apply this framework to estimate the treatment effects of buprenorphine dispensing and naloxone distribution on overdose mortality rates 
using a calibrated agent-based opioid epidemic model in Pennsylvania counties.
Our approach achieves less than $5\%$ average relative error using fewer than 2\% of the runs required for an exhaustive simulation. Our two-stage framework provides a computationally efficient approach to 
support 
policymakers, enabling an efficient evaluation of alternative resource-allocation strategies to mitigate the opioid epidemic in local communities. 
\end{abstract}

\begin{keywords}
metamodel; large-scale simulation; sequential design; model selection; epidemiological models; precision public health
\end{keywords}

\section{Introduction}
\vspace{-5pt}

The opioid crisis remains a major public health challenge, with more than 750,000 deaths in the United States over the past two decades \shortcite{garnett2024cdcdeaths}. A defining feature of this epidemic is its pronounced heterogeneity across the country. For example, the drug overdose mortality rates in 2023 ranged from 9.0 per 100,000 in Nebraska to 81.9 per 100,000 in West Virginia \shortcite{CDC2025overdosemap}. Moreover, the epidemic has been evolving constantly, shifting from prescription opioids to heroin, then to synthetic opioids like fentanyl, and more recently to co-stimulant use as the primary drivers of mortality \shortcite{volkow2021opioid, jenkins2021opioid, ciccarone2021epidemiology, ahmed2022epidemic}. These complex dynamic and geospatial patterns have contributed to the variation seen in the extent of the opioid crisis across states and counties in the US.

It has become increasingly clear that a multi-pronged intervention approach is needed to mitigate the harm by the opioid crisis \shortcite{blanco2020america}.
It requires a coordinated set of interventions spanning prevention, treatment and harm reduction strategies. 
These interventions include implementing policies that support non-opioid pain management therapies, enhancing prescriber education, reducing stigma through public education campaigns, and improving access to medications for opioid use disorder (MOUD) by removing barriers to treatment \shortcite{cheng2022bupre}. Additional efforts have focused on increasing initiation and retention in care, facilitating linkage to care with peer support, and expanding access to fentanyl test strips and naloxone kits \shortcite{d2015linkage, liebschutz2014linkage}. Among the most widely deployed and evidence-based interventions available to county health departments are naloxone distribution, a harm-reduction strategy that reverses opioid overdoses, and buprenorphine, a medication for opioid use disorder (MOUD) that reduces disease burden by supporting recovery \shortcite{cerda2021systematic, lim2022modeling}.

Critically, the effectiveness of any given intervention combination is not uniform across communities. Counties differ substantially in their epidemic trajectory, demographic composition, and urban or rural character, such that the same multi-pronged strategy may produce markedly different reductions in overdose mortality depending on local conditions \cite{cerda2024simulating, dodson2018spatial, marotta2019assessing, zheng2025opioid}.
With the wide range of evidence-based interventions available but limited resources, policymakers need to assess the priority of available interventions and develop an effective intervention package that consists of multiple intervention components, each at a proper level. More importantly, given the heterogeneity of the crisis and the varying phases of the epidemic across counties, the multi-level, multi-component interventions must be tailored to the unmet needs of local communities, such as those in the counties. A key to facilitating such decision making processes is to understand the localized treatment effects for each intervention component. That is, the same level of intervention may have different impacts on opioid overdoses and other substance-related outcomes in different counties.

While several modeling-based studies have utilized simulation to project the impact of various intervention strategies \shortcite{scheidell2024reducing, irvine2022estimating, lim2022modeling, zang2022comparing}, estimating the impact of multi-level, multi-component interventions at local level is a non-trivial task, even with a validated simulation model. A naive strategy is to exhaustively enumerate and evaluate all possible combinations of interventions for a local county, estimate the treatment effect in this county, and then repeat for each county.  Clearly, such full factorial design leads to a rapid expansion in the total number of design points to be evaluated (by simulation).
\rev{For example, 
with seven interventions each assigned five levels, the number of treatment combinations grows to $5^7 = 78{,}125$ per county. To estimate treatment effects across counties, the total number of required simulation runs further increases 
multiplicatively: with 50 counties, a full factorial design entails 
$50 \times 5^7 = 3{,}906{,}250$ configurations, and this count increases further when multiple replications are needed for stochastic simulations, posing a 
significant computational challenge.}

While in practice, fractional factorial, Latin hypercube, or other space-filling designs are commonly used to mitigate this burden \shortcite{sanchez2020work}, these approaches still face a trade-off: one can capture higher-order interactions 
and policy synergies, but only {with} substantial computational cost; or reduce the number of runs, but at the {price} of overlooking some of these effects. These trade-offs, combined with the spatial correlation across counties, the stochastic variability of epidemic simulation outputs, and the exponentially growing design space, make large-scale simulation modeling particularly challenging.

Metamodels, or surrogate models, offer a solution by learning statistical mappings from inputs to simulation outputs. Once trained, a metamodel can provide rapid predictions and quantify uncertainty for input {parameter} that were not directly simulated, 
thereby enabling efficient design evaluation and sensitivity analysis. Among various metamodeling techniques, Gaussian process regression (GPR) is widely used because it flexibly captures nonlinear functions and produces calibrated uncertainty estimates \shortcite{rasmussen2006gaussian, forrester2008surrogate, gramacy2020surrogates}. The foundational work of \citeNP{kennedy2001bayesian} and subsequent extensions by \citeNP{conti2010bayesian} demonstrate how GPR can emulate expensive computer models while propagating parameter uncertainty. To improve learning efficiency, metamodeling is often coupled with sequential design, which adaptively selects the most informative simulation runs based on current model predictions, focusing resources on the most uncertain regions 
of the input space \shortcite{frazier2018bayesian, wilson2018maximizing, balandat2020botorch}. Acquisition functions such as predictive variance, entropy reduction, or expected improvement guide this selection and are particularly useful when the number of input combinations is large or when simulation cost is high \shortcite{fisher2020predicting}.

Metamodeling, including GPR, has been applied to opioid epidemic settings, with some specifically focusing on estimating the treatment effect of intervention policies.
\citeNP{ahmedbhi} propose a regression-based greedy sampling strategy that allocates simulation effort across treatment conditions based on confidence interval width, achieving accuracy comparable to uniform sampling with substantially fewer simulations. However, their work is limited to a single county and cannot capture the spatial and socio-economic structures that induce the county-level heterogeneity that is crucial to understanding the opioid epidemic.

This spatial structure arises naturally in opioid simulation models, where outputs are summarized as coefficients describing how each treatment affects overdose death rates.
For a single county, this can be achieved using regression-based estimates. For example, \citeNP{ahmed2024selection} study the optimal selection of linear regression functions to estimate treatment effects given a limited number of simulations for a single county. However, on the larger geographic scales of entire states with many counties, these coefficients become interdependent and geographically structured: counties that are spatially proximate or demographically similar frequently share treatment-response patterns \shortcite{banerjee2003hierarchical}.  Ignoring this structure can lead to inefficient sampling and incoherent predictions. 

Despite the increasing use of GPR in public health applications, current practice mainly focuses on prediction, calibration, or spatial risk mapping in limited policy spaces \shortcite{senanayake2016aaai, zimmer2020inficml}. No existing framework integrates spatially-aware GPR with a sequential design that scales to large policy spaces, which is needed to address the challenge of estimating \rev{treatment effects of opioid intervention policies across counties, where each of the two intervention components can be applied at multiple 
dispensing rate levels}. 
In this paper, we propose a two-stage metamodel, \rev{so named because it consists of two modeling stages:} the GPR as the first stage to model contextual variability of treatment effects across counties, and a response function as the second stage to map the GPR predictions to overdose mortality outcomes. We design a two-step sequential design that allocates simulation runs across counties and treatment conditions to maximize information gain.


\subsection{Related Work}

GPR is commonly used in epidemiological settings and for emulating computationally intensive simulations in public health, as illustrated in the following paragraphs;  
yet its application to opioid epidemic remains largely unexplored.

For example, \citeNP{langmuller2024gaussian} develop GPR surrogates to emulate dengue outbreak simulations across an eight-dimensional parameter space, enabling efficient evaluation of outbreak probability and epidemic duration. Similarly, \citeNP{sawe2024gaussian} employ GPR to approximate multi-disease agent-based simulations in Kenya, reducing computation time more than ten-fold while preserving predictive accuracy. In the context of malaria, \citeNP{reiker2021emulator} use GPR-based Bayesian optimization to calibrate high-dimensional transmission models, highlighting its value for accelerating policy-relevant inference. Influenza forecasting studies further illustrate the utility of GPR for spatio-temporal epidemic prediction \shortcite{senanayake2016aaai,zimmer2020influenza}. In more recent work, \citeNP{ahmed2023inferring} 
demonstrate the potential of GPR for capturing differences in spatial distribution of outcomes for different diseases, which can be attributed to differences in their underlying epidemic dynamics, when other confounding factors such as population size, location, and contact rates are kept identical in simulations using synthetic populations. However, their approach relies on fitting a single-output GPR surrogate to a single county's population and is not  applicable for evaluating the heterogeneous effects of multi-component interventions in different counties.

A parallel stream of work applies spatial GPR to disease risk mapping and inference. Classic geostatistical methods \shortcite{banerjee2003hierarchical,moraga2023spatial} and large-scale malaria risk maps \shortcite{bhatt2017improved} show the strength of spatial kernels for interpolating across heterogeneous regions. However, these studies primarily focus on observational prediction and mapping rather than emulating county-level policy simulations. Spatial structure has rarely been integrated with surrogate modeling for exponentially large policy spaces.  Appendix \ref{app:related_work} provides more detailed related work on the application of GPR in epidemiological modeling and public health.

Another relevant strand of work centers on healthcare and biomedical applications of GPR to provide flexible function approximation and uncertainty quantification. For example, GPR has been applied to ICU monitoring \shortcite{cheng2020sparse}, pharmacology, and to predict dose-response curves \shortcite{gutierrez2024multi}. These studies demonstrate the advantages of modeling correlated outputs but do not address spatial heterogeneity or simulation-based policy learning. 

Additionally, sequential design and active learning methods are well established for simulation emulation and Bayesian optimization \shortcite{frazier2018bayesian,wilson2018maximizing}. Approaches such as expected improvement, predictive variance, and entropy reduction have been applied broadly, and recent work explores active learning with multi-output surrogates \shortcite{li2022safe}. Yet, in healthcare applications, these methods are generally used for calibration or parameter tuning, not for the hierarchical problem of allocating simulation runs across counties and treatment conditions. 

In summary, existing literature demonstrates the utility of GPR for epidemic simulation, spatial prediction, and correlated outputs. However, no framework to date integrates spatially-aware GPR with a hierarchical sequential design that efficiently explores exponentially many opioid intervention combinations across heterogeneous counties. Our work addresses this gap by developing a two-stage metamodeling framework that combines GPR with a response function and introduces a two-step sequential design for allocating simulations across both counties and interventions.

\subsection{Main Contributions}
In this work, we develop a novel, sample-efficient metamodeling framework for estimating opioid intervention effects across multiple counties, that integrates spatial GPR, linear outcome models, and a two-step sequential design strategy for selective simulation. While our analysis and results are tailored to the opioid epidemic in Pennsylvania, the proposed analytical framework readily generalizes to other states and intervention modalities.
Our main methodological contributions are as follows:

\textbf{(1) GPR-based modeling of spatially varying response-function coefficients.}
 We extend traditional GPR metamodeling to a spatially-structured setting that captures demographic heterogeneity and geographic correlation across counties. Each county is encoded by its centroid coordinates and a set of socio-economic features, such as income, racial composition, and population density, forming a high-dimensional input space. The model outputs county-specific {\em coefficients} for a response function, representing the {\em treatment effect}, which translates treatment levels into predicted overdose death rates. Each coefficient of the response function is modeled by its own Gaussian process defined over the same spatial and socio-economic input space, producing cross-sectional mortality predictions for a target time point (e.g., at the end of a five-year period over which interventions are planned).

In addition, we incorporate a heteroscedastic noise model in which the observation variance of each coefficient is estimated from the sample variance and the number of simulation replicates selected for each county. This formulation naturally captures county-specific uncertainty, arising from differences in population size, urban–rural structure, and other socio-economic factors, and allows the metamodel to weight observations according to their estimated precision, driven by the number of simulation replicates and the resulting regression-coefficient variance at each county, yielding more reliable posterior estimates than a homoscedastic specification. Our kernel design combines multiple radial basis function kernels, allowing the model to represent smooth spatial variation across regions.

\textbf{(2) Two-step sequential design for efficient sampling.} To 
efficiently sample the input space of multi-component interventions across multiple counties, we develop a two-step sequential design procedure: In the first step, we select which counties to simulate based on their posterior uncertainty in the GPR model. Specifically, we adapt the signal-to-noise ratio (SNR) criterion by using the ratio of posterior standard deviation to posterior mean as the acquisition function to prioritize counties with the most uncertain model estimates. In the second step, for the selected county, we choose the treatment condition whose predicted outcome has the most posterior uncertainty, as measured by the width of its credible interval. This sequential design enables us to efficiently target simulations to regions where they are most needed, thereby accelerating convergence while maintaining model accuracy.

The remainder of the paper is organized as follows. Section~\ref{sec:model} introduces the two-stage metamodel that 
uses Gaussian process regression to estimate the coefficients of a response function and learn their contextual dependencies on county-specific features.
Section~\ref{sec:sqd_snr} describes the two-step sequential design framework, 
detailing how county and treatment-condition sampling are integrated under a unified two-step procedure. Section~\ref{sec:res} presents empirical results on model performance, kernel design, and sample complexity, including learning curves and estimated treatment effects using a calibrated model of opioid use disorder for counties in Pennsylvania. Finally, Section~\ref{sec:conclude} concludes with a summary of our findings, implications for policy evaluation, and directions for future work.

\section{Problem Formulation and Modeling Framework}
\label{sec:model}
Policymakers responding to the opioid epidemic face a fundamental challenge: identifying which intervention combinations work best for each community. This goal, tailoring public health strategies to local conditions rather than applying uniform policies, reflects the emerging paradigm of precision public health. Achieving it requires evaluating how different communities respond to different combinations of interventions such as naloxone distribution and buprenorphine treatment access. \rev{We focus on naloxone and buprenorphine as they represent two widely deployed, 
evidence-based interventions for the opioid crisis. Naloxone is a harm-reduction strategy that reverses opioid overdoses and has been shown to be effective at reducing overdose mortality \shortcite{naumann2019naloxone, walley2013opioid}. Buprenorphine is a medication for opioid use disorder that reduces disease 
burden by supporting recovery \shortcite{wen2017bupre, nosyk2024buprenorphine}.}

\rev{To avoid ambiguity, we clarify the terminology used throughout the paper.  \emph{Multi-level, multi-component interventions} refers to the policy design space, in which each of several intervention components (here, naloxone and buprenorphine) is applied at one of several dispensing-rate levels, where each level denotes a fixed proportional increase relative to a county's baseline dispensing rate, representing a comparable scale-up effort across counties. The \emph{two-stage metamodel} refers to the two-stage structure of the surrogate: a contextual stage, where GPR learns county-specific coefficients, and an outcome stage, where the response function maps treatment conditions to predicted overdose mortality. Finally, the \emph{two-step sequential design} refers to the sampling procedure, which first selects a county and then selects a treatment condition for the chosen county.}

A brute-force approach would calibrate a simulation model for each county independently and exhaustively evaluate all possible policy combinations. This is computationally inefficient. \rev{When each of two interventions is applied at five levels, the result is $5 \times 5 = 25$ distinct treatment conditions  per county. More generally, for $J$ interventions each applied at $\ell$ levels, the number of treatment conditions grows exponentially as $\ell^J$. Each condition requires hundreds of simulation replicates to reduce stochastic variability, so evaluating all conditions in a state such as Pennsylvania with 67 counties would demand hundreds of thousands of runs.} The problem intensifies as the intervention space expands: six interventions at seven levels each would yield $7^6$ combinations.
This challenge generalizes beyond opioid modeling. Any setting involving simulation-based policy evaluation across heterogeneous subgroups (counties, demographic strata, healthcare facilities) with multi-dimensional treatment spaces faces the same combinatorial barrier: The subgroups differ in baseline characteristics, the interventions operate at multiple levels, and the treatment response varies across both subgroups and intervention combinations.

Our goal is to develop a modeling framework that efficiently estimates subgroup-level treatment effects across a high-dimensional intervention space without exhaustively simulating every subgroup-treatment combination. 
Specifically, the framework must: (1) generalize across subgroups by learning how baseline characteristics shape treatment response, (2) interpolate across treatment levels to predict outcomes for less simulated and unsimulated conditions, and (3) quantify predictive uncertainty to guide sequential allocation of simulation effort.


\subsection{Two-stage Metamodel: Response Function and GPR Modeling}
\label{subsec:metamodel}
{We propose a two-stage metamodel to map county-specific features to outcomes across various treatment conditions. Unlike conventional approaches that learn the relationship for each of the $\ell^J$ treatment conditions separately and thus suffer from the curse of dimensionality, our framework, in the contextual stage,  employs GPR to learn county-specific coefficients for a response function; and then, in the outcome stage, it computes county-level outcomes for any treatment condition using the response function with the county-specific coefficients learned in the contextual stage.}
We denote counties by \rev{$c \in \mathcal{C}$, where $\mathcal{C}$ is the set of all counties,} and treatment conditions by $(n,b)$, where $n$ and $b$ take integer values in $\{1,2,\ldots,\ell\}$ and encode naloxone and buprenorphine levels. The implementation and numerical results in Section~\ref{sec:res} are developed for the set $\mathcal{C}$ of $67$ Pennsylvania counties and with $\ell = 5$ levels at each intervention arm. \rev{The choice of $\ell=5$ levels was guided by domain experts from our modeling team to span a realistic range of policy-relevant scale-up scenarios}. 

Our outcome of interest is the opioid overdose death rate, measured as deaths per 100{,}000 people. Rather than modeling this outcome separately for every treatment condition, we express it through a parametric response function whose coefficients vary across counties. 
We use the following linear response function:
\begin{equation}
z(n,b \mid c) = \mu_0(\mathbf{x}_c) + \mu_n(\mathbf{x}_c) \cdot n + \mu_b(\mathbf{x}_c) \cdot b.
\label{eq:response_surface2}
\end{equation} 
In our two-stage metamodel framework, we refer to the response function $z(n,b \mid c)$ as the {\em outcome stage}, which maps treatment condition $(n,b)$ to the outcome of interest (predicted overdose mortality) for each county $c$. We adopt a main-effects specification after examining factorial plots, which indicate no interaction between naloxone and buprenorphine across counties; details are provided in Appendix~\ref{sec:app:model-select}.

The coefficients of the response function are learned using a GPR model defined over spatial and socio-economic county features. The coefficients are expressed as a function of $\mathbf{x}_c \in\mathbb{R}^d$, which is a feature vector of spatial and socio-economic covariates for county $c$. For each county $c$, the coefficient vector in \eqref{eq:response_surface2},
$$
\boldsymbol{\mu}(\mathbf{x}_c) = [
\mu_0(\mathbf{x}_c), \ 
\mu_n(\mathbf{x}_c), \ 
\mu_b(\mathbf{x}_c)]^\top,$$
\rev{represents the {\em contextual stage}} and is learned as the posterior mean functions of three GPRs evaluated at $\mathbf{x}_c$.

This construction yields a two-stage metamodel: a {\em contextual stage}, which models how response-function parameters depend on county characteristics, and an {\em outcome stage}, which maps treatment levels $(n,b)$ to predicted overdose mortality using the response function.

\subsection{Contextual Modeling of Subgroup Heterogeneity using GPR}
\label{sec:model_def}
In the first stage, we use GPR to learn the contextual dependencies of $\mu_0$, $\mu_n$, and $\mu_b$ on county-specific spatial and socio-economic features $\mathbf{x}_c$ in Equation~\eqref{eq:response_surface2}. 
Specifically, to each coefficient $\mu_m,m\in\{0,n,b\}$ of the response function we associate a Gaussian process $\mathcal{GP}\!\left(
{\mu_m}(\cdot),
k_m(\cdot,\cdot)
\right)$, whose mean function evaluated at $\mathbf{x}_c$ gives the coefficient value for county $c$. The kernel function $k_m(\cdot,\cdot)$ determines the variance-covariance relations between county estimates based on their spatial and socio-economic features. Common kernels such as radial basis function have a width hyperparameter that controls similarities between county responses based on their covariates and is optimized separately (using maximum likelihood or other fit criteria). More complex kernels can be constructed as a composition of simpler kernels, for example, through addition to capture independent effects or multiplication to encode interactions among input features \cite[Chapter~4]{rasmussen2006gaussian}. In our implementation, the kernel encodes spatial and demographic similarity via a combination of multiple radial basis functions, defined in Appendix Equations \eqref{eq:RBF} and \eqref{eq:kernelcomposition}. Further discussion of the kernel design for our study is provided in Appendix~\ref{sec:app:kernel-design}.

GPR uses Bayes’ rule to yield a posterior distribution over the response-function coefficients. The posterior provides both mean estimates and uncertainty for each coefficient, with uncertainty contracting as additional simulation runs are collected for a county. These posterior summaries are subsequently propagated through the outcome-stage response function to generate uncertainty-aware predictions of overdose mortality. 

Specifically, for a new county $\mathbf{x}_{c^*}$, the posterior mean and variance are given by $\mu_m(\mathbf{x}_{c^*}) = \mathbf{k}_{c^*}^\top(\mathbf{K}_m + \mathbf{T}_m)^{-1}\mathbf{y}_m$ and $\sigma^2_m(\mathbf{x}_{c^*}) = k_m(\mathbf{x}_{c^*}, \mathbf{x}_{c^*}) - \mathbf{k}_{c^*}^\top(\mathbf{K}_m + \mathbf{T}_m)^{-1}\mathbf{k}_{c^*}$, where $\mathbf{K}_m$ is the kernel matrix evaluated at sampled counties, $\mathbf{k}_{c^*}$ is the vector of covariances (kernel evaluations) between $\mathbf{x}_{c^*}$ and all the sampled counties (training points), $\mathbf{y}_m$ is the vector of observed coefficient values for the sampled counties, and $\mathbf{T}_m = \mbox{diag}([\tau^2_m(\mathbf{x}_{c_1}), \ldots, \tau^2_m(\mathbf{x}_{c_N})])$ is the heteroscedastic noise matrix for output $m$ over the $N$ counties. 
As a new sample is collected, the posterior is updated by augmenting the training set and recomputing the above expressions, with hyperparameters re-optimized using marginal likelihood maximization \cite[Chapter~2]{rasmussen2006gaussian}.

\rev{To train the Gaussian process, we derive the observations of the response-function coefficients from the simulation outputs. We first evaluate the outcome for given treatment conditions $(n,b)$ and a given county $c$ using the simulation model. For example, county-level opioid overdose mortality can be simulated using the Framework for Reconstructing Epidemiological Dynamics (FRED), an open-source, agent-based simulation platform that models disease spread across census-derived synthetic populations ( see \citeNP{grefenstette2013fred}; full details of the platform and our opioid use disorder model are provided in Appendix~\ref{sec:fred-oud}). For a fixed county $c$, we fit a linear regression model to the simulated outcome based on the selected treatment conditions $(n,b)$:} 
\begin{align}
r(n,b \mid c) \;=\; \beta_{0,c} + \beta_{n,c} \cdot n + \beta_{b,c} \cdot b ,
\label{eq:regression-estimates}
\end{align}
where $\beta_{0,c}$, $\beta_{n,c}$, and $\beta_{b,c}$ are county-specific regression coefficients. Each set of simulation runs for county $c$ therefore yields updated estimates of these coefficients. 


Given the regression-based coefficient estimates as observations, GP training involves two distinct phases. First, the kernel hyperparameters $\boldsymbol{\theta}$ are optimized by maximizing the marginal 
log-likelihood over the training inputs $\mathbf{X} = [\mathbf{x}_{c_1}, \ldots, \mathbf{x}_{c_N}]^\top \in \mathbb{R}^{N \times d}$,
\begin{equation*}
\log p(\mathbf{y}_m \mid \mathbf{X}, \boldsymbol{\theta}) = 
-\frac{1}{2}\mathbf{y}_m^\top(\mathbf{K}_m + 
\mathbf{T}_m)^{-1}\mathbf{y}_m 
- \frac{1}{2}\log|\mathbf{K}_m + \mathbf{T}_m| 
- \frac{N}{2}\log 2\pi,
\end{equation*}
via \rev{the limited-memory Broyden--Fletcher--Goldfarb--Shanno (L-BFGS) algorithm} through \texttt{BoTorch}'s \texttt{fit\_gpytorch\_mll} 
routine \rev{\shortcite{balandat2020botorch} (version 0.17.0)}, and this re-optimization is performed at every 
\rev{iteration of the sequential design (introduced later in Section \ref{sec:sqd_snr})} as new county--condition observations are added to the training 
set. Second, given the optimized hyperparameters, the GPR posterior over 
the response-function coefficients is updated analytically via Bayes' 
rule. For each county $c$ with feature vector $\mathbf{x}_c \in 
\mathbb{R}^d$, the posterior mean $\mu_m(\mathbf{x}_c)$ and variance 
$\sigma^2_m(\mathbf{x}_c)$ for each coefficient $m \in \{0, n, b\}$ are given by $\mu_m(\mathbf{x}_c) = \mathbf{k}_c^\top 
(\mathbf{K}_{m} + \mathbf{T}_m)^{-1} \mathbf{y}_m$ 
and $\sigma^2_m(\mathbf{x}_c) = k_m(\mathbf{x}_c, \mathbf{x}_c) - 
\mathbf{k}_c^\top (\mathbf{K}_m + 
\mathbf{T}_m)^{-1} \mathbf{k}_c$,
where $\mathbf{K}_m$ is the kernel matrix evaluated over all training/sampled counties, $\mathbf{T}_m$ is the diagonal heteroscedastic noise matrix derived from the regression variances, $\mathbf{k}_c$ is the vector of kernel evaluations between $\mathbf{x}_c$ and all training points, and $\mathbf{y}_m$ is the vector of observed regression coefficients for output $m$. Together, these posterior means form the updated coefficient vector $\boldsymbol{\mu}(\mathbf{x}_c) = [\mu_0(\mathbf{x}_c),\, 
\mu_n(\mathbf{x}_c),\, \mu_b(\mathbf{x}_c)]^\top$.

We adopt an independent-output formulation, in which each output component  $\mu_m(\mathbf{x}_c)$ is modeled by its GP over the same $d$-dimensional spatial–socioeconomic feature space $\mathbf{x}_c\in\mathbb{R}^d$. 
This choice simplifies the implementation while remaining well-justified: the GP models the coefficients $\beta_0, \beta_n, \beta_b$ of a linear regression refit at each iteration from all observed simulation runs for the selected county, so the coefficients are inherently correlated 
through the joint regression fit and are further combined jointly through the response function~\eqref{eq:response_surface2} at the outcome stage. Consequently, although the three GPs are fit independently at the contextual stage, the sequential design operates on joint outcome predictions rather than individual coefficients in isolation. Empirical evidence from the analysis of pairwise correlations among iterative coefficient updates is provided in Section~\ref{sec:res}.

The regression coefficients obtained from simulation runs are modeled as noisy observations of the GP outputs. 
For county $c$, the observation model for estimating coefficient $\mu_m(\mathbf{x}_c)$ is 
$ {y} \mid {f}(\mathbf{x}_c)
\sim
\mathcal{N}\!\left(
{f}(\mathbf{x}_c),
\tau_m^2(\mathbf{x}_c)
\right),
$
where ${f}(\mathbf{x}_c)$ is the output drawn from $\mathcal{GP}\!\left(
{\mu_m}(\cdot),
k_m(\cdot,\cdot)
\right)$ at point $\mathbf{x}_c \in \mathbb{R}^d$. The output noise
$\tau^2_m(\mathbf{x}_c)$ models the heteroscedastic uncertainty arising
from finite simulation replicates and the variability of the regression-based estimates obtained from Equation~\eqref{eq:regression-estimates}. 
Specifically, for each county $c$ and coefficient $\mu_m(\mathbf{x}_c)$, the noise variance is set equal to the estimated variance of the corresponding regression coefficient, $
\tau^2_m(\mathbf{x}_c) \;\propto\;
\mathrm{Var}(\widehat{\beta}_{m,c})\,/\,R_c,$ where $\widehat{\beta}_{m,c}$ is the estimated regression coefficient for output $m$ in county $c$, $\mathrm{Var}(\widehat{\beta}_{m,c})$ is its regression-based variance, and $R_c$ denotes the number of simulation replicates selected for county $c$.

As additional replicates are collected, $\tau^2_m(\mathbf{x}_c)$ decreases, enabling the GPR to represent heterogeneity across counties in a variance-aware manner, rather than treating all counties as equally informative. In contrast to a homoscedastic specification, which assumes a constant noise level across counties and coefficients, the heteroscedastic formulation adopted here explicitly links observation noise to the number of simulation replicates used to estimate each regression coefficient. As a result, uncertainty from the regression-based estimation propagates coherently through the GPR and into the final outcome-stage predictions.

\noindent

Given that these simulations are computationally expensive, how can we efficiently estimate effects across high-dimensional spaces under a limited simulation budget? We address this challenge by adopting a sequential design approach that leverages model uncertainty to guide sampling decisions. In the next section, we describe our two-step sequential design that jointly allocates simulation effort across both counties and treatment conditions.

\section{Two-Step Sequential Design for Joint Sampling of Counties and Treatment Conditions}
\label{sec:sqd_snr}

\begin{figure}[h!]
    \centering
    \includegraphics[width=0.95\linewidth]{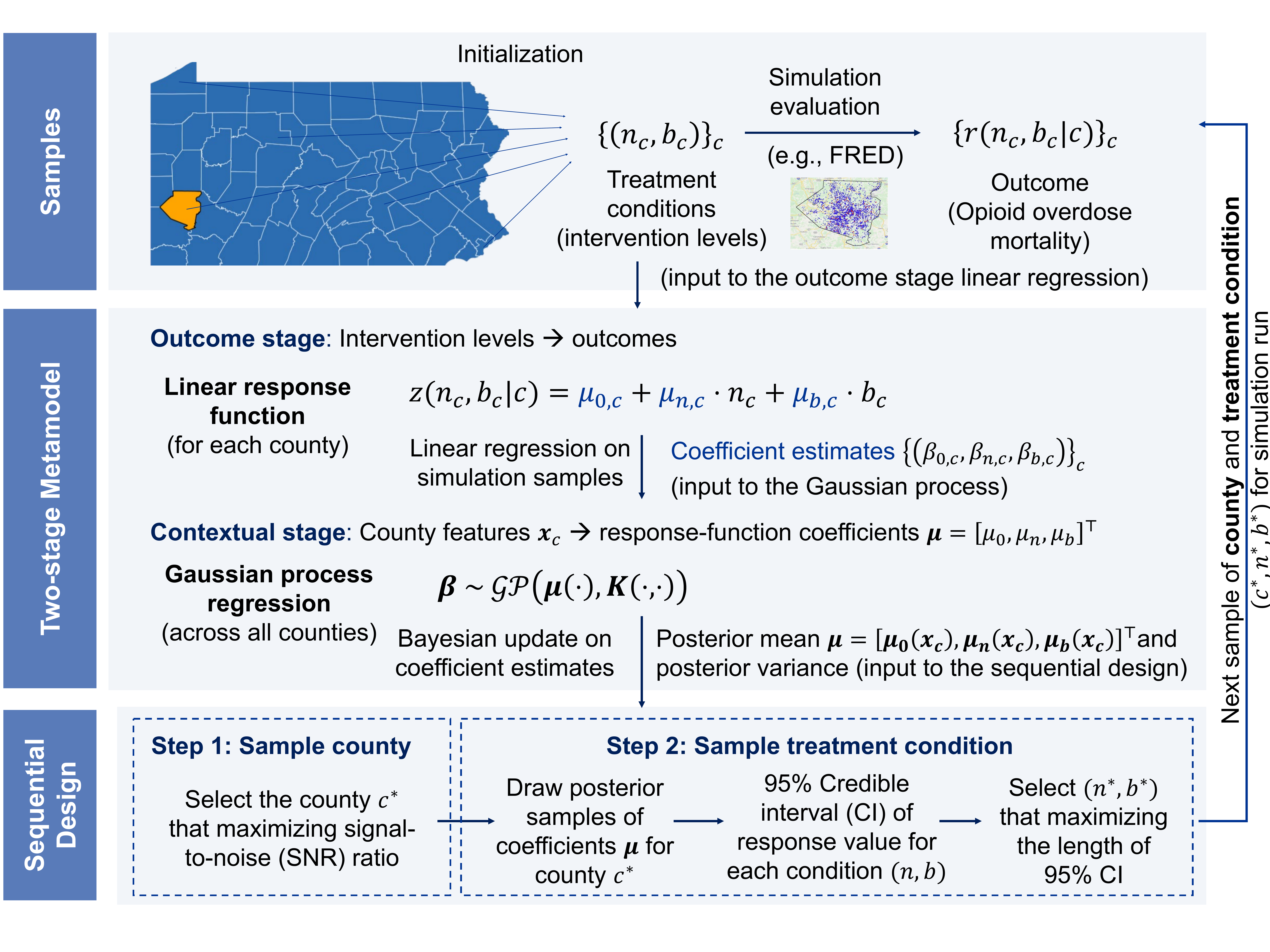}
    \caption{
    \rev{\textbf{An overview of the proposed two-stage metamodeling and two-step sequential design framework for evaluating opioid interventions across Pennsylvania counties.} Simulation outcomes (top) are summarized into response-function coefficients by linear regression, which serve as observations for the contextual-stage GPR. The GPR posterior drives a two-step sequential design that selects the next county (step~1, by the SNR-based acquisition function) and treatment condition (step~2, by widest credible interval), whose simulation closes the loop. }
    }
    \label{fig:overview-figure}
\end{figure}

In large-scale simulation settings, where evaluating every input configuration 
\rev{to construct a metamodel for the simulation} is computationally prohibitive, sequential design strategies enable efficient learning by guiding the sampling process to the most informative regions of the input space. 
\rev{Figure \ref{fig:overview-figure} provides a high-level overview for the overall framework of learning the two-stage metamodel with a two-step sequential design. Samples of treatment conditions for a given county are evaluated using the county-specific simulation model (here the FRED model) to obtain the outcome (here the opioid overdose deaths). These outcomes are then used to obtain the estimates of the coefficients in the response function (i.e., the outcome stage of the two-stage metamodel) by fitting a linear regression model. The estimated coefficients for each county are then provided to the GPR (i.e., the contextual stage) as training observations. The GPR uses county features through its kernel to borrow information across similar counties when predicting the response-function coefficients. Based on the posterior of the updated GPR model, the sequential design selects the next sample of county (step 1) and treatment condition (step 2) for the simulation to evaluate, focusing on the regions in the input space that are more uncertain and thus could benefit the most from additional observations. With the new sampled county-treatment combination and its simulation outcome, we refit the linear model to update the estimates of the response-function coefficients for the selected county, which are then used to update the GPR model, and the process repeats. In the following, we elaborate on the step one of  the sequential design (sampling counties) in Section \ref{subsec:acquisition} and step two (treatment condition selection) in Section \ref{subsec:treatment_select}, and the overall algorithmic workflow in Section \ref{subsec:workflow}.}


\subsection{First-Step Sequential Design for Sampling Counties}
\label{subsec:acquisition}

To allocate simulation resources efficiently, we prioritize sampling from counties where the metamodel exhibits high predictive uncertainty relative to the expected outcome. The GPR kernel guides both the similarity structure and uncertainty estimation, thereby influencing the sequential sampling trajectory for counties. We use the signal-to-noise ratio (SNR) criterion for our acquisition function to determine which counties to sample from. This strategy is designed to improve global model accuracy across a high-dimensional spatial domain by focusing computational resources on regions where the model is least certain relative to the scale of the predicted effect.

\textbf{Acquisition function formulation.} For a given county $c$, the GPR posterior yields estimates of the response-function parameters, consisting of the posterior mean vector $\boldsymbol{\mu}(\mathbf{x}_c) = [\mu_0(\mathbf{x}_c), \mu_n(\mathbf{x}_c), \mu_b(\mathbf{x}_c)]^\top$ and the associated diagonal matrix of posterior variances $\boldsymbol{\Sigma}(\mathbf{x}_c) = \mbox{diag}([\sigma^2_0(\mathbf{x}_c), \sigma^2_n(\mathbf{x}_c), \sigma^2_b(\mathbf{x}_c)]^\top)$. In our sequential design framework, the acquisition function is calculated directly from the posterior mean and covariance.

To compute a single acquisition value from these three posteriors, we employ a scalarized posterior transform. This transform applies a fixed weight vector $\mathbf{w} = [1/3, 1/3, 1/3]^\top$ to the posterior distribution, effectively averaging the predictions across all three parameters:
\begin{equation*}
\mu = \mathbf{w}^\top \boldsymbol{\mu}(\mathbf{x}_c) = \frac{1}{3}\mu_0(\mathbf{x}_c) + \frac{1}{3}\mu_n(\mathbf{x}_c) + \frac{1}{3}\mu_b(\mathbf{x}_c), \quad \sigma^2 = \mathbf{w}^\top \boldsymbol{\Sigma}(\mathbf{x}_c) \mathbf{w},
\end{equation*}where $\mu$ and $\sigma^2$ are the combined mean and variance. More generally, $\mathbf{w}$ may be specified to emphasize particular components of the response function, depending on modeling objectives or policy focus. The SNR-based acquisition function is then defined as:
\begin{equation}
\alpha_{\text{SNR}}(c) = \frac{\sigma}{\mu} = \frac{\sqrt{ \mathbf{w}^\top \boldsymbol{\Sigma}(\mathbf{x}_c) \mathbf{w} }}{\mathbf{w}^\top \boldsymbol{\mu}(\mathbf{x}_c)}.
\label{eq:snr}
\end{equation}
Note that $\alpha_{\text{SNR}}(c)$ in \eqref{eq:snr} is the coefficient of variation of the scalarized prediction (the reciprocal of the signal-to-noise ratio $\mu/\sigma$); maximizing this reciprocal form prioritizes counties whose predictive uncertainty is largest relative to the expected outcome.

\textbf{Sequential sampling procedure for counties.} 
At iteration $t$, the GPR posterior for county $c$ provides $\boldsymbol{\mu}_t(\mathbf{x}_c)$ and $\boldsymbol{\Sigma}_t(\mathbf{x}_c)$, where $\boldsymbol{\mu}_t(\mathbf{x}_c)$ and $\boldsymbol{\Sigma}_t(\mathbf{x}_c)$ denote the mean vector and the diagonal covariance matrix evaluated at the county feature vector $\mathbf{x}_c$ after incorporating all data available up to iteration $t$. These quantities are scalarized as $\mu_t(c) = \mathbf{w}^\top \boldsymbol{\mu}_t(\mathbf{x}_c)$ and
$\sigma_t(c) = (\mathbf{w}^\top \boldsymbol{\Sigma}_t(\mathbf{x}_c)\mathbf{w})^{1/2}$. 
The SNR-guided acquisition $\alpha_{\mathrm{SNR}}(c)=\sigma_t(c)/\mu_t(c)$
selects the next county using $c^*=\arg\max_{c\in\mathcal{C}}\alpha_{\mathrm{SNR}}(c)$. 
\rev{Ties in the argmax are resolved by selecting the lowest-indexed county, following the default behavior of the discrete acquisition optimizer; in practice exact ties do not arise, as the acquisition values are continuous}. The first-step sequential design is summarized below:

\begin{algorithm}[H]
\caption{First-Step Sequential Design for Sampling Counties (SNR-Guided)}\label{alg:county_sqd}
\begin{algorithmic}
    \Require Current posterior means $\boldsymbol{\mu}_t(\mathbf{x}_c)$ and posterior variance $\boldsymbol{\Sigma}_t(\mathbf{x}_c)$ 
    \State Compute scalarized mean and variance for each county $c$:
    \State \quad $\mu_t = \mathbf{w}^\top \boldsymbol{\mu}_t(\mathbf{x}_c)$
    \State \quad $\sigma_t = \sqrt{ \mathbf{w}^\top \boldsymbol{\Sigma}_t(\mathbf{x}_c) \mathbf{w} }$
    \State Compute $\alpha_{\,\text{SNR}}(c)=\sigma_t(c)/\mu_t(c)
    $
    \State Select county $c^* = \arg\max{_c} \; \alpha_{\text{SNR}}(c)$
\end{algorithmic}
\end{algorithm}

Based on this selection, we draw $S$ posterior samples in county $c^*$ and use them to construct credible intervals for treatment effects in county $c^*$, explained next in Section~\ref{subsec:treatment_select}. 

\subsection{Second-Step Sequential Design for Sampling Treatment Conditions}
\label{subsec:treatment_select}

At each iteration, the first-step sequential design identifies a county $c^*$ for additional sampling, represented by its feature vector $\mathbf{x}_{c^*}$.
However, evaluating all $ \ell \times \ell$ treatment combinations of $n$ and $b$ levels for the selected county is computationally inefficient. The second step of the sequential design helps us direct simulation effort toward the treatment condition whose predicted response has the widest $95\%$ credible interval. Given the selected county in the first step, the second-step sequential design chooses the most uncertain treatment condition within that county. To identify the most uncertain treatment condition at county $c$, we use the GPR posterior. 
 Specifically, instead of using $\boldsymbol{\mu}(\mathbf{x}_{c^*}) = [\mu_0(\mathbf{x}_{c^*}) \ \mu_n(\mathbf{x}_{c^*}) \ \mu_b(\mathbf{x}_{c^*})]^\top$ to estimate the overdose death rate for a specific treatment condition $(n,b)$ in county ${c^*}$ in equation~\eqref{eq:response_surface2}, we draw $S$ posterior samples
$\bigl\{\, f_0^{(s)}(\mathbf{x}_{c^*}),\, f_n^{(s)}(\mathbf{x}_{c^*}),\, f_b^{(s)}(\mathbf{x}_{c^*}) \,\bigr\}_{s=1}^S$,  and use each sample  as coefficients in the response function to generate posterior samples for the treatment effect estimates at each treatment condition $(n,b)$ in county ${c^*}$:
\begin{align}
\zeta^{(s)}(n,b \mid {c^*}) = f^{(s)}_0(\mathbf{x}_{c^*}) + f^{(s)}_n(\mathbf{x}_{c^*}) \cdot n + f^{(s)}_b(\mathbf{x}_{c^*}) \cdot b.
\label{eq:treatmnet_condition_posterior_samples}
\end{align} Here, we use the $S$ values $\bigl\{\zeta^{(s)}(n,b \mid {c^*})\bigr\}_{s=1}^S$ as posterior predictive samples for  $z(n,b|{c^*})$ in equation~\eqref{eq:response_surface2} 
to evaluate predictive uncertainty.
To that end, we compute empirical $95\%$ credible intervals:
\begin{align}
\text{CI}_{95}(n,b \mid {c^*}) = 
\bigl[
\text{Quantile}_{2.5\%}\{\zeta^{(s)}\}_{s=1}^S,
\text{Quantile}_{97.5\%}\{\zeta^{(s)}\}_{s=1}^S
\bigr],
\label{eq:posteirorCI}
\end{align}
with interval width
\begin{align}
\mbox{width}(\text{CI}_{95}(n,b \mid {c^*})) = \text{Quantile}_{97.5\%}\{\zeta^{(s)}\}_{s=1}^S - \text{Quantile}_{2.5\%}\{\zeta^{(s)}\}_{s=1}^S,
\label{eq:CIwidth}
\end{align}
 and choose the treatment condition with the widest credible interval (Algorithm~\ref{alg:tc_sqd}):
\[
(n^*, b^*) = \arg \max_{(n,b)} \{ \mbox{width}(\text{CI}_{95}(n,b \mid {c^*}))\}.
\]

\begin{algorithm}[h]
\caption{Second-Step Sequential Design for Sampling Treatment Conditions}
\label{alg:tc_sqd}
\begin{algorithmic}
\Require feature vector $\mathbf{x}_{c^*}$ for the county $c^*$ selected by Algorithm~\ref{alg:county_sqd}
    \State Draw $S$ posterior samples $\{f_0^{(s)}(\mathbf{x}_c^*),  f_n^{(s)}(\mathbf{x}_c^*),  f_b^{(s)}(\mathbf{x}_c^*)\}_{s=1}^S$
    \For{each treatment condition $(n,b)$}
        \State Compute predicted outcomes $\zeta^{(s)}(n,b \mid c^*)$ according to equation~\eqref{eq:treatmnet_condition_posterior_samples}
        \State Estimate $95\%$ credible interval $\text{CI}_{95}(n,b \mid c^*)$ using equation~\eqref{eq:posteirorCI}
        \State Compute $\mbox{width}(\text{CI}_{95}(n,b \mid c^*))$ using equation~\eqref{eq:CIwidth}
    \EndFor
    \State Select condition $(n^*, b^*) = \arg\max_{(n,b)} \mbox{width}(\text{CI}_{95}(n,b \mid c^*))$
\end{algorithmic}
\end{algorithm}
Together, Algorithms \ref{alg:county_sqd} and \ref{alg:tc_sqd} constitute our two-step sequential design framework that is grounded in GPR posterior sampling and credible-interval selection and enables efficient allocation of simulations to the most informative counties and treatment conditions to obtain a superior metamodel fit under tight computational constraints. 

\subsection{Two-stage Metamodel Workflow and Simulation Output Integration}
\label{subsec:workflow}

During initialization, we allocate relatively more simulation replications to the baseline treatment condition $(n,b)=(0,0)$ than to other conditions within the initial batch. \rev{Specifically, the initial batch uses one replication per county-condition, increased to five replications for the baseline condition.} This condition establishes baseline overdose mortality and contributes disproportionately to early prediction error, so improved estimation at this point stabilizes subsequent learning of treatment effects.
After each simulation batch, the resulting outputs are summarized at the county level by fitting a linear regression to estimate the response-function coefficients. These regression-based coefficient estimates are then incorporated into the contextual-stage GPR, which is implemented using the \texttt{BoTorch} framework \shortcite{balandat2020botorch}. 


At the end of each iteration, the GPR posterior is updated with the new regression-based coefficient estimates (as explained in Section \ref{sec:model_def}). Each element of $\boldsymbol{\mu}(\mathbf{x}_c)$ corresponds to the posterior mean of a response-function coefficient for the county characterized by feature vector $\mathbf{x}_c$. In the second stage, these coefficients are plugged into the response function to get predicted overdose death rates for any treatment condition. 
\noindent
Sequential design proceeds in two steps, with the first step selecting the next county using an SNR-based acquisition rule, evaluated from the GPR posterior mean and variance. For that specific county $c^*$, the second step uses posterior samples from the GPR to form credible intervals over all treatment conditions, then selects the single condition with the widest interval to run additional simulation samples. The simulation outcomes from these runs are then used to fit the linear regression in Equation~\ref{eq:regression-estimates}, yielding updated coefficients for the baseline overdose mortality rate and naloxone and buprenorphine effects. These regression coefficients are then appended to the training dataset to update the GPR posterior using the same \texttt{BoTorch} implementation. This loop continues until the simulation sample budget is exhausted or the estimates stabilize. 
A complete summary of notations used in this workflow is provided in Table~\ref{tab:notation}.

\begin{algorithm}[H]
\caption{Two-stage Metamodel Workflow}
\label{alg:workflow}
\begin{algorithmic}
\State Initialize GPR prior with spatial–socio-economic kernel
\State Initialize with a small set of county–treatment simulations
\Repeat
    \State \textbf{county selection:} Evaluate the SNR-based acquisition function over candidate counties, select the county c* using Algorithm \ref{alg:county_sqd}
    \State \textbf{posterior sampling:} For county $c^*$ with feature vector \(\mathbf{x}_{c^*}\), draw $S$ posterior samples $\bigl\{\, f_0^{(s)}(\mathbf{x}_{c^*}),\, f_n^{(s)}(\mathbf{x}_{c^*}),\, f_b^{(s)}(\mathbf{x}_{c^*}) \,\bigr\}_{s=1}^S$ from the GPR
    \State \textbf{treatment selection:} For each treatment pair \((n,b)\), compute the empirical 95\% credible interval; select \((n^*, b^*)\) with the widest interval using Algorithm \ref{alg:tc_sqd}
    \State \textbf{Simulation and augmentation:} Run the simulation at \((c^*, n^*, b^*)\) for a number of replications and append the new observations to the training set
    \State \textbf{Linear regression:} Fit $r(n^*,b^* \mid {c^*}) = \beta_{0,{c^*}} + \beta_{n^*,{c^*}}\,n + \beta_{b^*,{c^*}}\,b$ using all observed runs for county ${c^*}$, and obtain the coefficient vector $\boldsymbol\beta_{c^*} = [\beta_{0,{c^*}} , \beta_{n^*,{c^*}}, \beta_{b^*,{c^*}}]$
    \State \textbf{model update:} update the \texttt{BoTorch} GPR using $\boldsymbol\beta_{c^*}$ as the observation
    
\Until{the sample budget is reached or the relative error stabilizes}
\end{algorithmic}
\end{algorithm}

\begin{table}[h!]
\centering
\caption{ {\bf Summary of notation used in the paper.} By convention, bold lowercase letters (e.g., $\mathbf{x}$) denote vectors, bold uppercase letters (e.g., $\mathbf{K}$) denote matrices, and Greek letters (e.g., $\mu,\tau, \Sigma$) denote GPR parameters. 
}
\begin{tabular}{ll}
\hline
Symbol & Description \\
\hline
$J$ & Number of interventions (factors) \\
$\ell$ & Number of dispensing rate levels per intervention \\
$\mathcal{C}$ & Set of all counties, with $|\mathcal{C}| = N$ total counties \\
$t$ & Iteration index in the sequential design algorithm \\
$(n,b)$ & Treatment condition defined by naloxone level $n$ 
\\ & and buprenorphine level $b$ \\
$\mathbf{x}_c \in \mathbb{R}^d$ & Feature vector for county $c$ (dimension $d$) \\
$\beta_0, \beta_n, \beta_b$ & Linear regression coefficients \\
$r(n,b \mid c)$ & Linear regression function $\beta_0 + \beta_n n + \beta_b b$ for county $c$ \\
$z(n,b \mid c)$ & Response function $\mu_0 + \mu_n n + \mu_b b$ for county $c$ \\
$\zeta^{(s)}(n,b \mid c)$ & Posterior samples of $z(n,b \mid c)$ (indexed by $s$) \\
$\sigma^2_m(\mathbf{x}_c)$ & Posterior variance of coefficient $m$ for county $c$ \\
$\tau^2_m(\mathbf{x}_c)$ & Observation-noise variance of coefficient $m$ for county $c$ \\
$\boldsymbol{\Sigma}(\mathbf{x}_c) = \mbox{diag}[\sigma^2_0(\mathbf{x}_c), \sigma^2_n(\mathbf{x}_c), \sigma^2_b(\mathbf{x}_c)]^\top$ & Diagonal matrix of posterior variances\\ & for county $c$ in the heteroscedastic GPR \\
$\boldsymbol{\mu}(\mathbf{x}_c)=[\mu_0(\mathbf{x}_c),\mu_n(\mathbf{x}_c),\mu_b(\mathbf{x}_c)]^\top$ & Response function coefficients modeled by \\ & GPR posterior means \\
$\mathbf{T}_m = \mbox{diag}([\tau^2_m(\mathbf{x}_{c_1}), \ldots, \tau^2_m(\mathbf{x}_{c_N})])$ & \small{Diagonal observation-noise matrix for output $m$ over $N$ counties} \\
\hline
\end{tabular}
\label{tab:notation}
\end{table}

\section{Implementation and Numerical Results}
\label{sec:res}



Our outcome of interest is the cumulative number of overdose deaths over a five-year horizon for each county and treatment condition. We calibrated our OUD model to reproduce county-level opioid mortality patterns over a five-year pre-pandemic study period (2015-2019), for which we had county-level outcomes and covariate data available. Six counties, Allegheny, Philadelphia, Dauphin, Erie, Columbia, and Clearfield, were calibrated using incremental mixture importance sampling \shortcite{menzies2017bayesian}. These counties were then used as prototypes to generalize model parameters across the remaining Pennsylvania counties using a feature-based matching procedure. 

The matching transfers calibrated model parameters governing outcome progression and behavioral dynamics to counties with similar historical overdose mortality and dispensing-rate trajectories, while county-specific naloxone and buprenorphine dispensing rates are preserved and applied independently as local inputs. This approach avoids the infeasibility of full calibration for all 67 counties while retaining county-level heterogeneity in treatment exposure. Further details of the FRED simulation platform and the OUD model structure are provided in Appendix~\ref{sec:fred-oud}, calibration \rev{and county matching} details are given in Appendix~\ref{sec:app:calibration}, and the computational infrastructure and implementation details are described in Appendix~\ref{sec:implementation}.

To rigorously test the metamodel framework, we conducted an exhaustive numerical experiment consisting of $25$ treatment conditions (five naloxone levels $\times$ five buprenorphine levels) for each of the $67$ counties. Each condition was replicated 1000 times to average out stochastic variation in the agent-based simulation, resulting in more than 1.6 million simulation runs in total. This large baseline experiment serves two purposes: first, to establish a benchmark for comparing the proposed metamodel against brute-force simulation; and second, to highlight that further scaling of the design space is computationally prohibitive without surrogate modeling. \rev{Generating this exhaustive baseline required approximately two weeks of wall-clock time on 32 cores, roughly 10{,}000 core-hours (CPU cores multiplied by hours of computation). By contrast, the proposed metamodel attains comparable accuracy using fewer than 2\% of these simulation runs.}
Training the full two-stage metamodel, including sequential design and GPR fitting, required approximately two hours of wall-clock time on a Google Colab environment equipped with 8 vCPUs (AMD EPYC, $\sim$2.25 GHz), and 12 GB of RAM, highlighting the practical computational efficiency of the proposed approach relative to exhaustive simulation.
 
Model accuracy is evaluated using relative error, defined for each county and treatment condition as the absolute difference between the response function prediction and the average held-out simulation output, divided by that average, and then averaged across all counties and treatment conditions. The test set consists of 20\% of the simulation replications per county-treatment condition, held out before any model training. Since each replication is an independent draw from the stochastic simulator under a fixed condition, the replications are exchangeable, and the holdout requires no ordering or stratification.

\subsection{Sampling Efficiency and Learning Curves}
We organize the subsequent empirical evaluation around three interrelated questions that govern the performance of the proposed two-stage metamodel. First, how many simulation runs are required to achieve reliable predictive accuracy under a sequential design? Second, how much complexity is needed in the response function to capture treatment effects without sacrificing interpretability? Third, how should the kernel be designed to support this model complexity, capturing spatial and socio-economic heterogeneity across counties while avoiding overfitting and unstable posterior behavior? The results that follow examine these questions in turn and demonstrate how sample allocation, response-function structure, and kernel design must be jointly balanced to achieve accuracy, efficiency, and robustness.

The cumulative allocation of simulation runs across counties is highly uneven, reflecting the adaptive behavior of the sequential design. As shown in Figure~\ref{fig:samples_pa}, more than two-thirds of counties require fewer than 150 simulation runs, while a small subset of north-central counties, highlighted in green and yellow, receive substantially more samples, with up to 600 runs per county. These allocations correspond to the state of the model after a total of 10,000 simulation runs. 

\begin{figure}[p]
    \centering
    
    \begin{subfigure}[b]{0.48\textwidth}
        \centering
        \includegraphics[width=\textwidth]{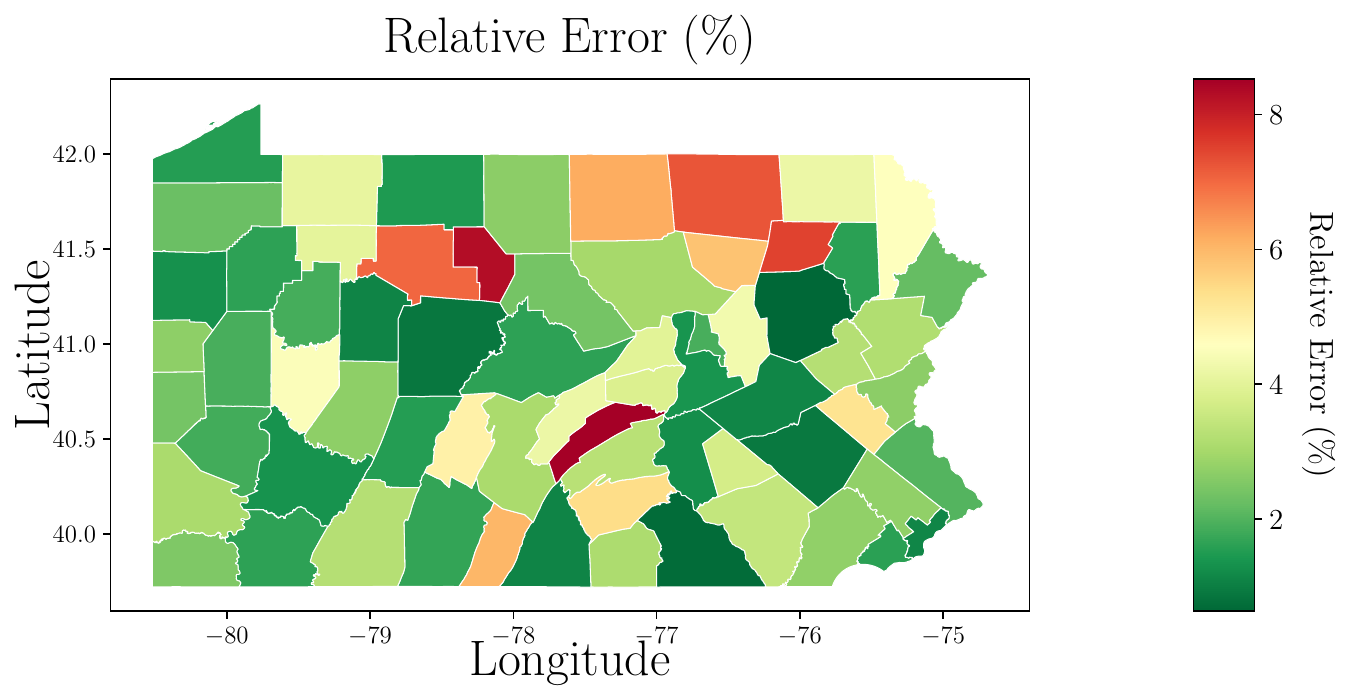}
        \caption{County-level predictive accuracy}
        \label{fig:pa_rel_error}
    \end{subfigure}
    \hfill
    \begin{subfigure}[b]{0.48\textwidth}
        \centering
        \includegraphics[width=\textwidth]{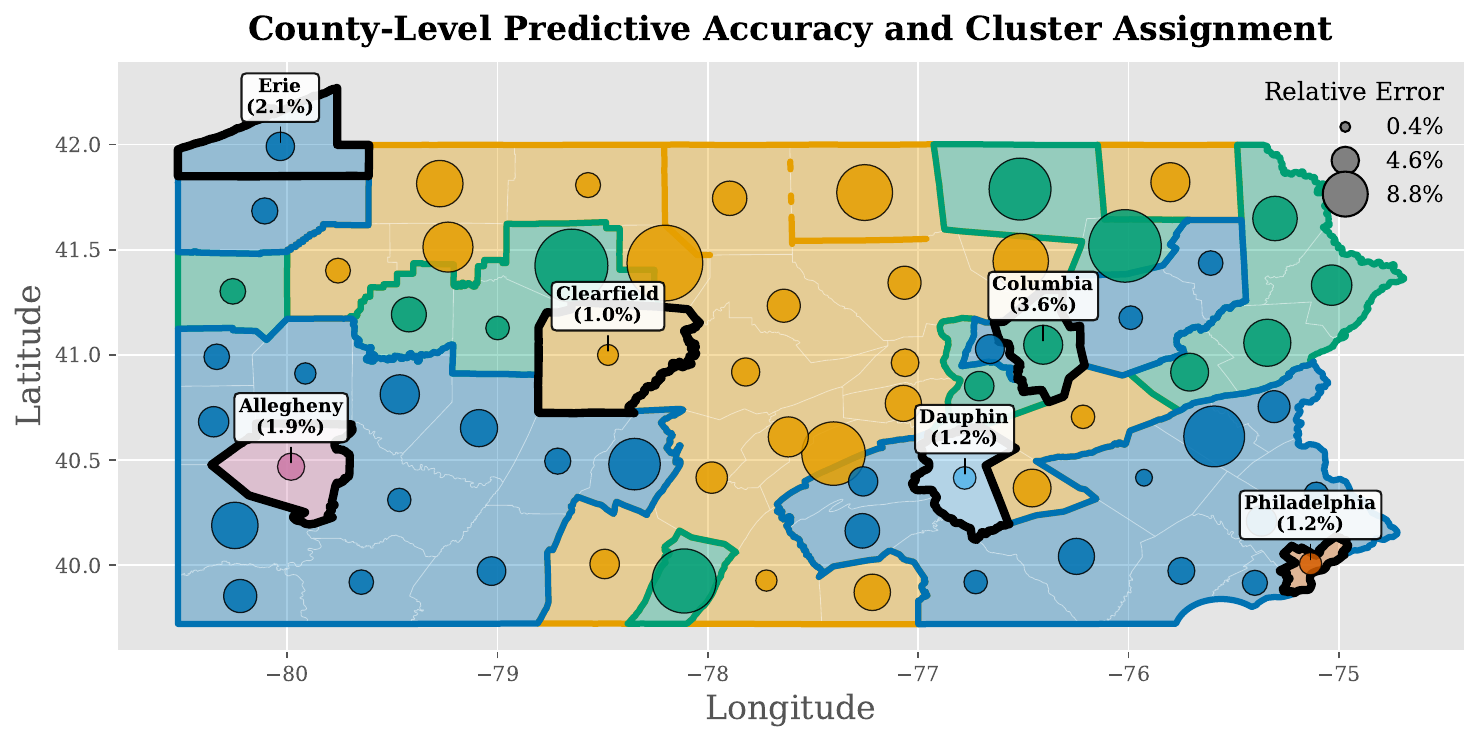}
        \caption{County-level relative error and cluster assignment}
        \label{fig:re_clustering}
    \end{subfigure}
    \hfill
    
    \vspace{0.3cm}

    \begin{subfigure}[b]{0.48\textwidth}
        \centering
        \includegraphics[width=\textwidth]{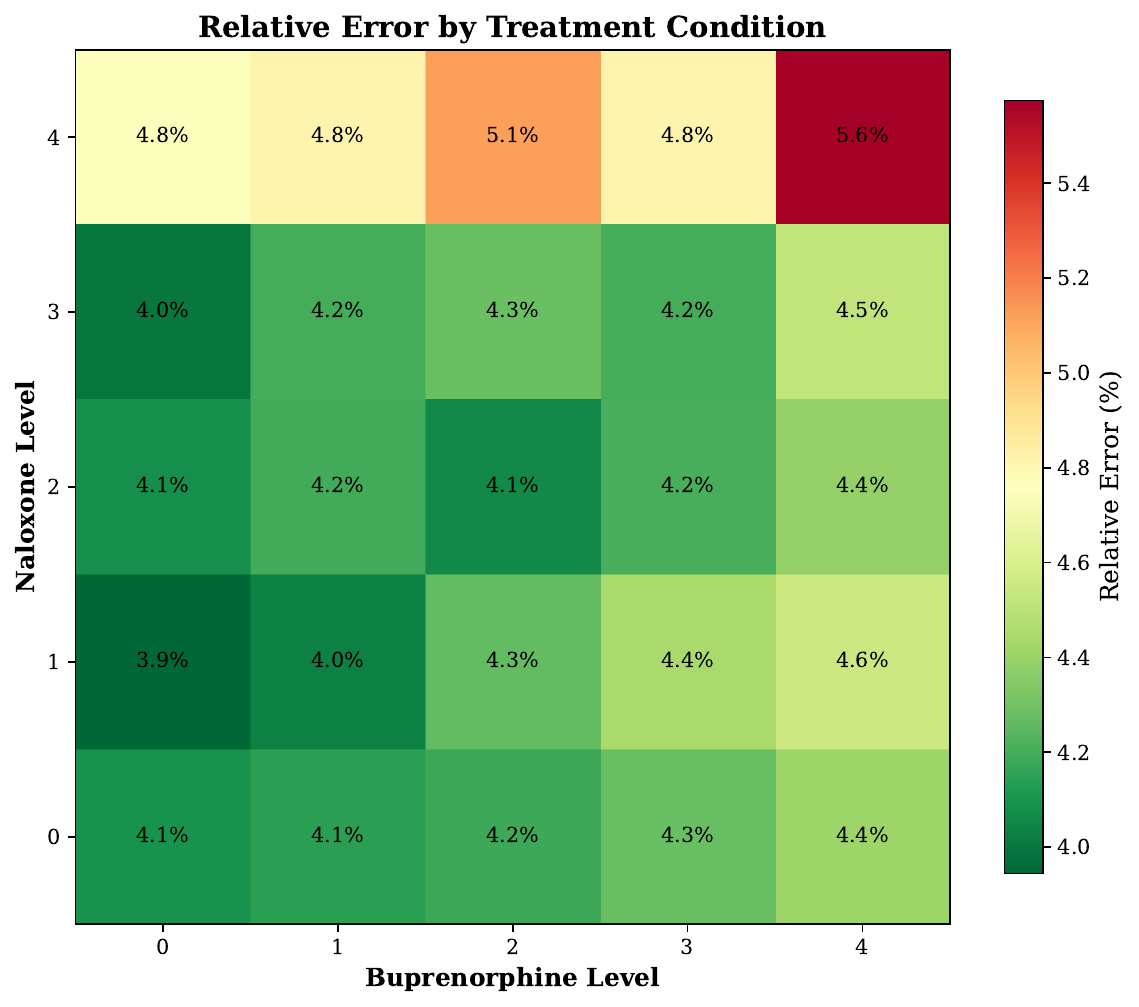}
        \caption{Relative error by treatment condition}
        \label{fig:heatmap}
    \end{subfigure}
    \hfill
    \begin{subfigure}[b]{0.48\textwidth}
        \centering
        \includegraphics[width=\textwidth]{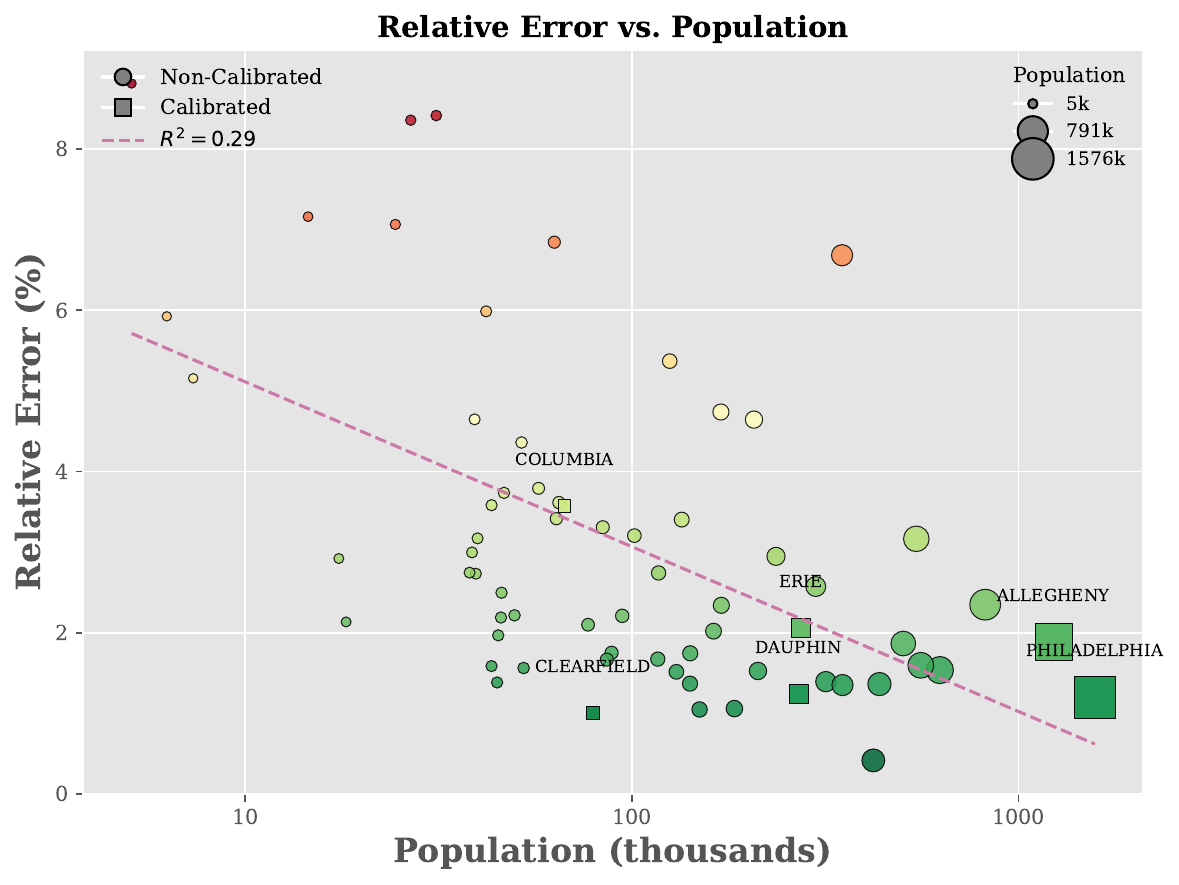}
        \caption{Relative error vs. county population size}
        \label{fig:re_population}
    \end{subfigure}

    \vspace{0.3cm}

    \begin{subfigure}[b]{0.48\textwidth}
        \centering
        \includegraphics[width=\textwidth]{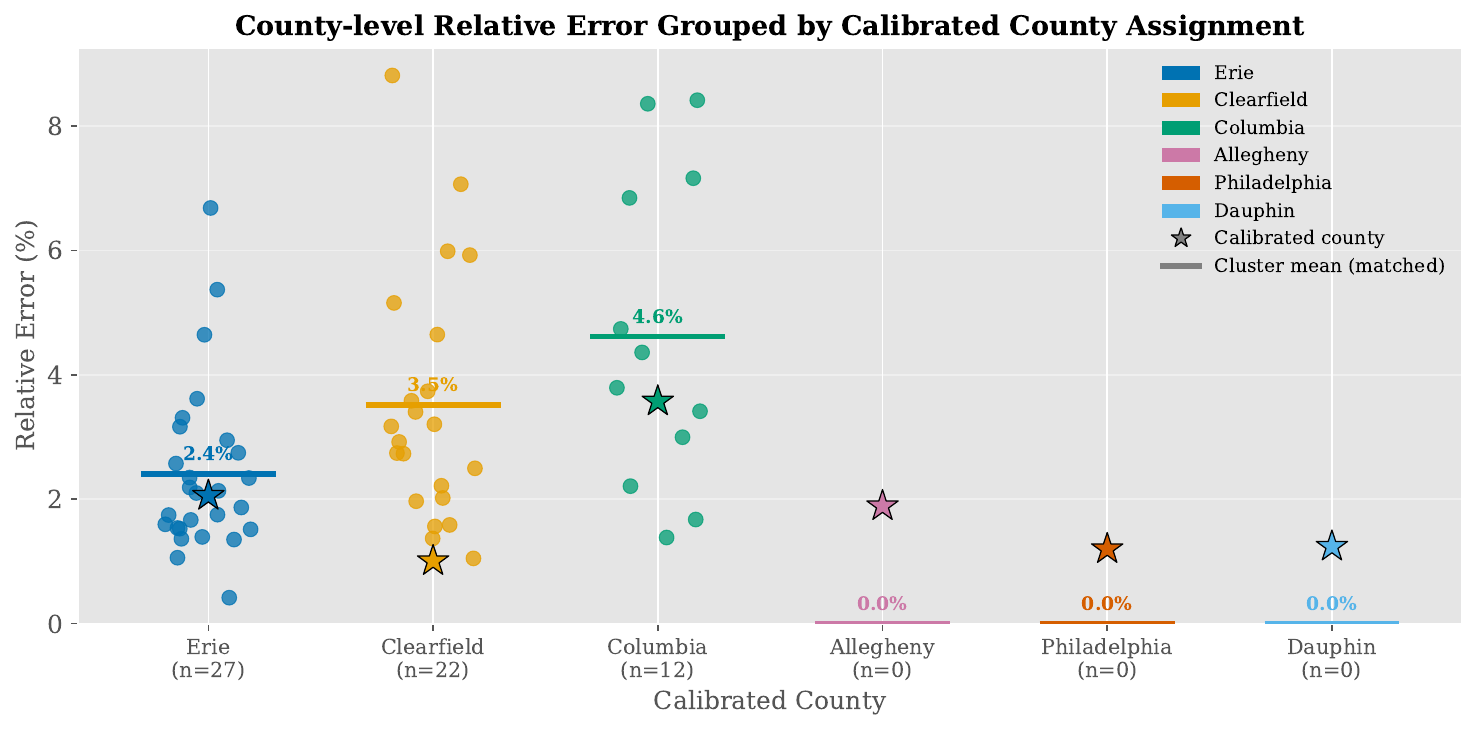}
        \caption{Relative error grouped by calibration assignments}
        \label{fig:calib_vs_noncalib}
    \end{subfigure}
    \hfill
    \begin{subfigure}[b]{0.48\textwidth}
        \centering
        \includegraphics[width=\textwidth]{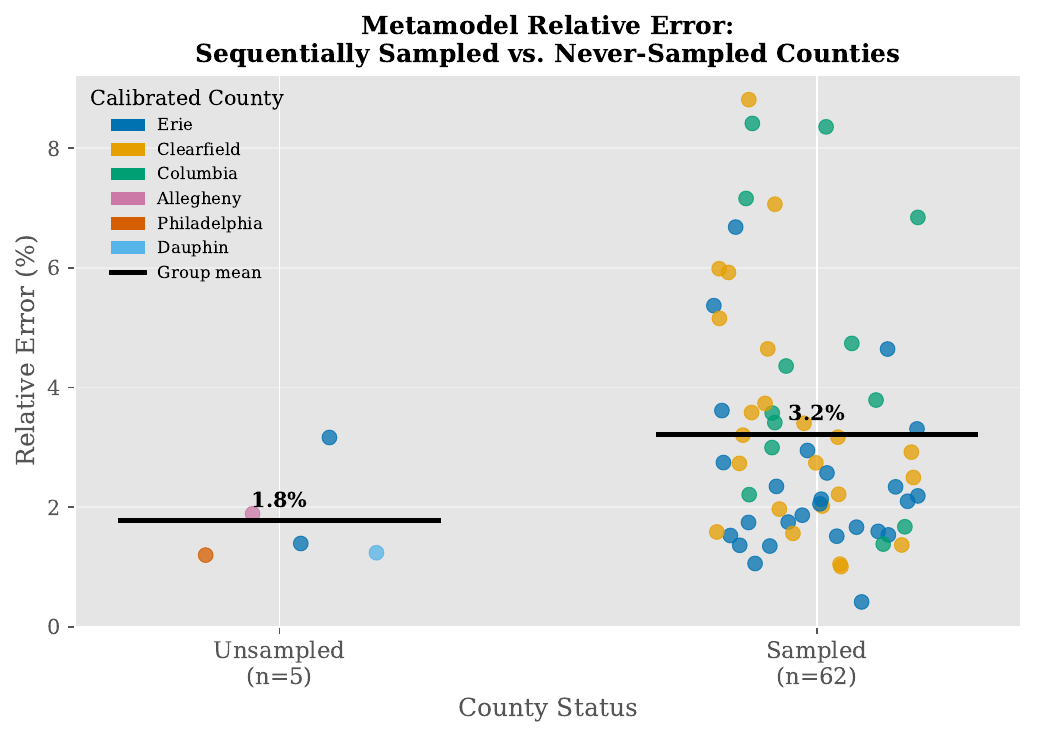}
        \caption{Relative error for sampled vs.\ unsampled counties}
        \label{fig:selected_counties}
    \end{subfigure}
    
    \caption{\textbf{Empirical evaluation of the heteroscedastic noise modeling and sequential design strategies for improving the sample efficiency of the proposed two-stage modeling framework.}\\ \textbf{Panel (a)} shows the county-level predictive accuracy of the metamodel after 10,000 total samples.\\ 
    \textbf{Panel (b)} shows county-level relative error and cluster assignment, where circle size reflects the magnitude of relative error and fill color indicates the assigned cluster. Counties are grouped into six clusters, each represented by a calibrated county (Allegheny, Philadelphia, Dauphin, Erie, Columbia, and Clearfield) shown with bold black borders. Each county is shaded with the color of its assigned calibrated county.\\
    \textbf{Panel (c)} shows the mean relative error across the $5 \times 5$ intervention grid, with slightly higher error at extreme treatment conditions.\\ 
    \textbf{Panel (d)} shows a negative association between county population and relative error ($R^2 = 0.29$), indicating that the metamodel achieves higher accuracy in more populous counties.\\
    \textbf{Panel(e)} reports county-level relative error grouped by calibrated county assignment. 
    \\
    \textbf{Panel(f)} compares relative error between the 62 sequentially sampled counties and the 5 never-selected counties (Philadelphia, Dauphin, Allegheny, Luzerne, Lancaster).
    }
    \label{fig:grid_curves1}
\end{figure}

\begin{figure}[p]
    \begin{subfigure}[b]{0.48\textwidth}
        \centering
        \includegraphics[width=\textwidth]{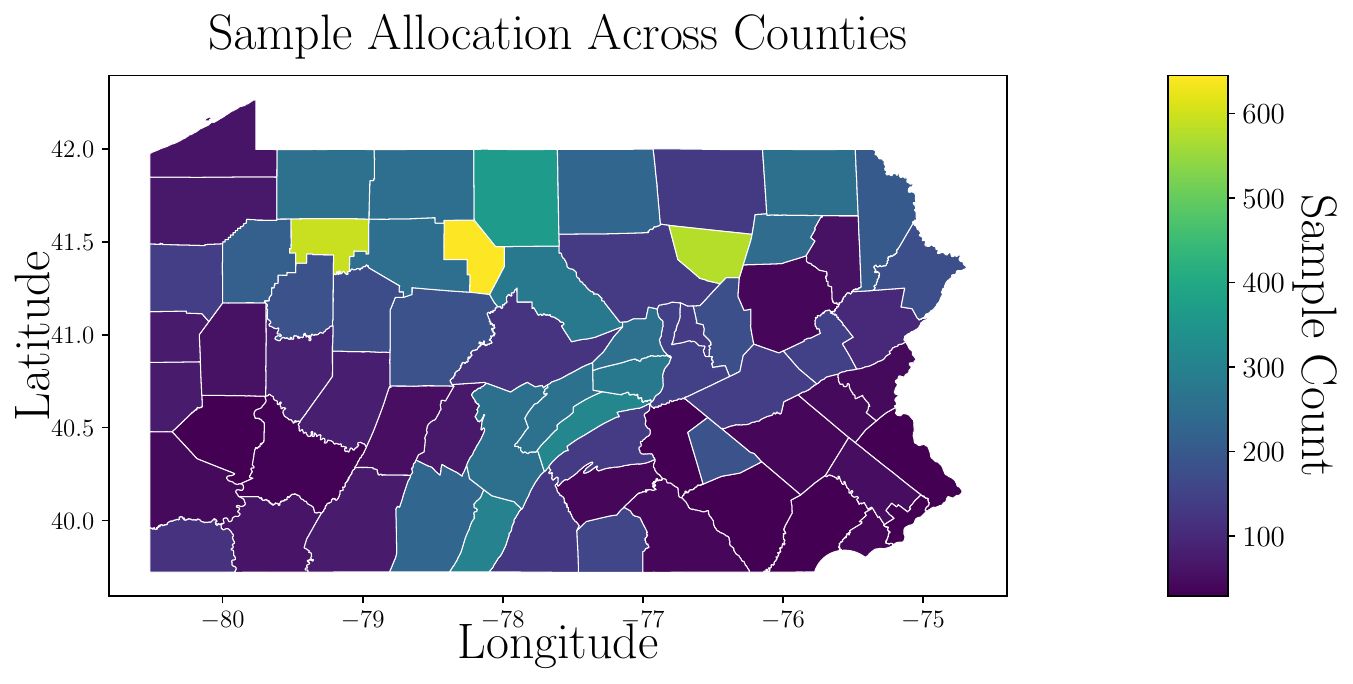}
        \caption{Adaptive sample allocation across counties}
        \label{fig:samples_pa}
    \end{subfigure}
    \hfill
    \begin{subfigure}[b]{0.48\textwidth}
        \centering
        \includegraphics[width=\textwidth]{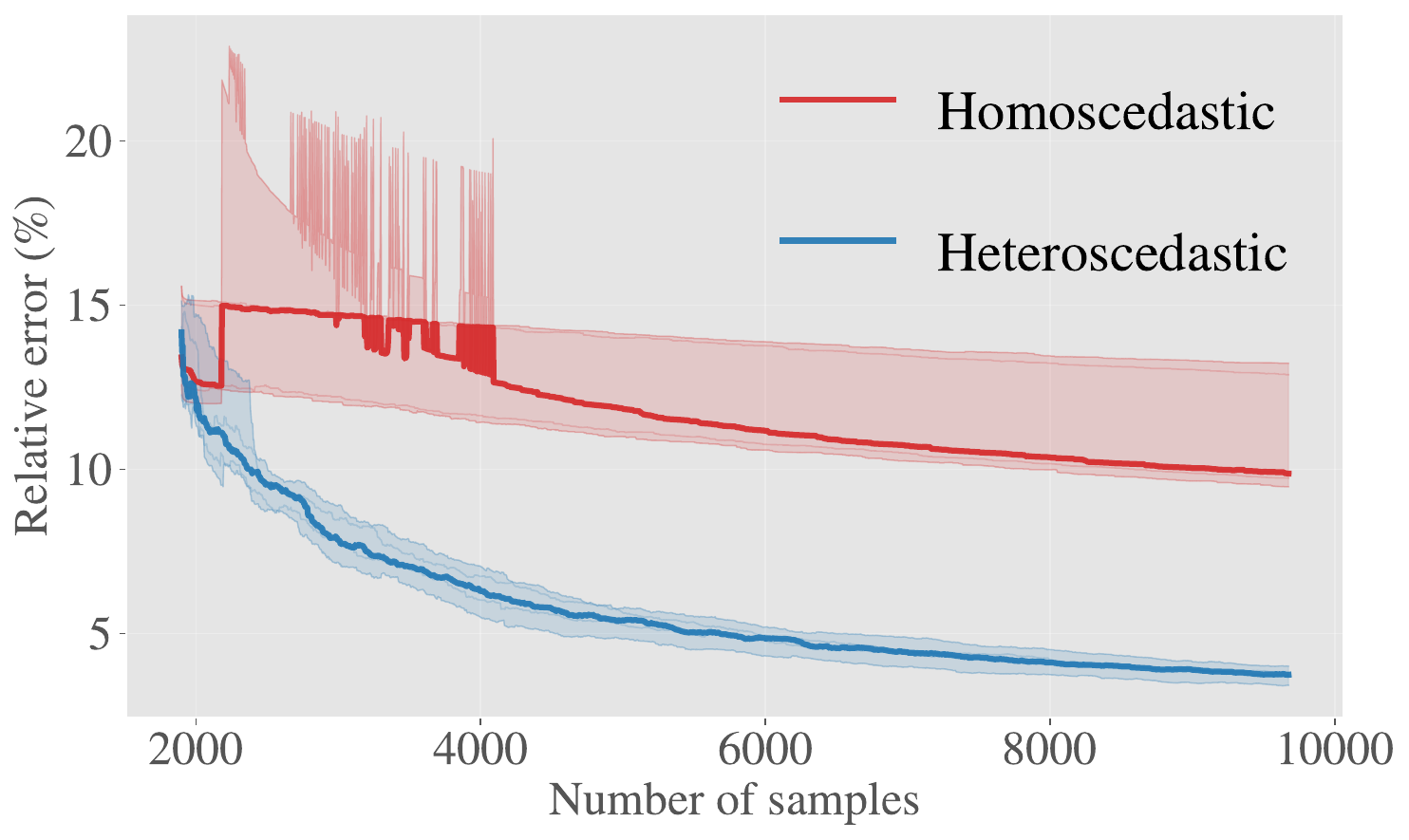}
        \caption{Noise modeling using heteroscedastic GPR}
        \label{fig:hetero_homo}
    \end{subfigure}

    \begin{subfigure}[b]{0.48\textwidth}
        \centering
        \includegraphics[width=\textwidth]{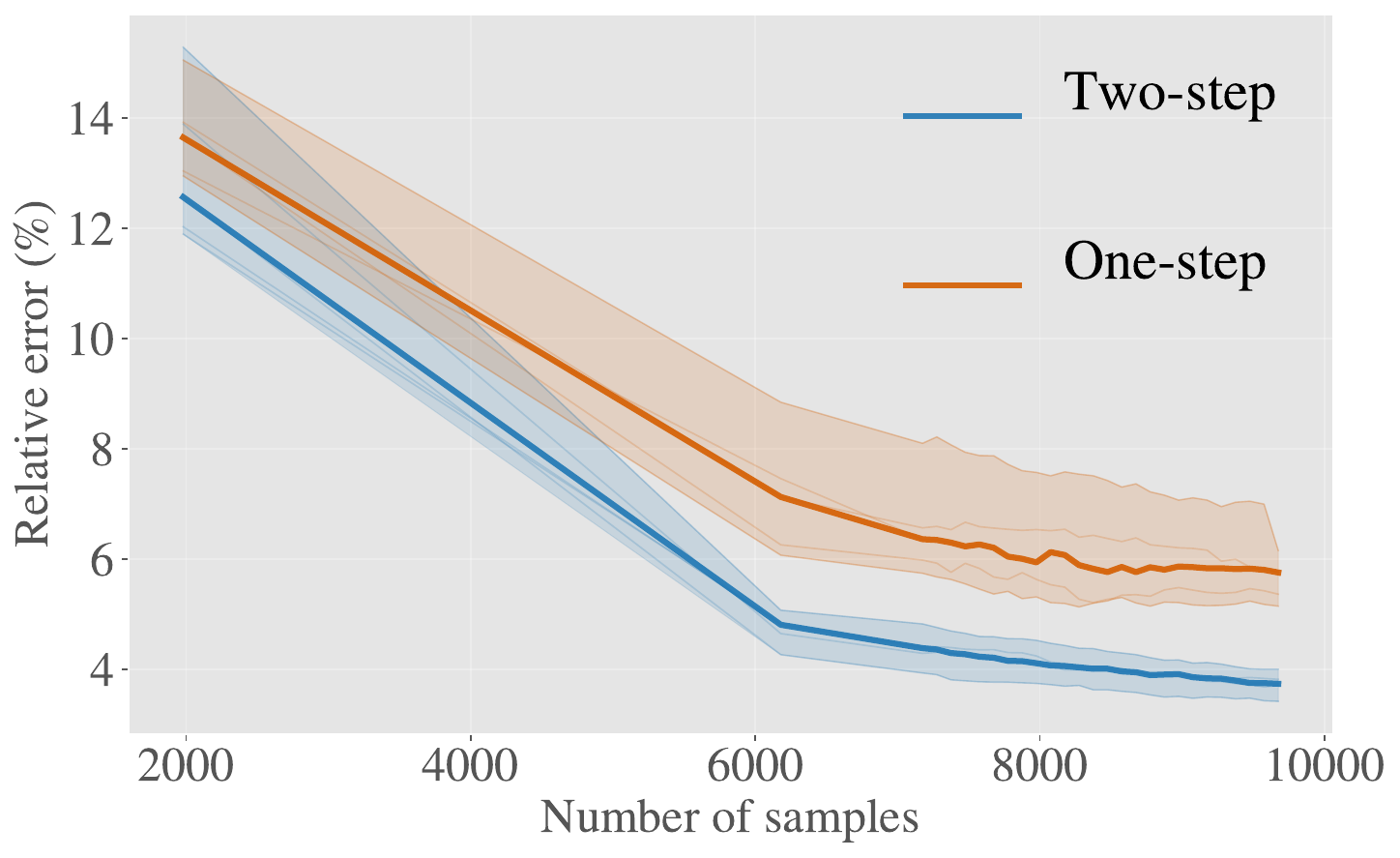}
        \caption{Efficiency of the two-step sequential design}
        \label{fig:seqdes_comparison}
    \end{subfigure}
    \hfill
    \begin{subfigure}[b]{0.48\textwidth}
        \centering
        \includegraphics[width=\textwidth]{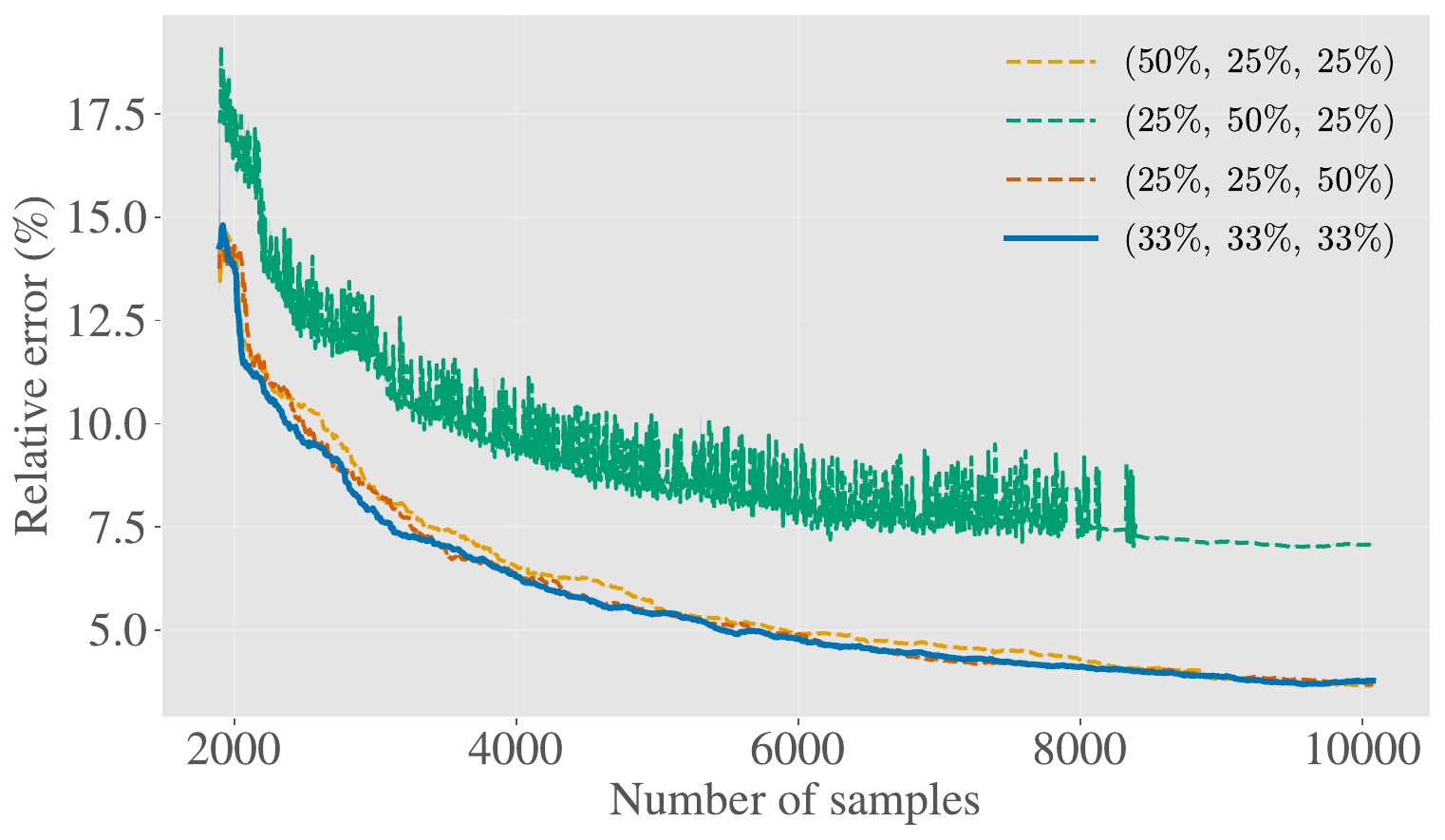}
        \caption{Relative error by changing weights}
        \label{fig:re_weights}
    \end{subfigure}
    \vspace{0.3cm}

    \begin{subfigure}[b]{0.48\textwidth}
        \centering
        \includegraphics[width=\textwidth]{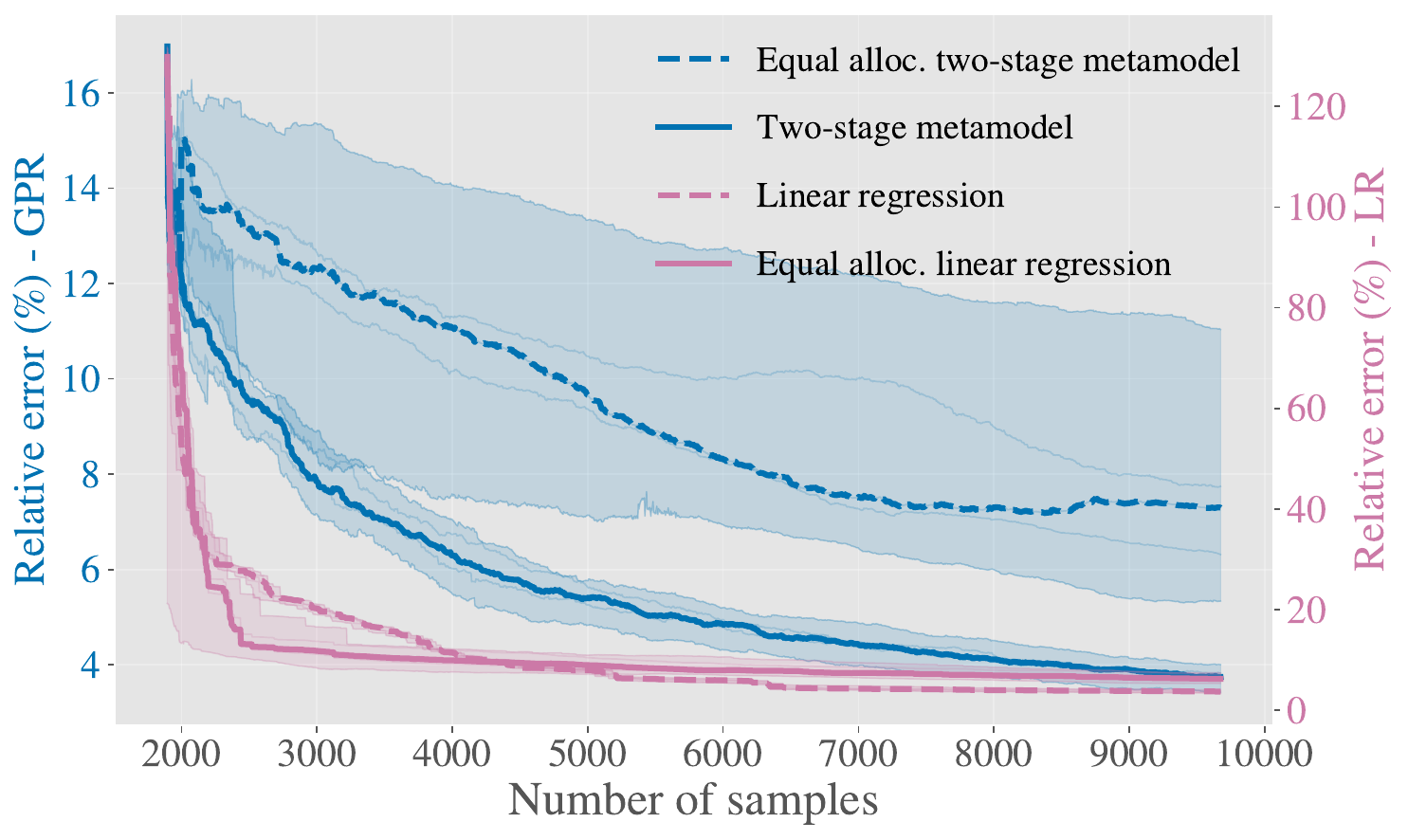}
        \caption{Learning curves for different baselines}
        \label{fig:baseline_compare}
    \end{subfigure}
    \hfill
    \begin{subfigure}[b]{0.48\textwidth}
        \centering
        \includegraphics[width=\textwidth]{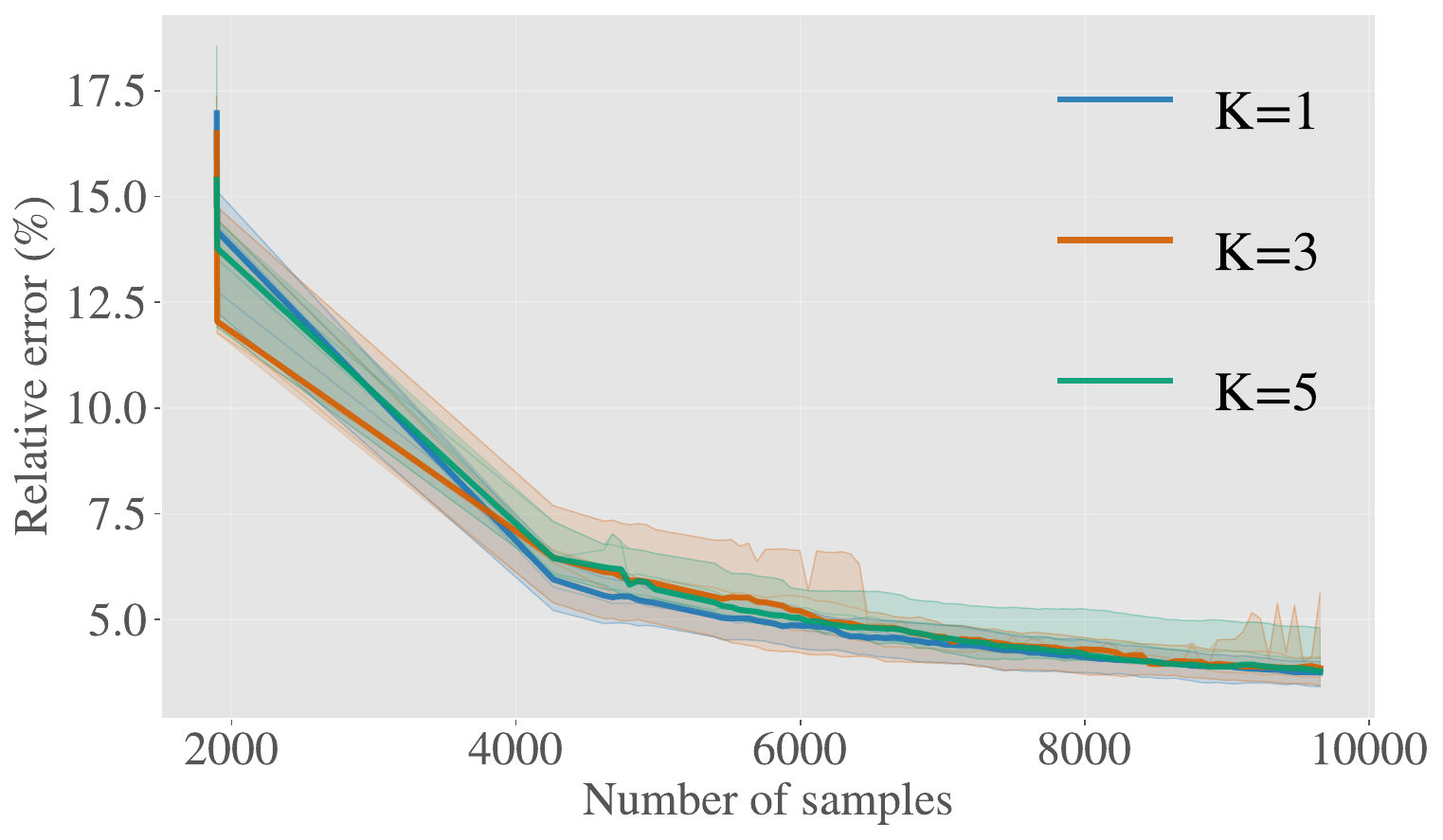}
        \caption{Effect of batch size on learning curve }
        \label{fig:batch_compare}
    \end{subfigure}
    \hfill
    
    \caption{\textbf{Sample efficiency of the two-stage metamodel under alternative noise specifications, sequential design strategies, baselines, and batch sizes.} 
    \textbf{Panel (a)} shows the adaptive allocation of simulation runs across 
    counties. The sequential design concentrates effort on counties with higher posterior uncertainty, as reflected by the uneven 
    distribution of sample counts.
    \textbf{Panel (b)} shows ignoring county-level heteroscedasticity when specifying observation noise in the GPR model results in a slow, unstable and inefficient learning behavior (the red learning curve). \textbf{Panel (c)} compares two sampling strategies: (i) a one-step sequential design, which selects counties adaptively and then exhaustively simulates all treatment conditions within the selected county, and (ii) the proposed two-step design that additionally selects treatment conditions based on posterior uncertainty.
     \textbf{Panel (d)} reports a sensitivity analysis over five scalarization weight configurations $\mathbf{w} = (w_0, w_n, w_b)$ in the SNR acquisition function, varying the emphasis placed on the intercept ($\mu_0$), naloxone ($\mu_n$), and buprenorphine ($\mu_b$) coefficients. 
    All configurations with the exception of $\mathbf{w} = (25\%, 25\%, 50\%)$ converge to comparable relative errors by 10,000 samples, confirming that equal weighting is a robust default and that the sequential design is generally not sensitive to this choice.
     \textbf{Panel (e)} 
    compares the sequential two-stage metamodel against three baselines under identical simulation budgets: {(i)} a sequential linear regression baseline, which assigns each sampled county its own OLS estimate but cannot extend these estimates to unsampled counties or exploit the spatial and socio-economic correlation structures to learn across counties; {(ii)} an equal allocation two-stage metamodel, which uses the same two-stage metamodel framework but distributes simulation runs uniformly across counties and treatment conditions rather than adaptively with sequential design; and (iii) an equal allocation linear regression baseline. \rev{The left vertical axis reports the relative error of the two-stage metamodels, and the right vertical axis reports the relative error of the linear regression baselines, which is plotted on a different scale.}
     \textbf{Panel (f)} compares the learning curves under batch sizes $K = 1$, $K = 3$, and $K = 5$, where $K$ denotes the number of counties selected per sequential design iteration. 
    }
    \label{fig:grid_curves2}
\end{figure}

Across most counties, the relative error of overdose mortality predictions is below 5\%, indicating high predictive accuracy of the metamodel. Counties with higher relative error coincide with those receiving greater simulation effort, reflecting the adaptive behavior of the sequential design, which concentrates sampling where the model is most difficult to learn until uncertainty is reduced to an acceptable level. This spatial error pattern is shown in Figure~\ref{fig:pa_rel_error}, with the corresponding sample allocation illustrated in Figure~\ref{fig:samples_pa}. Figure~\ref{fig:re_clustering} further shows that prediction error varies across county clusters, with the Clearfield and Columbia clusters exhibiting higher mean relative error. \rev{Figure~\ref{fig:heatmap} shows that error is largely 
uniform across treatment conditions but is somewhat higher for naloxone than 
buprenorphine and largest at the highest naloxone level, reflecting the larger 
cross-county range of the naloxone coefficient $\mu_n$, whose estimation error is 
amplified proportionally to the naloxone level in the response function.  Figure~\ref{fig:re_population} confirms that error decreases with county population size. }

As a further check of model performance, we examine prediction error across 
two additional comparisons: calibrated versus non-calibrated counties, and sequentially sampled versus never-selected counties. Figure~\ref{fig:calib_vs_noncalib} shows that higher errors within the Columbia and Clearfield clusters are concentrated in small-population counties (Cameron, Wyoming, Sullivan), where simulation variance is inherently higher 
due to small population size. Figure~\ref{fig:selected_counties} demonstrates 
that the five never-selected counties achieve lower mean relative error (1.8\% vs.\ 3.2\%), indicating that the SNR acquisition function correctly identified them as low-uncertainty regions and that the GPR generalizes accurately without direct simulation runs.

Compared with the approximately 1.6 million simulations required for exhaustive enumeration, the two-stage metamodel framework, combining Gaussian process regression with outcome-stage response-function modeling, achieves comparable statewide accuracy using only a fraction of the total simulation budget, demonstrating its suitability for large-scale policy analysis under tight computational constraints.

\FloatBarrier
\clearpage
\thispagestyle{empty}

\FloatBarrier
\clearpage
\thispagestyle{empty}
\clearpage

Explicitly modeling heterogeneous observation variance leads to substantially more stable and accurate learning behavior in the GPR. Under the heteroscedastic specification, relative error decreases smoothly as additional samples are incorporated, reflecting consistent posterior updating as uncertainty contracts in counties with increasing numbers of simulation replicates. In contrast, the homoscedastic formulation exhibits noticeably less stable learning, with oscillatory error trajectories at the beginning when sample sizes are small. These fluctuations arise from the constant-variance assumption, under which early or sparsely sampled observations can exert disproportionate influence on the posterior, resulting in mischaracterized uncertainty and irregular updates. Consequently, the homoscedastic model converges more slowly and displays greater variability across sample sets. This contrast is illustrated in Figure~\ref{fig:hetero_homo}, where the limited overlap between the shaded min–max bands further indicates that the advantage of the heteroscedastic model is robust rather than driven by a small number of favorable runs. Overall, these results demonstrate that linking observation variance to sample size and regression uncertainty improves both estimation accuracy and learning stability in sequential simulation settings.

Having characterized how the specification of the observation noise influences learning behavior in Figure~\ref{fig:hetero_homo}, we next examine how the sequential design procedure impacts the rate at which the metamodel improves. Figure~\ref{fig:seqdes_comparison} compares two variants of the metamodel for a 25-condition array: one that employs the variance-oriented sequential design for selecting the next treatment condition (blue curve) and a baseline that samples all treatment conditions in the selected county (orange curve). The two-step sequential strategy achieves a noticeably steeper decline in relative error during the early phases of training and maintains a consistently lower error across the full range of simulated samples. By concentrating additional runs on the single most informative intervention level within each county, the procedure accelerates convergence and reduces the total number of simulations required to reach a given accuracy threshold. This confirms that adaptive allocation of treatment conditions complements county-level sampling, yielding further gains in overall sample efficiency.

Building on the benefits of the two-step sequential design, we next examine the sensitivity of the framework to three modeling choices: the scalarization weights in the acquisition function, the contribution of spatial interpolation, and the effect of batch size on learning performance. Figure~\ref{fig:re_weights} reports a sensitivity analysis over seven scalarization 
weight configurations $\mathbf{w}$ in the SNR-based acquisition function. All configurations 
converge to comparable relative errors by 10,000 samples, confirming that equal weighting is a robust default and that the framework is not sensitive to this choice. The exception is the configuration that places higher weight on the buprenorphine coefficient ($\mathbf{w} = (25\%, 25\%, 50\%)$), which exhibits higher and more 
unstable relative error throughout training. This is consistent with buprenorphine 
having a smaller effect size than the intercept and naloxone coefficients, so over-weighting it directs sampling toward counties that are uncertain about a low-magnitude coefficient rather than about the overall outcome.

To benchmark the two-stage metamodel, Figure~\ref{fig:baseline_compare} compares the sequential two-stage metamodel against \rev{three baselines: a sequential linear regression baseline, an equal allocation two-stage metamodel, and an equal allocation linear regression baseline. The left vertical axis reports the relative error of the two-stage metamodels, while the right vertical axis reports the relative error of the linear regression baselines, whose errors are substantially larger and are therefore plotted on a different scale.}
The sequential two-stage metamodel achieves substantially lower relative error than all three baselines throughout the learning curve, demonstrating that efficiency gains stem from both GPR-based spatial interpolation across counties and the adaptive sample allocation of the sequential design.
\rev{The linear regression baselines cannot extend their estimates to unsampled counties, since each county is assigned its own independent OLS estimate. This also clarifies where the advantage of the framework may not generalize: the gains depend on the presence of exploitable spatial and socio-economic structure among subgroups (i.e., counties). In settings where subgroups are effectively independent, with little spatial correlation to borrow strength from, the benefit of GPR-based interpolation over per-subgroup regression would diminish, and the framework would offer smaller gains relative to simpler baselines.}

Figure~\ref{fig:batch_compare} examines whether selecting $K > 1$ counties per iteration improves performance. Increasing the batch size to $K = 3$ or $K = 5$ does not reduce relative error compared to the single-county update $K = 1$, and experiments with larger batch sizes confirm the same conclusion. This confirms that one-at-a-time posterior updating is the most sample-efficient 
strategy. This is consistent with the sequential nature of the design: each selection benefits from the full posterior update induced by all previous observations, an advantage that diminishes when multiple counties are selected simultaneously on the same posterior surface.

\begin{figure}[H]
    \centering
    \begin{subfigure}[b]{0.48\textwidth}
        \centering
        \includegraphics[width=\textwidth]{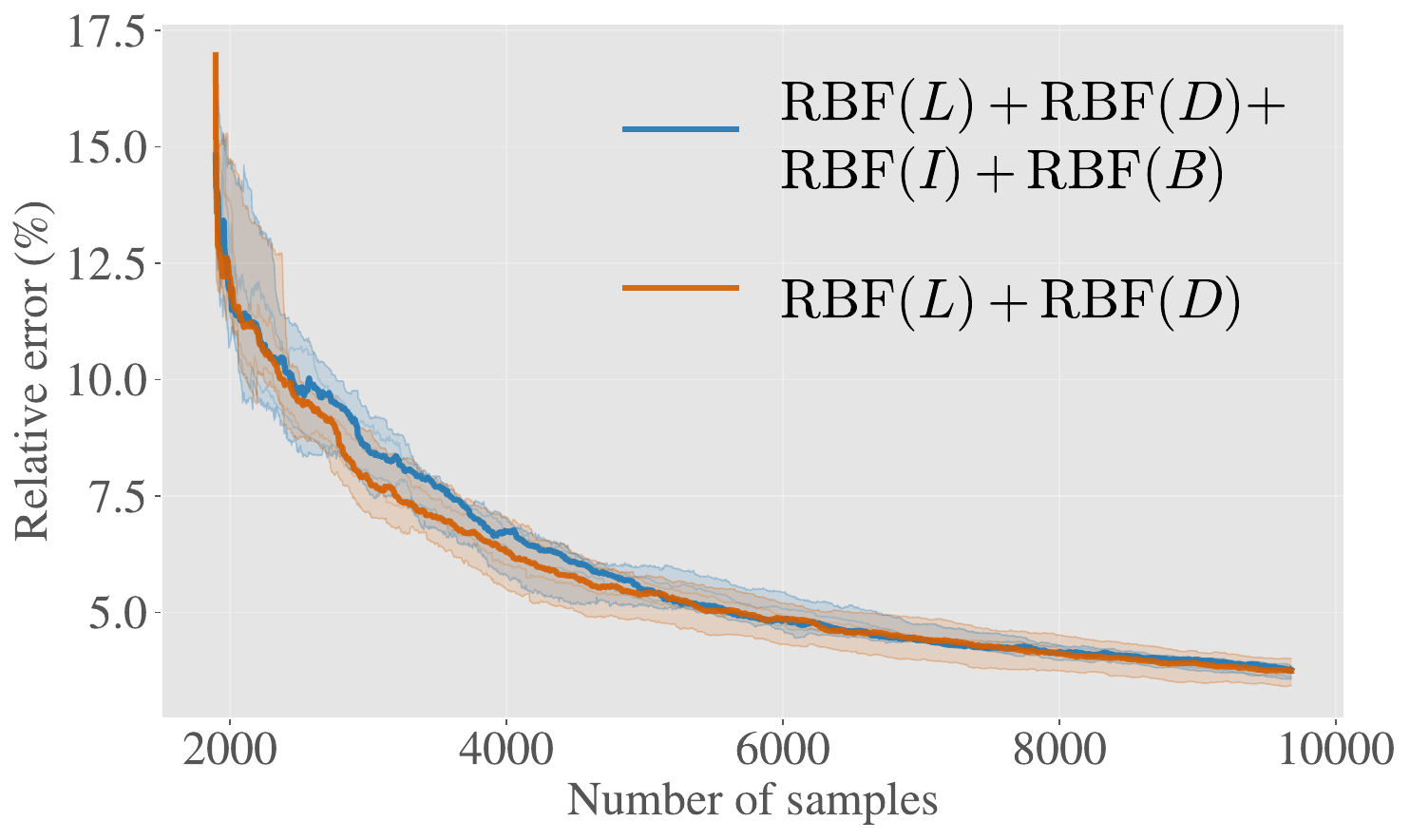}
        \caption{Kernel complexity}
        \label{fig:kernel_des}
    \end{subfigure}~
    \hfill
    \begin{subfigure}[b]{0.48\textwidth}
        \centering
        \includegraphics[width=\textwidth]{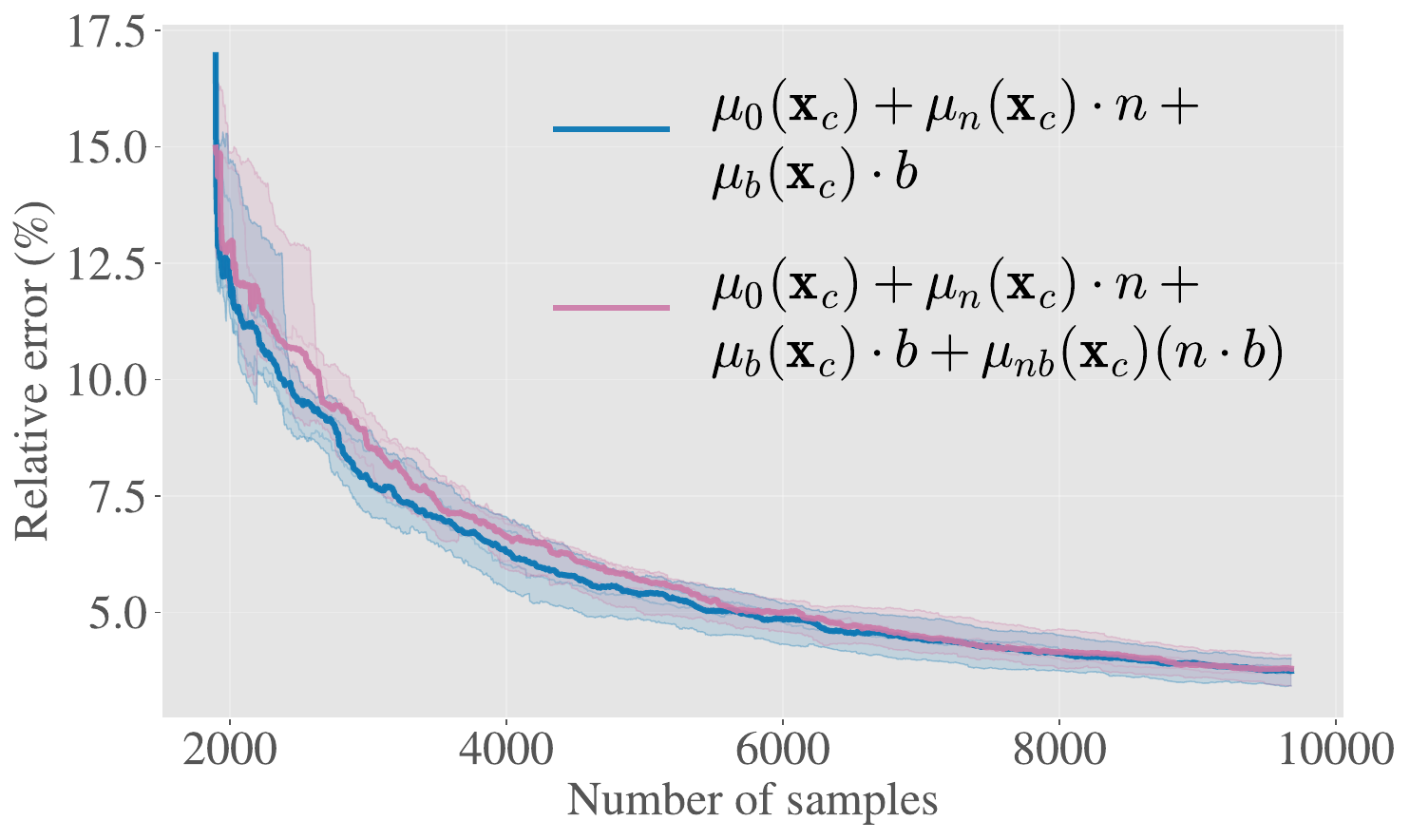}
        \caption{Response function complexity}
        \label{fig:model_comp}
    \end{subfigure}~
    
    \vspace{0.3cm}
    \begin{subfigure}[b]{0.48\textwidth}
        \centering
        \includegraphics[width=\textwidth]{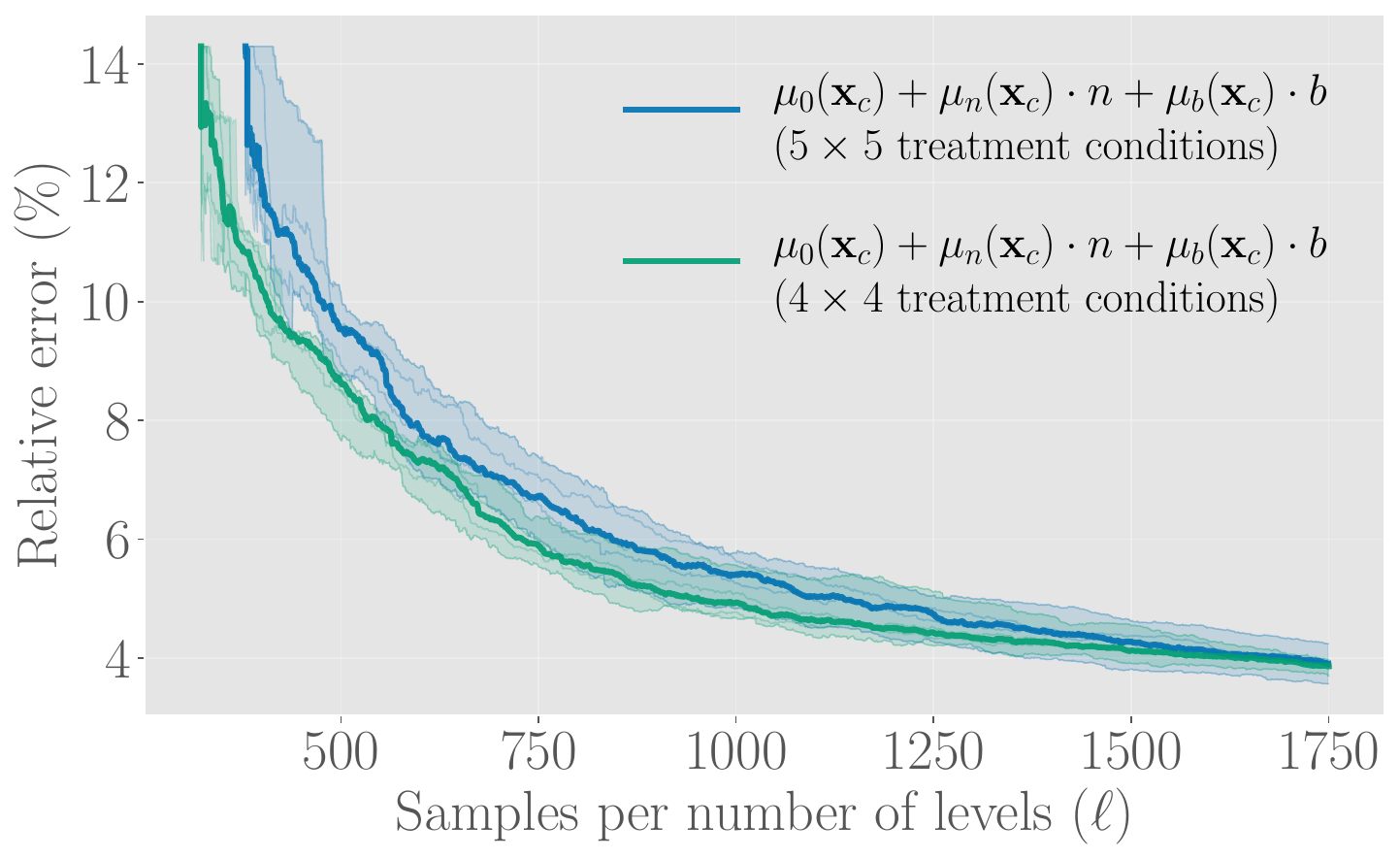}
        \caption{Design space complexity}
        \label{fig:des_space}
    \end{subfigure}
    \hfill
     \begin{subfigure}[b]{0.48\textwidth}
        \centering
        \includegraphics[width=\textwidth]{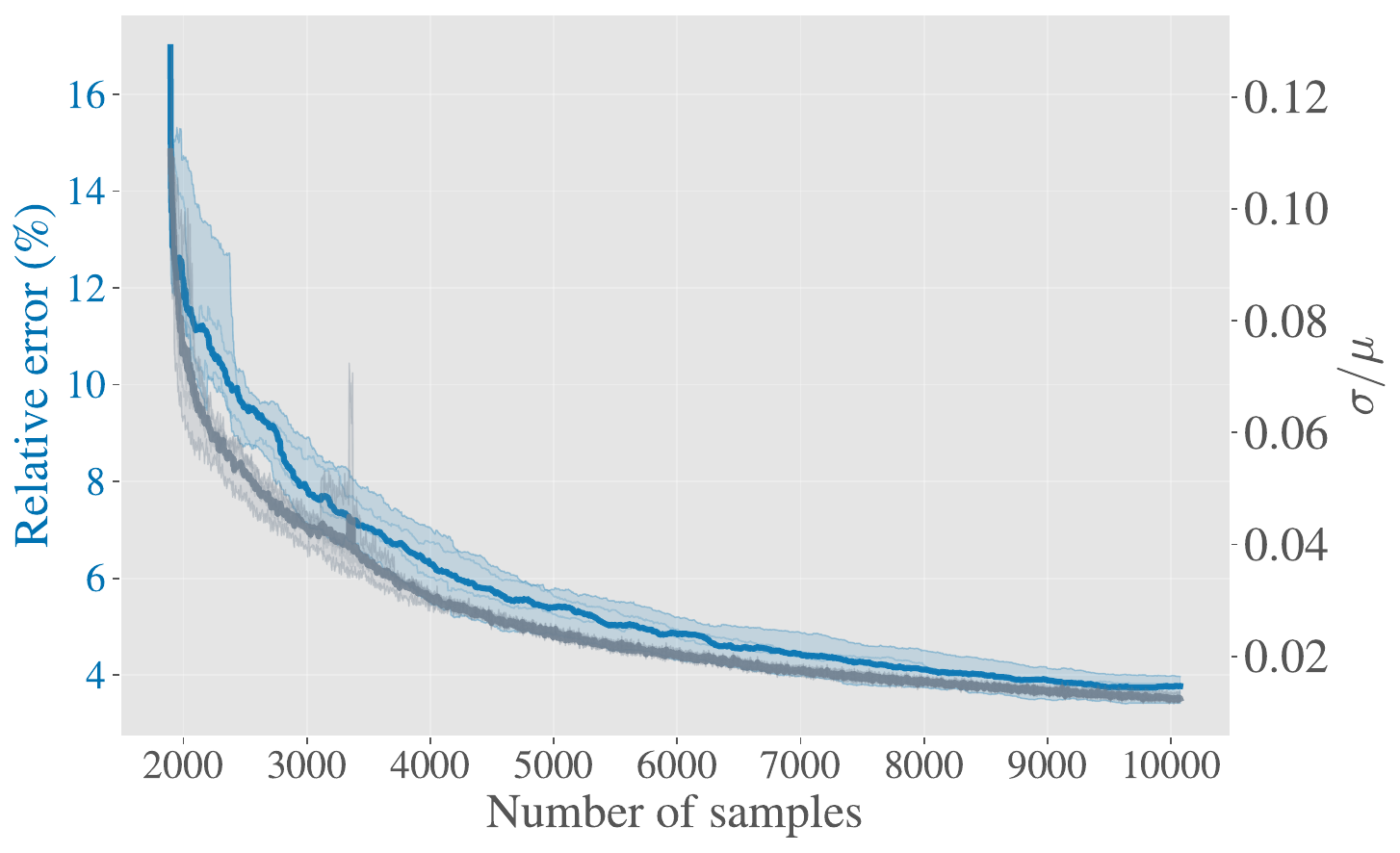}
        \caption{SNR trajectory during sequential sampling}
        \label{fig:snr_trajectory}
    \end{subfigure}
    
    \caption[]{\textbf{Effect of different types of model complexity on sample efficiency of the learning curves.} The baseline model shown in blue remains the same across the four panels.\\ \textbf{Panel (a)} Learning is slower for the more complex Kernel that combines more features: the blue baseline uses location ($L$), population density ($D$), median household income ($I$), and percent back population ($B$) versus only $L$ and $D$ used in the orange.\\ \textbf{Panel (b)} Learning is slower for the more complex response function that models the interaction between the two interventions (in red), compared to the blue baseline that only includes the main effects.\\ \textbf{Panel (c)} Learning is slower when modeling outcomes in a larger intervention grid (\ $5{\times}5$ in blue vs $4{\times}4$ in green); however the sample complexity scales with number of levels $\ell$ for each intervention rather than the grid size $\ell^2$
    , consistent with \citeNP[Theorem 1]{ahmed2024selection}. \\
    \textbf{Panel (d)} 
    shows the outcome SNR criterion, averaged across all counties and treatment conditions, decaying in parallel with the relative error throughout the sequential sampling process.
    }
    \label{fig:grid_curves3}
\end{figure}

Having established the benefits of modeling heteroscedastic uncertainty and the two-step sequential design in Figure~\ref{fig:grid_curves2}, we next examine the role of model complexity in shaping metamodel performance, i.e., sample efficiency and accuracy. In the following paragraphs, we discuss the effect of GPR kernel and the response-function complexity, as well as the size of the intervention grid on learning performance, shown in Figure~\ref{fig:grid_curves3}. 


{\bf Response-function complexity.} To investigate the effect of response function complexity and the trade-offs between sample efficiency and representational capacity, we compared the performance of the main-effects response function in Equation~\ref{eq:response_surface2} and an interaction-augmented specification within the metamodel framework. The more complex interaction-augmented response function model is specified as follows (compare with the main-effects-only model in Equation~\eqref{eq:response_surface2}):
\begin{align}
    z^{\mbox{(int)}}(n,b|c) & = \mu_0(\mathbf{x}_c) + \mu_n(\mathbf{x}_c)n + \mu_b(\mathbf{x}_c)b  \label{eq:response-function-with-interaction}
    + \mu_{nb}(\mathbf{x}_c)(n\cdot b).
\end{align}
    
Figure~\ref{fig:model_comp} shows the learning curves of two response-function specifications evaluated under the same kernel design,  $k(\cdot,\cdot) = \text{RBF}(L) + \text{RBF}(D)+\text{RBF}(I)+\text{RBF}(B)$, where $L$, $D$, $I$, and $B$ denote location (L), population density (D), income (I), and black population percentage (B; see Equation~\eqref{eq:kernelcomposition} in Appendix~\ref{sec:app:model-select}).  Across all sample sizes, the main-effects model attains substantially lower median relative error and exhibits a much narrower performance range. In contrast, the 
interaction specification shows persistently higher error and greater variability, indicating that the additional interaction term increases model complexity without improving predictive accuracy at the available sample sizes. This behavior reflects a well-known principle: when data are limited, a simpler response function can provide more stable and reliable estimates than more complex alternatives. In our setting, the main-effects model offers the best trade-off between interpretability, precision, 
and sample efficiency, while the interaction specification would require a larger simulation budget to be estimated reliably. 

{\bf Kernel complexity.} 
We next examined how the choice of kernel influences metamodel performance.
Figure~\ref{fig:kernel_des} reports relative-error learning curves for two kernel specifications applied to the same response function. The simpler specification, $\text{RBF}(L)+\text{RBF}(D)$, captures spatial and demographic variation using only location ($L$) and population density ($D$). The more expressive specification, $\text{RBF}(L)+\text{RBF}(D)+\text{RBF}(I)+\text{RBF}(B)$, augments this structure with additional socio-economic features, increasing kernel flexibility.
At smaller sample sizes, the higher-dimensional kernel exhibits slightly higher
error and greater variability, reflecting the increased variance and hyperparameter uncertainty associated with more complex covariance structures under limited data. As additional samples are collected, this disadvantage diminishes: the richer
kernel steadily closes the gap and ultimately achieves marginally lower relative error and smoother convergence than the simpler alternative.

{\bf Design space and intervention grid size.} To assess outcome-stage learning behavior under changes in the intervention design space, we evaluated the performance of the metamodel across treatment grids of different sizes. 
Figure~\ref{fig:des_space} plots the learning curves of the two-stage metamodel for two intervention grids: a $4{\times}4$ array (16 treatment conditions, green curve) and a $5{\times}5$ array (blue curve). Despite the 56\,\% increase in design points, the larger grid achieves approximately the same relative error after a comparable number of sequential design samples.  Both curves fall rapidly during the first $\sim\!200$ simulation runs (per treatment condition) and level off near a relative error of 4\% by 350 simulations (per treatment condition), indicating that the county-condition sequential design successfully targets high-uncertainty regions regardless of grid size. 
\rev{We emphasize that 
this comparison characterizes how the metamodel behaves as the design space grows in complexity, rather than a comparison between two grids of specific policy interest. Because the horizontal axis reports samples per intervention level $\ell$ rather than total samples, the larger $5{\times}5$ grid must characterize more treatment conditions from the same per-level budget; the modest increase in relative error therefore reflects the greater complexity of the larger design space rather than a deficiency of the metamodel, consistent with the result that sample complexity scales with $\ell$ rather than $\ell^2$ 
\shortcite[Theorem~1]{ahmed2024selection}.}

Figure~\ref{fig:snr_trajectory} overlays the relative error with the outcome SNR criterion, computed as $(\sigma/\mu)$ for the response function predictions, averaged across all counties and treatment conditions at each sampling iteration. In early iterations, when few counties have been sampled, outcome variations are high, indicating that the GPR posterior remains uncertain relative to the predicted treatment effects. As the sequential design preferentially allocates simulation runs to counties with the highest SNR-based acquisition criteria, posterior uncertainty contracts rapidly. 

The outcome SNR criterion closely tracks the relative error decline throughout the sampling process, falling steeply during the same early phase and gradually stabilizing as predictions converge. This parallel behavior highlights an important operational advantage of the GPR framework: unlike black-box surrogates, the GPR posterior provides well-characterized uncertainty estimates that serve as a reliable proxy for prediction accuracy, enabling system designers to monitor learning curve progress and anticipate convergence without access to ground truth simulation outcomes. The declining outcome SNR serves as an online stopping criterion, allowing decision-makers to determine when sufficient simulation effort has been allocated.

\subsection{Localized Treatment Effects}

\begin{figure}[p]
    \centering
    
    \includegraphics[width=0.9\textwidth]{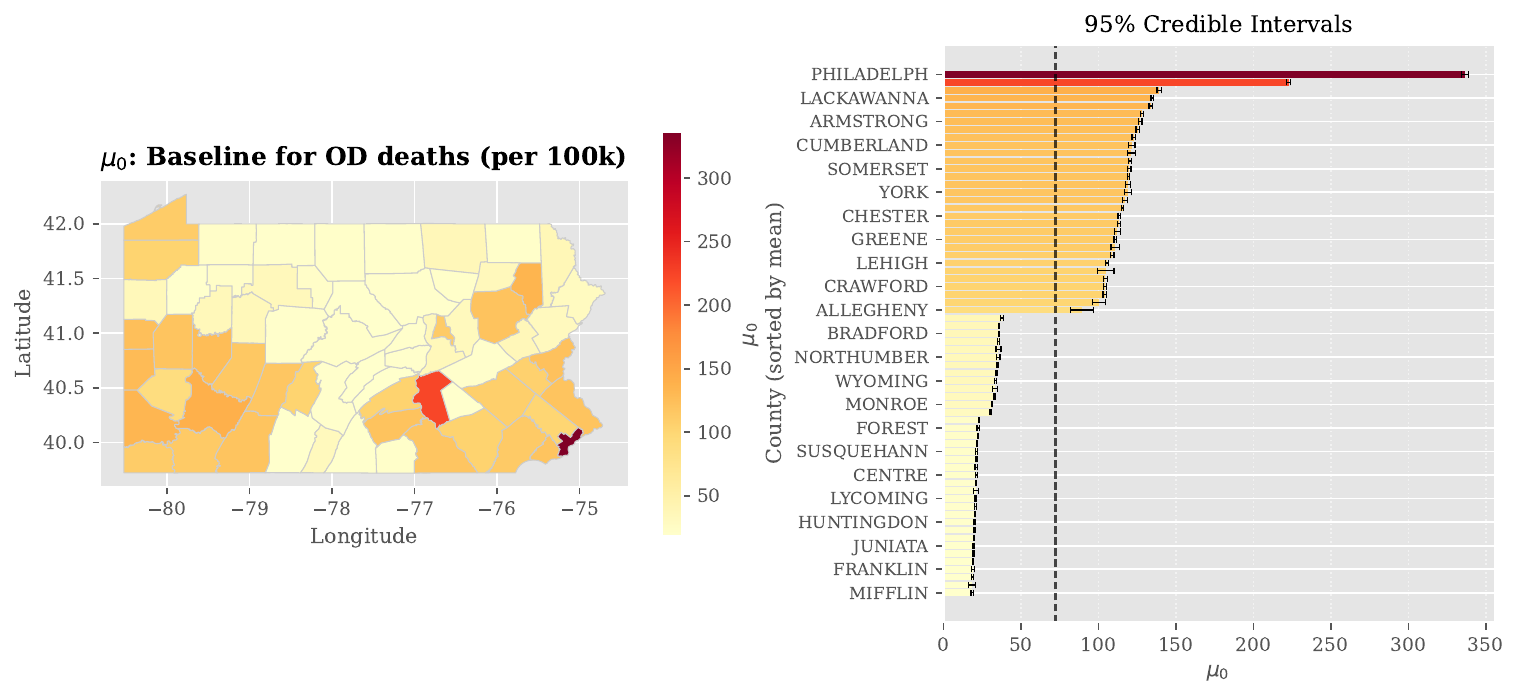}
    
    (a) Estimated baseline overdose mortality ($\mu_0$) per 100,000 people 
    
    \vspace{0.5em}
    
    \includegraphics[width=0.9\textwidth]{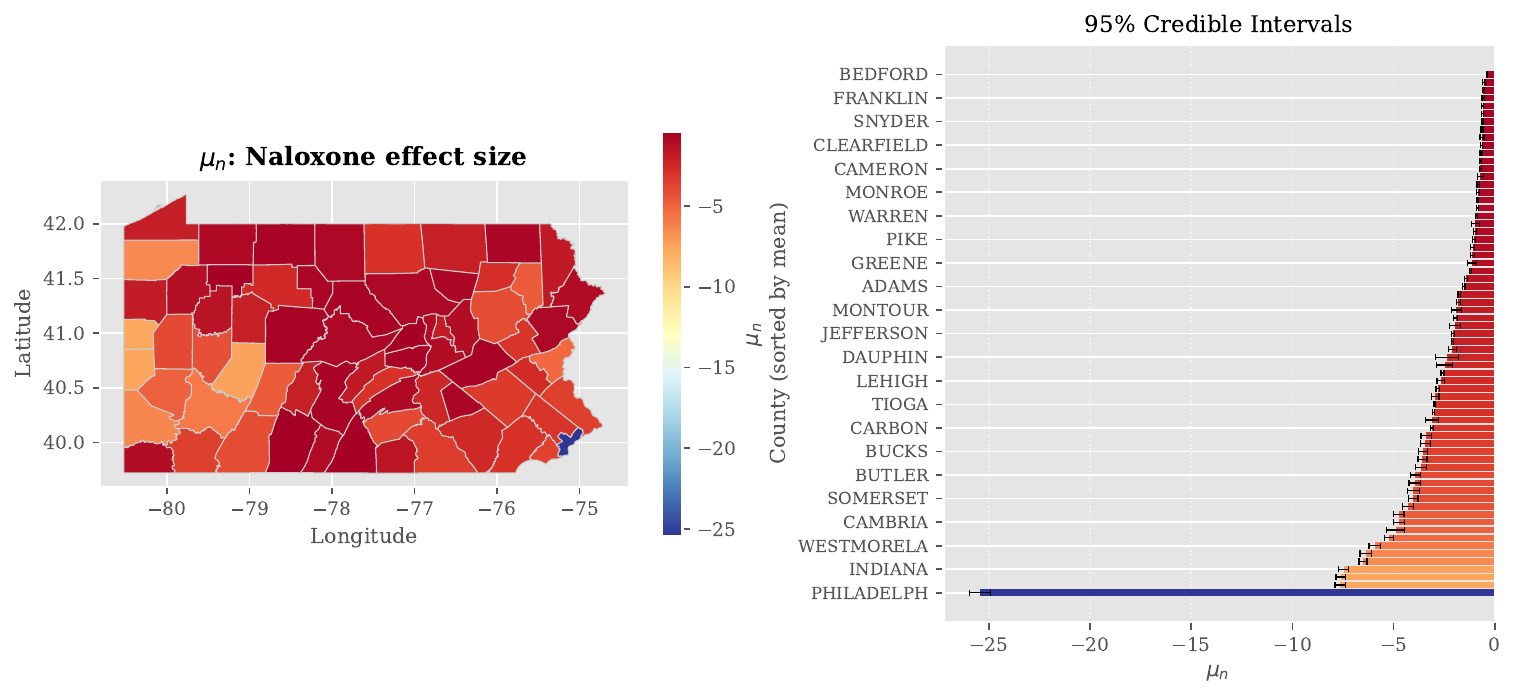}
    
    (b) County-specific estimates of naloxone effect sizes ($\mu_n$) 
    
    \vspace{0.5em}
    
    \includegraphics[width=0.9\textwidth]{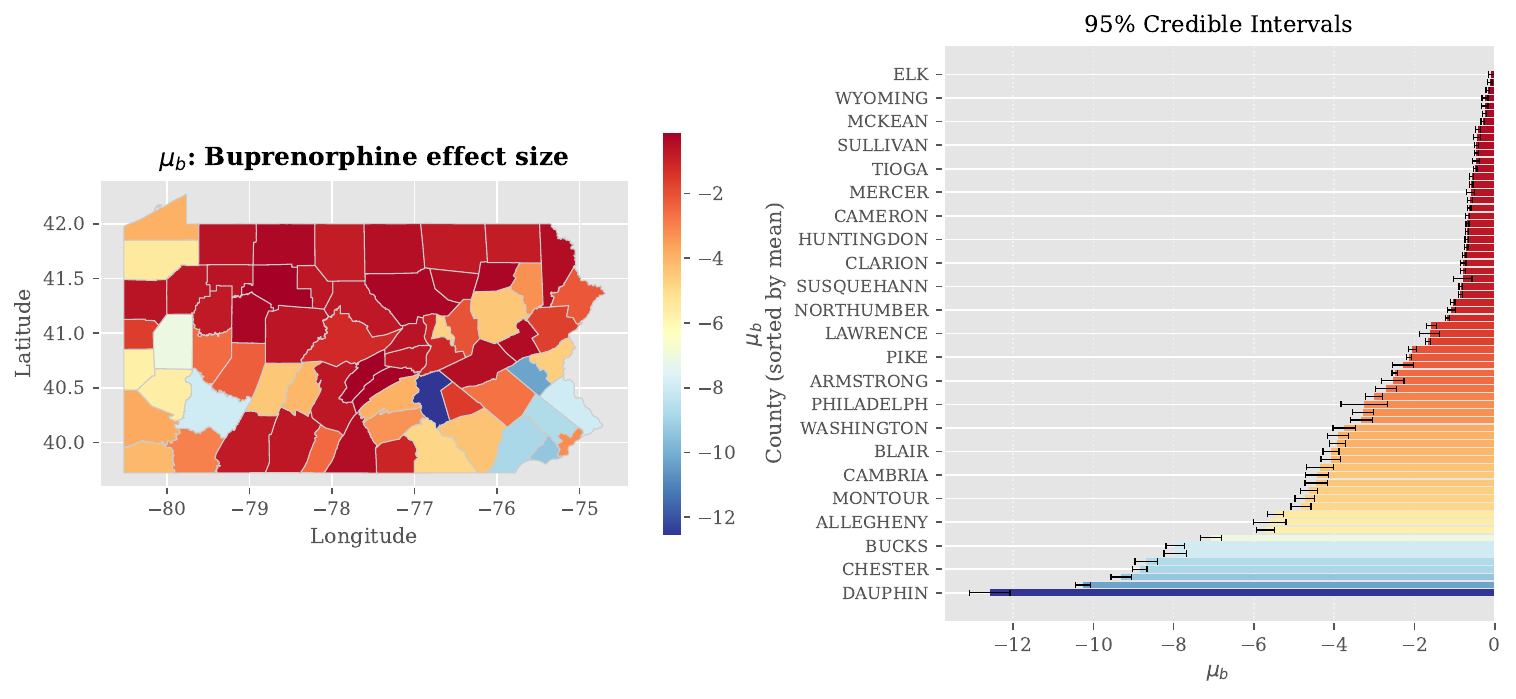}
    
    (c) County-specific estimates of buprenorphine effect sizes ($\mu_b$) 
    
    \caption{\textbf{Posterior summaries of the GPR-estimated response-function coefficients across Pennsylvania counties.} Each panel shows the posterior mean (left) and the corresponding 95\% credible intervals sorted by posterior mean (right). The three coefficients represent: (a)~$\mu_0$: intercept, (b)~$\mu_n$, the change in overdose deaths per 100{,}000 people associated with a one-level (25\%) increase in naloxone dispensing relative to the county’s baseline naloxone dispensing rate; and (c)~$\mu_b$, the corresponding change associated with a one-level (25\%) increase in buprenorphine dispensing relative to the county’s baseline buprenorphine dispensing rate. Together, these coefficients enable the estimation of overdose deaths per 100{,}000 people for any specified combination of naloxone and buprenorphine treatment levels in all counties.}
    \label{fig:effect_res}
\end{figure}

We observed substantial heterogeneity in both baseline overdose mortality and intervention outcome estimates across Pennsylvania counties. Baseline levels differ considerably across regions, as evidenced by large variation in the intercept term $\mu_0$, while the estimated treatment effects for naloxone and buprenorphine are predominantly negative yet vary considerably in magnitude across counties. This pattern indicates that increases in either intervention are generally associated with reductions in overdose deaths, but that the strength of these associations is highly county dependent. In addition, uncertainty in the estimated effects is uneven across regions, with wider credible intervals in counties with more limited simulation or calibration data, reflecting differential information content across the spatial domain. Collectively, these findings suggest that uniform statewide intervention policies are unlikely to be efficient and instead motivate the need for county-specific strategies that account for local baseline mortality and heterogeneous treatment responsiveness, as shown in Figure~\ref{fig:effect_res}.

To better understand the spatially localized and heterogeneous treatment effects identified above, in Table~\ref{tab:coefficients_ci} we report posterior means and 95\% credible intervals for the response-function coefficients in the six calibrated counties that we used as prototypes in our modeling framework. Philadelphia exhibits the highest baseline mortality ($\mu_0$), per 100,000 people, but also the strongest naloxone effect, whereas smaller counties such as Columbia and Clearfield show lower baseline and more modest treatment effects.

\begin{table}[ht]
\centering
\small
\caption{Posterior mean and 95\% credible intervals for response-function coefficients across selected calibrated counties (predicting overdose deaths per 100,000 people over five years).}
\begin{tabular}{lccc}
\hline
\textbf{County} & $\mu_0$ (95\% CI) & $\mu_n$ (95\% CI) & $\mu_b$ (95\% CI) \\ 
\hline
Allegheny 
& $88.96\;[88.36,\;89.56]$
& $-4.26\;[-4.48,\;-4.04]$
& $-5.65\;[-5.87,\;-5.43]$ \\

Philadelphia
& $338.71\;[335.68,\;341.74]$
& $-25.85\;[-26.14,\;-25.56]$
& $-3.91\;[-4.22,\;-3.60]$ \\

Dauphin
& $216.17\;[213.20,\;219.14]$
& $-2.09\;[-2.38,\;-1.79]$
& $-9.65\;[-9.93,\;-9.37]$ \\

Erie
& $109.58\;[107.33,\;111.84]$
& $-3.41\;[-3.66,\;-3.15]$
& $-2.48\;[-2.73,\;-2.23]$ \\

Columbia
& $34.34\;[33.43,\;35.25]$
& $-0.96\;[-1.14,\;-0.77]$
& $-2.22\;[-2.40,\;-2.03]$ \\

Clearfield
& $22.28\;[21.67,\;22.89]$
& $-0.93\;[-1.02,\;-0.85]$
& $-0.80\;[-0.90,\;-0.70]$ \\
\hline
\end{tabular}
\label{tab:coefficients_ci}
\vspace{-5pt}
\end{table}

\begin{figure}[H]
    \centering
    \begin{subfigure}[b]{1\textwidth}
    \centering
    \includegraphics[width=\textwidth]{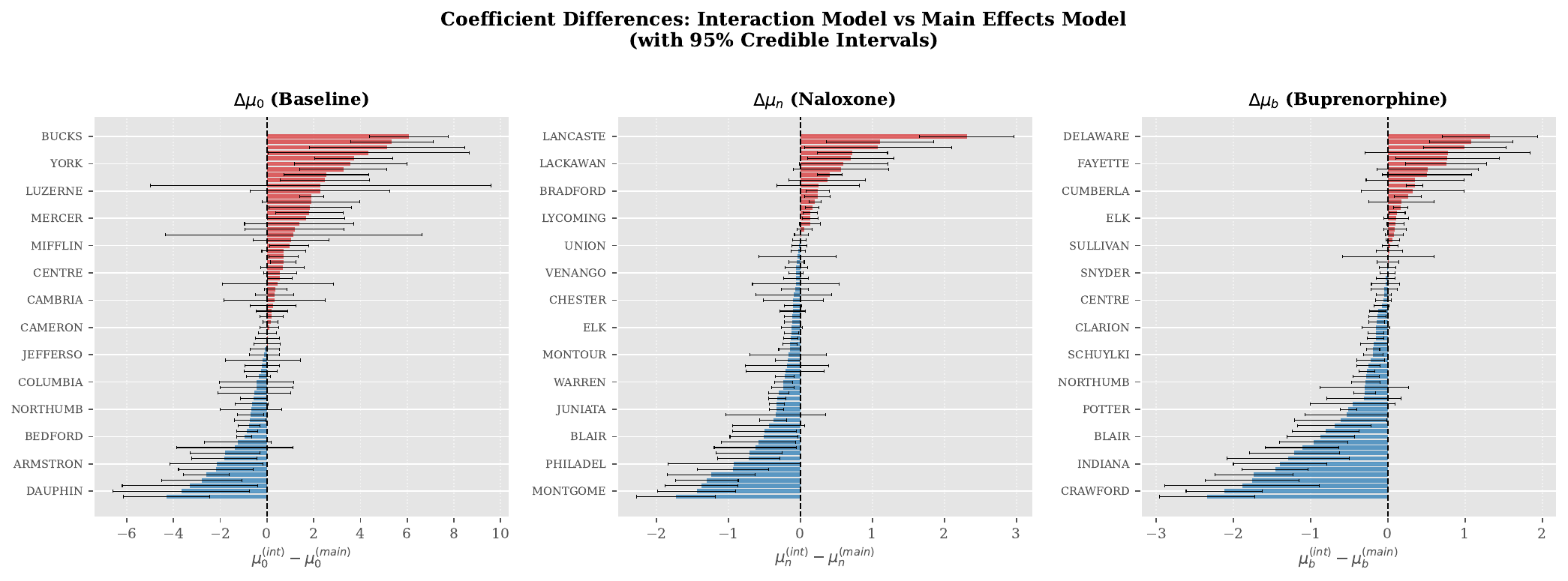}
    \caption{{\small Coefficient differences between the interaction model and the 
    main-effects model, computed as 
    $\Delta\mu = \hat{\mu}^{(\text{int})} - \hat{\mu}^{(\text{main})}$, 
    with 95\% credible intervals for baseline mortality $\Delta\mu_0$ on the left, naloxone effect $\Delta\mu_n$ in the middle, and buprenorphine effect $\Delta\mu_b$ on the right. Counties are sorted by the magnitude of the differences and only the top 14 are shown. 
    }}
    \label{fig:coefficinet-diff}
  \end{subfigure}
    \begin{subfigure}[b]{0.99\textwidth}
    \centering
    \includegraphics[width=\textwidth]{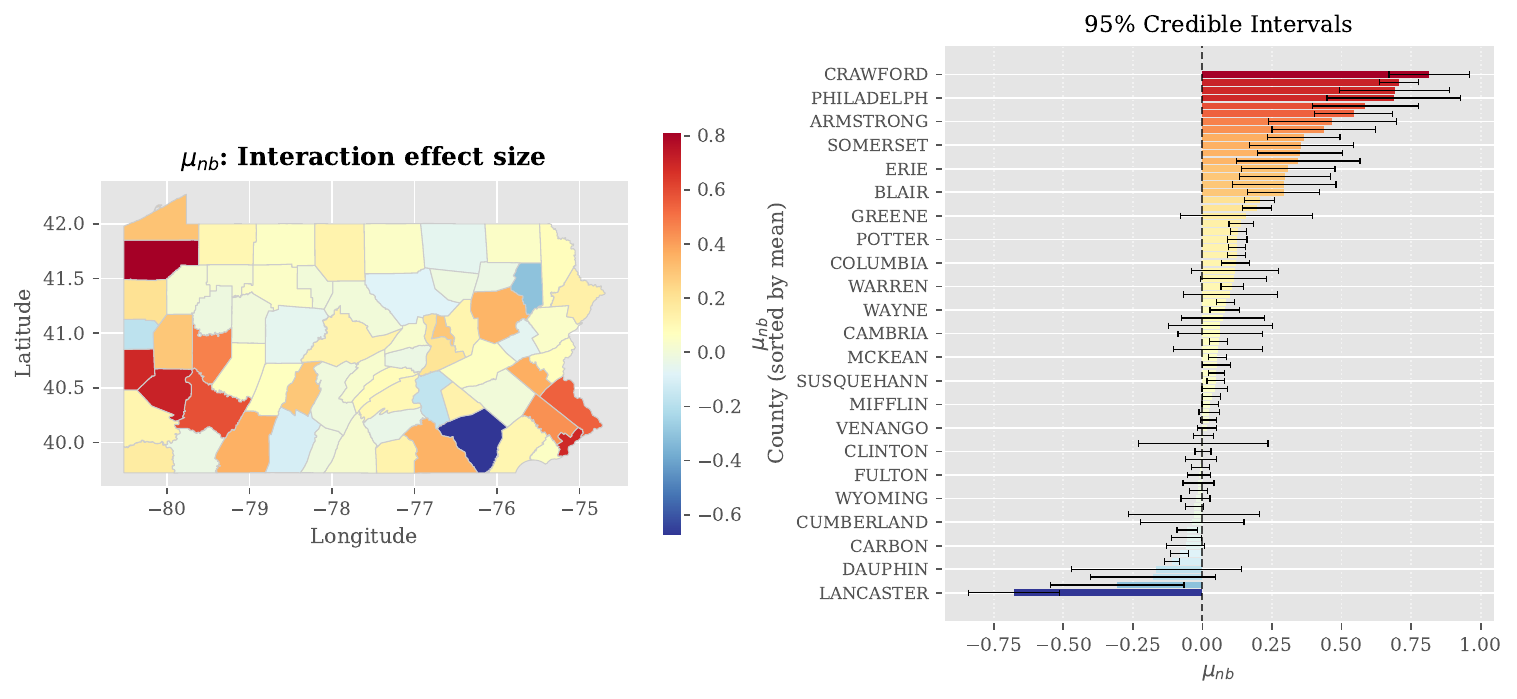}
    \caption{{\small Posterior estimates of the interaction coefficient $\mu_{nb}$ across Pennsylvania counties, 
    showing the spatial distribution of posterior means on the left and 95\% credible intervals on the right, shown only for the 23 largest magnitudes. 
    }}
    \label{fig:interaction-coefficients}
    \end{subfigure}

  \caption{{\bf Robustness analysis comparing estimates obtained for the main-effects model in Equation~\eqref{eq:response_surface2} and the interaction-augmented response function in Equation \eqref{eq:response-function-with-interaction}.}  
    Panel~(a) confirms that adding the interaction term does not substantially alter the main effect 
    estimates, as the majority of coefficient differences have credible intervals spanning zero.
    Panel~(b) shows that the estimated interaction effects are small in magnitude.
    }
  \label{fig:int_effect} 
\end{figure}

\textbf{Robustness to outcome-stage model specification.} To test the stability of the county-level estimates, we compared the naloxone and buprenorphine effect sizes obtained from the two-stage framework under different response function specifications in Equation~\eqref{eq:response-function-with-interaction}, with and without the interaction term. On average, the interaction model produced slightly larger-magnitude estimates than the main-effects-only model, but the differences between the estimates from the two model specifications were consistently small in all counties: the  mean difference (interaction minus main effects) in the estimates for $\mu_n$ was $-0.13$ (SD = $0.61$) and for $\mu_b$ was $-0.25$ (SD = $0.73$). 
In Figure~\ref{fig:coefficinet-diff}, we report the differences between the main effect estimates in the two models (with and without the interaction term) with credible intervals for the top 14 counties with the largest-magnitude differences. The majority of intervals span zero, confirming that the main-effects estimates obtained from the combined GPR and response function modeling framework are robust to the specification of an interaction term in the response function. 

Figure~\ref{fig:interaction-coefficients} shows the county-level estimates of the interaction coefficient ($\mu_{nb}$). The interaction effects are small in magnitude with mean = $0.12$ (SD $=0.24$, 10th–90th percentile: $[-0.06,\,0.45]$) compared to the main effects for naloxone and buprenorphine, whose means are $-2.86$ (SD $=3.40$, 10th–90th percentile: $[-5.50,\,-0.62]$) and $-2.80$ (SD $=2.89$, 10th–90th percentile: $[-7.43,\,-0.37]$), respectively. Most credible intervals for estimated interactions include zero, supporting the main-effects specification used in the primary analysis, and consistent with our observations of parallel trends in the factorial analysis in Figure~\ref{fig:model_selection_slices}. 

To further validate the independent-output GPR formulation, we examined 
the pairwise correlations among iterative coefficient updates $\Delta\beta_m = \beta_m^{(t)} - \beta_m^{(t-1)}$ across all sequential design iterations. The results show that $\text{corr}(\Delta\beta_n, \Delta\beta_b) = -0.04$, confirming that the two treatment effect coefficients update nearly independently throughout training. The stronger 
correlations observed between the intercept and treatment coefficients, $\text{corr}(\Delta\beta_0, \Delta\beta_n) = -0.63$ and $\text{corr}(\Delta\beta_0, \Delta\beta_b) = -0.58$, reflect the well-known intercept-slope trade-off in OLS regression.
\rev{While these low empirical correlations justify the independent-output simplification adopted here, they do not preclude potential gains from a correlated multi-output GP that explicitly models cross-output dependencies, which we identify as a direction for future work.
}\\

\textbf{From localized treatment effects to locally-tailored policies.} Our results show that uniform, state-wide intervention strategies are unlikely to be effective, as baseline overdose mortality and treatment effect magnitudes vary substantially across counties, necessitating locally tailored policies. Recent simulation-based studies of opioid interventions find that the combined effects of harm-reduction strategies (i.e., naloxone distribution), medications for opioid use disorder (i.e., buprenorphine treatment), and diversion-based policies depend on local conditions, and that these interventions can exhibit synergistic effects when deployed jointly, such that the most effective policy combinations vary across communities \shortcite{cerda2024simulating, white2025diversionevaluating}. This conclusion is consistent with work in operations and policy analytics demonstrating that policies optimized at an aggregate level can perform poorly when applied uniformly across heterogeneous regions, whereas explicitly accounting for spatial and contextual heterogeneity improves resource allocation efficiency and policy performance \shortcite{luo2024frontiers}. Complementary epidemiological evidence further shows that opioid overdose risk and intervention effectiveness vary across geographic areas due to differences in local drug environments, population characteristics, and healthcare access, and that population-averaged analyses may not accurately reflect these local patterns \shortcite{dodson2018spatial}. Together, these findings motivate policy evaluation frameworks that support locally adaptive strategies informed by heterogeneous baseline mortality and treatment effects.

\rev{\textbf{Policy interpretation for Pennsylvania.} The estimated county-level effects provide direct evidence for the localized treatment effects motivated above, and illustrate how the framework can guide resource allocation. Because the naloxone and buprenorphine effects vary substantially across counties (Table~\ref{tab:coefficients_ci}, Figure~\ref{fig:effect_res}), the framework identifies where each intervention yields the largest marginal reduction in overdose mortality per proportional increase in dispensing relative to each county's baseline. Counties with large-magnitude naloxone effects, such as Philadelphia, would be expected to benefit most from expanded naloxone distribution, whereas counties where the buprenorphine effect dominates, such as Dauphin, indicate larger returns from expanding medication-based treatment access. For a decision-maker allocating a fixed budget, such as opioid settlement funds, these estimates provide a principled basis for directing resources to the counties and interventions with the greatest expected marginal benefit, rather than distributing funds uniformly, and could also serve as inputs to a cost-effectiveness analysis, though we do not include one here. }

\rev{The interaction estimates in Figure~\ref{fig:int_effect} further inform how the two interventions may be combined. For most counties, the interaction between naloxone and buprenorphine is small and indistinguishable from zero, indicating that the two interventions can be scaled largely independently; a smaller number of counties show a positive interaction, suggesting a compounding benefit when the interventions are deployed jointly; and in a few counties, the estimates suggest that focusing resources on the single most effective intervention may be efficient. This heterogeneity reinforces that the most effective intervention mix is itself county-specific. We emphasize that this study is primarily methodological: it identifies, under the calibrated simulation model, which counties are most responsive to each intervention and with what uncertainty, but it does not address operational questions such as implementation capacity, or cost. }

\FloatBarrier
\clearpage
\thispagestyle{empty}
\clearpage

\textbf{Computational complexity and scalability.}
The $O(N^3)$ cost of GPR posterior inference, arising from the inversion of the $N \times N$ kernel matrix $\mathbf{K}$ \shortcite[Chapter~2]{rasmussen2006gaussian}, applies with respect to $N$, the number of counties incorporated into the training set, rather than the number of simulation replicates. In our setting, $N = 67$ Pennsylvania counties across all iterations: the acquisition function selects among the same fixed set of locations at every step, so the kernel matrix retains its $67 \times 67$ dimensions and the $O(N^3)$ cost remains a constant per-iteration overhead. Scaling to all U.S.\ counties 
($N \approx 3{,}143$) would increase the per-iteration inference cost 
substantially as it grows as $O(N^3)$. In such massive, national-scale implementations, sparse GP 
approximations \shortcite{titsias2009variational, katzfuss2021general} 
would be necessary, replacing the full kernel matrix with a smaller set of $M \ll N$ inducing points (auxiliary locations chosen to summarize the information in the full training set) and reducing the inference cost to $O(M^2 N)$.

Separately, larger county populations require more simulation replicates to achieve the same variance reduction in the regression coefficients, increasing the simulation cost independently of the GP. Regarding kernel structure, the composite RBF 
kernel encoding spatial proximity and socioeconomic similarity is expected to remain applicable across U.S.\ counties since the same county-level features (location, income, population density, racial composition) are available nationwide. However, extending to multiple states would require careful consideration of regional heterogeneity, such as state-level policy environments and drug supply patterns, which may necessitate additional kernel components to capture between-state variation not explained by the current feature set.

\section{Conclusions}
\label{sec:conclude}
This study presents a novel two-stage metamodeling framework to address the computational challenges of evaluating multi-level, multi-component intervention policies across a high-dimensional, spatially heterogeneous domain. The methodology integrates a Gaussian process regression (GPR) surrogate with a response function to efficiently emulate a complex opioid use disorder (OUD) simulation model outcome. The contextual-stage GPR, equipped with a custom composite kernel that incorporates geographic and socio-economic features, learns the simulation outcome by capturing spatial correlations and county-level heterogeneity across communities. 
The outcome-stage response function then provided an interpretable and computationally efficient mechanism for estimating outcomes under any intervention combination within the evaluated grid.

The two-stage metamodel, when coupled with the proposed two-step sequential design, achieves high predictive accuracy and sampling efficiency. Across all counties, the metamodel achieves relative errors of approximately 5\% or less while requiring fewer than 2\% of the simulation runs required for exhaustive enumeration of the county–intervention space. The GPR component enables efficient learning of spatial structure and cross-county heterogeneity in the response-function coefficients, while the outcome-stage model translates these learned relationships into accurate predictions across the full intervention grid. As a result, the framework not only reproduces simulation outcomes with high fidelity but also recovers meaningful county-specific differences in baseline overdose mortality and intervention effect sizes for naloxone and buprenorphine. These empirical findings confirm that the two-step sequential design effectively concentrates computational effort where it is most informative, allowing the metamodel to scale to large, high-dimensional policy spaces without sacrificing interpretability or accuracy.

Beyond methodological advances, the broader impact of this framework lies in its potential to inform policy decisions at the county and state levels. We envision that the proposed framework could serve as the computational backbone of an interactive decision-support tool designed to support policy exploration workflows used by public health agencies, such as the Centers for Disease Control and Prevention (CDC). Such a tool would enable decision makers to explore projected outcomes under alternative treatment allocations across multiple counties and years without requiring exhaustive simulation. 
By leveraging the metamodel, county-level projections of overdose mortality and OUD prevalence could be generated for arbitrary intervention combinations, along with associated credible intervals to support uncertainty-aware comparisons across strategies.
\rev{We emphasize that this framework operates under a model-based prediction paradigm: the estimated localized treatment effects are derived from a calibrated simulation model rather than directly from observational outcomes. As such, their accuracy and policy relevance depend on the validity of the underlying simulation model, including its calibration and structural assumptions.
}



\vspace{-10pt}
\paragraph*{Limitations and Future Work}\mbox{}\\
\rev{We conclude by outlining the main limitations of the study, focused on the 
Pennsylvania case study, together with the future directions they motivate.}
\rev{First, while the current framework already quantifies uncertainty in the estimated coefficients through the GPR posterior, a promising direction is to build on this posterior as an added Bayesian step that extracts information more efficiently.} 
In such a formulation, the GPR posterior would define informative prior distributions over the response-function coefficients, which could then be updated using additional simulation or observational data within a Bayesian model. The Bayesian model would then combine these priors with observed simulation data through Bayes’ theorem to obtain posterior distributions for each coefficient. This hierarchical formulation propagates uncertainty from the GPR into the final coefficient estimates, supports sequential updating as additional data are collected, and provides a principled basis for model assessment. Overall, the framework transforms the metamodel from a predictive surrogate into a probabilistic inference framework that yields comprehensive uncertainty quantification for policy evaluation.

\rev{Second, the current implementation models each response-function coefficient 
using an independent Gaussian process}, a choice that simplifies the implementation while remaining well-justified given that the three coefficients are always combined jointly at the outcome stage through the response function. In future work, the GPR framework could be extended to incorporate cross-output covariance through coregionalization or related multi-output GPR constructions. Such structures would allow the GPR to learn explicit dependencies among the latent functions and may further improve sample efficiency and predictive performance in regions where treatment effects exhibit shared spatial or socio-economic structure.

\rev{Third, the present metamodel is based on cross-sectional data}, 
treating counties as static units, characterized by geographic centroids and socio-economic features over the five-year study period. 
In reality, epidemic trajectories are shaped not only by the underlying disease dynamics but also by policy shifts and emerging drug trends which evolve over time. Subsequently, intervention outcomes can be better optimized when modeled over time and allowed to adapt to the changing dynamics. Incorporating the time dimension would transform the model into a spatio-temporal GPR, where each county’s coefficient vector depends not only on spatial–socio-economic features but also on time. 

A natural starting point would be to extend the GPR kernel to jointly model spatial and temporal variation, capturing smooth evolution of the response-function coefficients over time while preserving the efficiency of the current framework. The heteroscedastic noise model and two-step sequential design would extend naturally to this setting, with the first step selecting the most uncertain county-year combination at each iteration rather than a county alone, enabling the metamodel to capture how treatment effectiveness evolved as the opioid epidemic changed over the 2015--2019 period and to propagate temporal uncertainty into county-level policy recommendations.

\vspace{-10pt}
\section*{Data and Code Availability}
\vspace{-5pt}
Our analysis relies on multiple data streams that collectively capture OUD dynamics at the county level. Monthly county‐level dispensing rates for prescription opioids, naloxone, and buprenorphine were obtained from the IQVIA dataset. These data reflect prescriptions dispensed across retail, mail‐order, and long‐term care pharmacies. This data is not publicly available. Access can be requested through IQVIA at \url{https://www.iqvia.com/insights/the-iqvia-institute/available-iqvia-data}. County‐level overdose death counts were collected from the CDC Wide‐Ranging Online Data for Epidemiologic Research, identified using International Classification of Diseases, 10th Revision (ICD–10) codes for opioid‐related poisoning (X40–X44, X60–X64, X85, Y10–Y14, T40.0–T40.4, T40.6). In addition, fentanyl seizure rates were obtained from the National Forensic Laboratory Information System (NFLIS) and can be accessed at \url{https://www.nflis.deadiversion.usdoj.gov/}.
\rev{The simulation outcomes 
analyzed in this study were generated using FRED, an open-source agent-based 
platform available at 
\url{https://github.com/PublicHealthDynamicsLab/FRED}, which constructs 
census-based synthetic populations }
\ifanonymous
  All code used to implement the GPR metamodel and generate the figures presented in this paper is available at \url{https://zenodo.org/records/19352368}.
\else
  All code used to implement the GPR metamodel and generate the figures presented in this paper is publicly available at \url{https://github.com/abdulrahmanfci/gpr-metamodel}.
\fi

\vspace{-10pt}




\bibliographystyle{apacite}
\bibliography{bib}

@misc{fredwebsite,
    author = "PHDL",
    year = "2024",
  title = "OUD model Phase I",
  URL = "https://fred.publichealth.pitt.edu/",
  note = "[Accessed: 2024-08-31]"
}

@misc{rti,
  title={US Synthetic Population Database 2005--2009: Quick Start Guide. RTI International},
    year={2012},
  author={Wheaton, WD}
}

@article{guclu2016agent,
  title={An agent-based model for addressing the impact of a disaster on access to primary care services},
  author={Guclu, Hasan and Kumar, Supriya and Galloway, David and Krauland, Mary and Sood, Rishi and Bocour, Angelica and Hershey, Tina Batra and van Nostrand, Elizabeth and Potter, Margaret},
  journal={Disaster medicine and public health preparedness},
  volume={10},
  number={3},
  pages={386--393},
  year={2016},
  publisher={Cambridge University Press}
}

@article{grefenstette2013fred,
  title={FRED (A Framework for Reconstructing Epidemic Dynamics): an open-source software system for modeling infectious diseases and control strategies using census-based populations},
  author={Grefenstette, John J and Brown, Shawn T and Rosenfeld, Roni and DePasse, Jay and Stone, Nathan TB and Cooley, Phillip C and Wheaton, William D and Fyshe, Alona and Galloway, David D and Sriram, Anuroop and others},
  journal={BMC public health},
  volume={13},
  number={1},
  pages={1--14},
  year={2013},
  publisher={BioMed Central}
}

@article{fearnhead2014inference,
  title={Inference for reaction networks using the linear noise approximation},
  author={Fearnhead, Paul and Giagos, Vasilieos and Sherlock, Chris},
  journal={Biometrics},
  volume={70},
  number={2},
  pages={457--466},
  year={2014},
  publisher={Wiley Online Library}
}

@article{zimmer2017likelihood,
  title={A likelihood approach for real-time calibration of stochastic compartmental epidemic models},
  author={Zimmer, Christoph and Yaesoubi, Reza and Cohen, Ted},
  journal={PLoS computational biology},
  volume={13},
  number={1},
  pages={e1005257},
  year={2017},
  publisher={Public Library of Science San Francisco, CA USA}
}

@article{ball2017heterogeneous,
  title={Heterogeneous network epidemics: real-time growth, variance and extinction of infection},
  author={Ball, Frank and House, Thomas},
  journal={Journal of Mathematical Biology},
  volume={75},
  number={3},
  pages={577--619},
  year={2017},
  publisher={Springer}
}

@misc{CDCOpioids,
    author = "CDC",
    year = "2024",
  title = "Understanding Drug Overdose and Deaths",
  URL = "https://www.cdc.gov/drugoverdose/epidemic/index.html",
  note = "[Accessed: 2024-08-10]"
}

@article{buckingham2018gaussian,
  title={Gaussian process approximations for fast inference from infectious disease data},
  author={Buckingham-Jeffery, Elizabeth and Isham, Valerie and House, Thomas},
  journal={Mathematical biosciences},
  volume={301},
  pages={111--120},
  year={2018},
  publisher={Elsevier}
}

@article{jalal2018opioiddynamics,
  title={Changing dynamics of the drug overdose epidemic in the United States from 1979 through 2016},
  author={Jalal, Hawre and Buchanich, Jeanine M and Roberts, Mark S and Balmert, Lauren C and Zhang, Kun and Burke, Donald S},
  journal={Science},
  volume={361},
  number={6408},
  pages={eaau1184},
  year={2018},
  publisher={American Association for the Advancement of Science}
}

@inproceedings{senanayake2016aaai,
  title={Predicting spatio-temporal propagation of seasonal influenza using variational Gaussian process regression},
  author={Senanayake, Ransalu and O'Callaghan, Simon and Ramos, Fabio},
  booktitle={Proceedings of the AAAI conference on artificial intelligence},
  volume={30},
  number={1},
  year={2016}
}

@inproceedings{zimmer2020inficml,
  title={Influenza forecasting framework based on Gaussian processes},
  author={Zimmer, Christoph and Yaesoubi, Reza},
  booktitle={International Conference on Machine Learning},
  pages={11671--11679},
  year={2020},
  organization={PMLR}
}

@inproceedings{balandat2020botorch,
  title={{BoTorch: A Framework for Efficient Monte-Carlo Bayesian Optimization}},
  author={Balandat, Maximilian and Karrer, Brian and Jiang, Daniel R. and Daulton, Samuel and Letham, Benjamin and Wilson, Andrew Gordon and Bakshy, Eytan},
  booktitle = {Advances in Neural Information Processing Systems 33},
  year={2020},
  url = {http://arxiv.org/abs/1910.06403}
}

@book{forrester2008surrogate,
  title={Engineering design via surrogate modelling: a practical guide},
  author={Forrester, Alexander and Sobester, Andras and Keane, Andy},
  year={2008},
  publisher={John Wiley \& Sons}
}

@book{gramacy2020surrogates,
  title={Surrogates: Gaussian process modeling, design, and optimization for the applied sciences},
  author={Gramacy, Robert B},
  year={2020},
  publisher={Chapman and Hall/CRC}
}

@article{kennedy2001bayesian,
  title={Bayesian calibration of computer models},
  author={Kennedy, Marc C and O'Hagan, Anthony},
  journal={Journal of the Royal Statistical Society: Series B (Statistical Methodology)},
  volume={63},
  number={3},
  pages={425--464},
  year={2001},
  publisher={Wiley Online Library}
}

@article{conti2010bayesian,
  title={Bayesian emulation of complex multi-output and dynamic computer models},
  author={Conti, Stefano and O’Hagan, Anthony},
  journal={Journal of statistical planning and inference},
  volume={140},
  number={3},
  pages={640--651},
  year={2010},
  publisher={Elsevier}
}

@article{banerjee2008gaussian,
  title={Gaussian predictive process models for large spatial data sets},
  author={Banerjee, Sudipto and Gelfand, Alan E and Finley, Andrew O and Sang, Huiyan},
  journal={Journal of the Royal Statistical Society: Series B (Statistical Methodology)},
  volume={70},
  number={4},
  pages={825--848},
  year={2008},
  publisher={Wiley}
}

@book{rasmussen2006gaussian,
  title={Gaussian Processes for Machine Learning},
  author={Rasmussen, Carl Edward and Williams, Christopher KI},
  year={2006},
  publisher={MIT press}
}

@article{brochu2010tutorial,
  title={A tutorial on Bayesian optimization of expensive cost functions, with application to active user modeling and hierarchical reinforcement learning},
  author={Brochu, Eric and Cora, Vlad M and De Freitas, Nando},
  journal={arXiv preprint arXiv:1012.2599},
  year={2010}
}

@article{imisraftery2010,
  title={Estimating and projecting trends in HIV/AIDS generalized epidemics using incremental mixture importance sampling},
  author={Raftery, Adrian E and Bao, Le},
  journal={Biometrics},
  volume={66},
  number={4},
  pages={1162--1173},
  year={2010},
  publisher={Oxford University Press}
}

@article{bhatt2017improved,
  title={Improved prediction accuracy for disease risk mapping using Gaussian process stacked generalization},
  author={Bhatt, Samir and Cameron, Ewan and Flaxman, Seth R and Weiss, Daniel J and Smith, David L and Gething, Peter W},
  journal={Journal of The Royal Society Interface},
  volume={14},
  number={134},
  pages={20170520},
  year={2017},
  publisher={The Royal Society}
}

@article{cerda2021systematic,
  title={A systematic review of simulation models to track and address the opioid crisis},
  author={Cerd{\'a}, Magdalena and Jalali, Mohammad S and Hamilton, Ava D and DiGennaro, Catherine and Hyder, Ayaz and Santaella-Tenorio, Julian and Kaur, Navdep and Wang, Christina and Keyes, Katherine M},
  journal={Epidemiologic reviews},
  volume={43},
  number={1},
  pages={147--165},
  year={2021},
  publisher={Oxford University Press}
}

@article{lim2022modeling,
  title={Modeling the evolution of the US opioid crisis for national policy development},
  author={Lim, Tse Yang and Stringfellow, Erin J and Stafford, Celia A and DiGennaro, Catherine and Homer, Jack B and Wakeland, Wayne and Eggers, Sara L and Kazemi, Reza and Glos, Lukas and Ewing, Emily G and others},
  journal={Proceedings of the National Academy of Sciences},
  volume={119},
  number={23},
  pages={e2115714119},
  year={2022},
  publisher={National Academy of Sciences}
}

@incollection{frazier2018bayesian,
  title={Bayesian optimization},
  author={Frazier, Peter I},
  booktitle={Recent advances in optimization and modeling of contemporary problems},
  pages={255--278},
  year={2018},
  publisher={Informs}
}

@article{wilson2018maximizing,
  title={Maximizing acquisition functions for Bayesian optimization},
  author={Wilson, James and Hutter, Frank and Deisenroth, Marc},
  journal={Advances in neural information processing systems},
  volume={31},
  year={2018}
}

@article{langmuller2024gaussian,
  title={Gaussian Process Emulation for Exploring Complex Infectious Disease Models},
  author={Langm{\"u}ller, Anna M and Chandrasekher, Kiran A and Haller, Benjamin C and Champer, Samuel E and Murdock, Courtney C and Messer, Philipp W},
  journal={medRxiv},
  pages={2024--11},
  year={2024},
  publisher={Cold Spring Harbor Laboratory Press}
}

@article{sawe2024gaussian,
  title={Gaussian process emulation to improve efficiency of computationally intensive multidisease models: a practical tutorial with adaptable R code},
  author={Sawe, Sharon Jepkorir and Mugo, Richard and Wilson-Barthes, Marta and Osetinsky, Brianna and Chrysanthopoulou, Stavroula A and Yego, Faith and Mwangi, Ann and Gal{\'a}rraga, Omar},
  journal={BMC Medical Research Methodology},
  volume={24},
  number={1},
  pages={26},
  year={2024},
  publisher={Springer}
}

@article{reiker2021emulator,
  title={Emulator-based Bayesian optimization for efficient multi-objective calibration of an individual-based model of malaria},
  author={Reiker, Theresa and Golumbeanu, Monica and Shattock, Andrew and Burgert, Lydia and Smith, Thomas A and Filippi, Sarah and Cameron, Ewan and Penny, Melissa A},
  journal={Nature communications},
  volume={12},
  number={1},
  pages={7212},
  year={2021},
  publisher={Nature Publishing Group UK London}
}

@inproceedings{zimmer2020influenza,
  title={Influenza forecasting framework based on Gaussian processes},
  author={Zimmer, Christoph and Yaesoubi, Reza},
  booktitle={International conference on machine learning},
  pages={11671--11679},
  year={2020},
  organization={PMLR}
}

@book{banerjee2003hierarchical,
  title={Hierarchical modeling and analysis for spatial data},
  author={Banerjee, Sudipto and Carlin, Bradley P and Gelfand, Alan E},
  year={2003},
  publisher={Chapman and Hall/CRC}
}

@book{moraga2023spatial,
  title={Spatial statistics for data science: theory and practice with R},
  author={Moraga, Paula},
  year={2023},
  publisher={Chapman and Hall/CRC}
}

@article{cheng2020sparse,
  title={Sparse multi-output Gaussian processes for online medical time series prediction},
  author={Cheng, Li-Fang and Dumitrascu, Bianca and Darnell, Gregory and Chivers, Corey and Draugelis, Michael and Li, Kai and Engelhardt, Barbara E},
  journal={BMC medical informatics and decision making},
  volume={20},
  number={1},
  pages={152},
  year={2020},
  publisher={Springer}
}

@article{gutierrez2024multi,
  title={Multi-output prediction of dose--response curves enables drug repositioning and biomarker discovery},
  author={Gutierrez, Juan-Jos{\'e} Giraldo and Lau, Evelyn and Dharmapalan, Subhashini and Parker, Melody and Chen, Yurui and {\'A}lvarez, Mauricio A and Wang, Dennis},
  journal={npj Precision Oncology},
  volume={8},
  number={1},
  pages={209},
  year={2024},
  publisher={Nature Publishing Group UK London}
}

@inproceedings{li2022safe,
  title={Safe active learning for multi-output gaussian processes},
  author={Li, Cen-You and Rakitsch, Barbara and Zimmer, Christoph},
  booktitle={International Conference on Artificial Intelligence and Statistics},
  pages={4512--4551},
  year={2022},
  organization={PMLR}
}

@inproceedings{sanchez2020work,
  title={Work smarter, not harder: A tutorial on designing and conducting simulation experiments},
  author={Sanchez, Susan M and Sanchez, Paul J and Wan, Hong},
  booktitle={2020 Winter Simulation Conference (WSC)},
  pages={1128--1142},
  year={2020},
  organization={IEEE}
}

@inproceedings{fisher2020predicting,
  title={Predicting the resource needs and outcomes of computationally intensive biological simulations},
  author={Fisher, Andrew and Adhikari, Bhisma and Zhai, Chao and Morgan, Joshua E and Mago, Vijay K and Giabbanelli, Philippe J},
  booktitle={2020 Spring Simulation Conference (SpringSim)},
  pages={1--12},
  year={2020},
  organization={IEEE}
}

@article{luo2024frontiers,
  title={Frontiers in operations: Equitable data-driven facility location and resource allocation to fight the opioid epidemic},
  author={Luo, Joyce and Stellato, Bartolomeo},
  journal={Manufacturing \& Service Operations Management},
  volume={26},
  number={4},
  pages={1229--1244},
  year={2024},
  publisher={INFORMS}
}

@article{dodson2018spatial,
  title={Spatial methods to enhance public health surveillance and resource deployment in the opioid epidemic},
  author={Dodson, Zan M and Enki Yoo, Eun-Hye and Martin-Gill, Christian and Roth, Ronald},
  journal={American journal of public health},
  volume={108},
  number={9},
  pages={1191--1196},
  year={2018},
  publisher={American Public Health Association}
}

@article{cerda2024simulating,
  title={Simulating the simultaneous impact of medication for opioid use disorder and naloxone on opioid overdose death in eight New York counties},
  author={Cerd{\'a}, Magdalena and Hamilton, Ava D and Hyder, Ayaz and Rutherford, Caroline and Bobashev, Georgiy and Epstein, Joshua M and Hatna, Erez and Krawczyk, Noa and El-Bassel, Nabila and Feaster, Daniel J and others},
  journal={Epidemiology},
  volume={35},
  number={3},
  pages={418--429},
  year={2024},
  publisher={LWW}
}

@article{white2025diversionevaluating,
  title={Evaluating diversion and treatment policies for opioid use disorder},
  author={White, Veronica M and Albert, Laura A},
  journal={IISE Transactions on Healthcare Systems Engineering},
  pages={1--28},
  year={2025},
  publisher={Taylor \& Francis}
}

@article{menzies2017bayesian,
  title={Bayesian methods for calibrating health policy models: a tutorial},
  author={Menzies, Nicolas A and Soeteman, Dj{\o}ra I and Pandya, Ankur and Kim, Jane J},
  journal={Pharmacoeconomics},
  volume={35},
  number={6},
  pages={613--624},
  year={2017},
  publisher={Springer}
}

@inproceedings{titsias2009variational,
  title={Variational learning of inducing variables in sparse Gaussian processes},
  author={Titsias, Michalis},
  booktitle={Artificial intelligence and statistics},
  pages={567--574},
  year={2009},
  organization={PMLR}
}

@article{katzfuss2021general,
  title={A general framework for Vecchia approximations of Gaussian processes},
  author={Katzfuss, Matthias and Guinness, Joseph},
  journal={Statistical Science},
  volume={36},
  number={1},
  pages={124--141},
  year={2021},
  publisher={JSTOR}
}

@article{naumann2019naloxone,
  title={Impact of a community-based naloxone distribution program on opioid overdose death rates},
  author={Naumann, Rebecca B and Durrance, Christine Piette and Ranapurwala, Shabbar I and Austin, Anna E and Proescholdbell, Scott and Childs, Robert and Marshall, Stephen W and Kansagra, Susan and Shanahan, Meghan E},
  journal={Drug and alcohol dependence},
  volume={204},
  pages={107536},
  year={2019},
  publisher={Elsevier}
}

@article{walley2013opioid,
  title={Opioid overdose rates and implementation of overdose education and nasal naloxone distribution in Massachusetts: interrupted time series analysis},
  author={Walley, Alexander Y and Xuan, Ziming and Hackman, H Holly and Quinn, Emily and Doe-Simkins, Maya and Sorensen-Alawad, Amy and Ruiz, Sarah and Ozonoff, Al},
  journal={Bmj},
  volume={346},
  year={2013},
  publisher={British Medical Journal Publishing Group}
}

@article{friedman2026charting,
  title={Charting the decline of the fourth wave: US overdose deaths by race, ethnicity and substance involvement},
  author={Friedman, Joseph R and Palamar, Joseph J and Ciccarone, Daniel and Gaines, Tommi L and Borquez, Annick and Shover, Chelsea L and Strathdee, Steffanie A},
  journal={Addiction},
  year={2026},
  publisher={Wiley Online Library}
}

@article{goedel2020association,
  title={Association of racial/ethnic segregation with treatment capacity for opioid use disorder in counties in the United States},
  author={Goedel, William C and Shapiro, Aaron and Cerd{\'a}, Magdalena and Tsai, Jennifer W and Hadland, Scott E and Marshall, Brandon DL},
  journal={JAMA network open},
  volume={3},
  number={4},
  pages={e203711},
  year={2020}
}

@article{cheng2022bupre,
  title={Expanding access to medications for opioid use disorder in primary care clinics: an evaluation of common implementation strategies and outcomes},
  author={Cheng, Hannah and McGovern, Mark P and Garneau, H{\'e}l{\`e}ne Chokron and Hurley, Brian and Fisher, Tammy and Copeland, Meaghan and Almirall, Daniel},
  journal={Implementation Science Communications},
  volume={3},
  number={1},
  pages={72},
  year={2022},
  publisher={Springer}
}

@article{wen2017bupre,
  title={Impact of Medicaid expansion on Medicaid-covered utilization of buprenorphine for opioid use disorder treatment},
  author={Wen, Hefei and Hockenberry, Jason M and Borders, Tyrone F and Druss, Benjamin G},
  journal={Medical care},
  volume={55},
  number={4},
  pages={336--341},
  year={2017},
  publisher={LWW}
}

@article{nosyk2024buprenorphine,
  title={Buprenorphine/naloxone vs methadone for the treatment of opioid use disorder},
  author={Nosyk, Bohdan and Min, Jeong Eun and Homayra, Fahmida and Kurz, Megan and Guerra-Alejos, Brenda Carolina and Yan, Ruyu and Piske, Micah and Seaman, Shaun R and Bach, Paxton and Greenland, Sander and others},
  journal={Jama},
  volume={332},
  number={21},
  pages={1822--1831},
  year={2024}
}

@article{d2015linkage,
  title={Emergency department--initiated buprenorphine/naloxone treatment for opioid dependence: a randomized clinical trial},
  author={D’Onofrio, Gail and O’Connor, Patrick G and Pantalon, Michael V and Chawarski, Marek C and Busch, Susan H and Owens, Patricia H and Bernstein, Steven L and Fiellin, David A},
  journal={Jama},
  volume={313},
  number={16},
  pages={1636--1644},
  year={2015}
}

@article{liebschutz2014linkage,
  title={Buprenorphine treatment for hospitalized, opioid-dependent patients: a randomized clinical trial},
  author={Liebschutz, Jane M and Crooks, Denise and Herman, Debra and Anderson, Bradley and Tsui, Judith and Meshesha, Lidia Z and Dossabhoy, Shernaz and Stein, Michael},
  journal={JAMA internal medicine},
  volume={174},
  number={8},
  pages={1369--1376},
  year={2014}
}

@misc{garnett2024cdcdeaths,
  title={Drug overdose deaths in the United States, 2003--2023},
  author={Garnett, Matthew F and Mini{\~n}o, Arialdi M},
  year={2024},
  institution={Centers for Disease Control and Prevention (US)}
}

@misc{CDC2025overdosemap,
  author       = {{Centers for Disease Control and Prevention}},
  title        = {Drug Overdose Mortality: Stats of the States},
  year         = {2025},
  institution  = {National Center for Health Statistics},
  url          = {https://www.cdc.gov/nchs/state-stats/deaths/drug-overdose.html},
  note         = {Accessed: March 26, 2026}
}

@article{volkow2021opioid,
  title={The changing opioid crisis: development, challenges and opportunities},
  author={Volkow, Nora D and Blanco, Carlos},
  journal={Molecular psychiatry},
  volume={26},
  number={1},
  pages={218--233},
  year={2021},
  publisher={Nature Publishing Group UK London}
}

@article{jenkins2021opioid,
  title={The fourth wave of the US opioid epidemic and its implications for the rural US: a federal perspective},
  author={Jenkins, Richard A},
  journal={Preventive medicine},
  volume={152},
  pages={106541},
  year={2021},
  publisher={Elsevier}
}

@incollection{ciccarone2021epidemiology,
  title={The epidemiology of the opioid overdose epidemic in the United States},
  author={Ciccarone, Daniel},
  booktitle={The Opioid Epidemic and Infectious Diseases},
  pages={1--10},
  year={2021},
  publisher={Elsevier}
}

@misc{ahmed2022epidemic,
  title={A changing epidemic and the rise of opioid-stimulant co-use},
  author={Ahmed, Saeed and Sarfraz, Zouina and Sarfraz, Azza},
  journal={Frontiers in psychiatry},
  volume={13},
  pages={918197},
  year={2022},
  publisher={Frontiers Media SA}
}

@article{scheidell2024reducing,
  title={Reducing overdose deaths among persons with opioid use disorder in connecticut},
  author={Scheidell, Joy D and Townsend, Tarlise N and Zhou, Qinlian and Manandhar-Sasaki, Prima and Rodriguez-Santana, Ramon and Jenkins, Mark and Buchelli, Marianne and Charles, Dyanna L and Frechette, Jillian M and Su, Jasmine I-Shin and others},
  journal={Harm reduction journal},
  volume={21},
  number={1},
  pages={103},
  year={2024},
  publisher={Springer}
}

@article{irvine2022estimating,
  title={Estimating naloxone need in the USA across fentanyl, heroin, and prescription opioid epidemics: a modelling study},
  author={Irvine, Michael A and Oller, Declan and Boggis, Jesse and Bishop, Brian and Coombs, Daniel and Wheeler, Eliza and Doe-Simkins, Maya and Walley, Alexander Y and Marshall, Brandon DL and Bratberg, Jeffrey and others},
  journal={The Lancet Public Health},
  volume={7},
  number={3},
  pages={e210--e218},
  year={2022},
  publisher={Elsevier}
}

@article{zang2022comparing,
  title={Comparing projected fatal overdose outcomes and costs of strategies to expand community-based distribution of naloxone in Rhode Island},
  author={Zang, Xiao and Bessey, Sam E and Krieger, Maxwell S and Hallowell, Benjamin D and Koziol, Jennifer A and Nolen, Shayla and Behrends, Czarina N and Murphy, Sean M and Walley, Alexander Y and Linas, Benjamin P and others},
  journal={JAMA network open},
  volume={5},
  number={11},
  pages={e2241174},
  year={2022}
}

@article{blanco2020america,
  title={America’s opioid crisis: the need for an integrated public health approach},
  author={Blanco, Carlos and Wiley, Tisha RA and Lloyd, Jacqueline J and Lopez, Marsha F and Volkow, Nora D},
  journal={Translational psychiatry},
  volume={10},
  number={1},
  pages={167},
  year={2020},
  publisher={Nature Publishing Group UK London}
}

@article{marotta2019assessing,
  title={Assessing spatial relationships between prescription drugs, race, and overdose in New York State from 2013 to 2015},
  author={Marotta, Phillip L and Hunt, Tim and Gilbert, Louisa and Wu, Elwin and Goddard-Eckrich, Dawn and El-Bassel, Nabila},
  journal={Journal of psychoactive drugs},
  volume={51},
  number={4},
  pages={360--370},
  year={2019},
  publisher={Taylor \& Francis}
}

@article{zheng2025opioid,
  title={Opioid use disorder prevalence in 57 New York counties from 2017 to 2019: A Bayesian evidence synthesis},
  author={Zheng, Tian and Keyes, Katherine and Ji, Shouxuan and Calderon, Anna and Wu, Elwin and Doogan, Nathan J and Villani, Jennifer and Chandler, Redonna and Barocas, Joshua A and Nguyen, Trang and others},
  journal={Drug and alcohol dependence},
  volume={267},
  pages={112548},
  year={2025},
  publisher={Elsevier}
}

@inproceedings{ahmed2023inferring,
  title={Inferring epidemic dynamics using Gaussian process emulation of agent-based simulations},
  author={{Ahmed}, Abdulrahman A and Rahimian, M Amin and Roberts, Mark S},
  booktitle={{Proceedings of Winter Simulation Conference ({WSC})}},
  year={2023},
  organization={IEEE}
}

@inproceedings{ahmedbhi,
  title={Estimating Treatment Effects Using Costly Simulation Samples from a Population-Scale Model of Opioid Use Disorder},
  author={{Ahmed}, Abdulrahman A and Rahimian, M Amin and Roberts, Mark S},
  booktitle={{2023 IEEE EMBS International Conference on Biomedical and Health Informatics ({BHI})}},
  pages={1--4},
  year={2023},
  organization={IEEE}
}

@inproceedings{ahmed2024selection,
  title={Optimized Model Selection for Estimating Treatment Effects from Costly Simulations of the US Opioid Epidemic},
  author={{Ahmed}, Abdulrahman A and Rahimian, M Amin and Roberts, Mark S},
  booktitle={2024 Annual Modeling and Simulation Conference (ANNSIM)},
  pages={1--13},
  year={2024},
  organization={IEEE}
}

@article{ahmed2026oud,
  title={County-Level Heterogeneity in Opioid Harm Reduction and Treatment Effects: A Calibrated Markov Model},
  author={Ahmed, Abdulrahman and Rahimian, M Amin and Chen, Qiushi and Kumar, Praveen},
  journal={medRxiv},
  pages={2026--05},
  year={2026},
  publisher={Cold Spring Harbor Laboratory Press}
}

\clearpage
\appendix

\section{Other Related Work}
\label{app:related_work}


Estimating a single treatment effect from a population-scale simulation is feasible. However, the challenge intensifies when multiple treatment effects must be evaluated, particularly across a wide design space. In such settings, the computational cost of running repeated simulations becomes a significant barrier. Metamodels offer a promising solution by approximating simulation outputs without requiring full evaluations at every point. Among various approaches, Gaussian Process regression (GPR) has emerged as a widely used metamodeling technique due to its flexibility, ability to quantify uncertainty, and strong performance in interpolating between sparse observations.

Several studies have demonstrated the utility of GPR in modeling epidemic dynamics and public health processes. For instance, \shortciteA{fearnhead2014inference} use GPR models to infer reaction rates in biological networks, showing how their method generalizes to a broad class of reaction systems relevant to epidemic modeling. \shortciteA{senanayake2016aaai} develop a spatiotemporal GPR framework to forecast influenza trends using large-scale data, incorporating multiple kernels to capture seasonality, non-stationarity, and long- and short-term patterns. \shortciteA{zimmer2020inficml} apply GPR regression to influenza forecasting using CDC data, benchmarking its performance against alternative methods. Similarly, \shortciteA{zimmer2017likelihood} employ GPR models to calibrate epidemic parameters such as the duration of infectiousness and expected future cases in real time. \shortciteA{ball2017heterogeneous} estimate the mean and variance of infections in an SIR network model using a GPR informed by a branching process covariance structure. \shortciteA{buckingham2018gaussian} propose a GPR-based framework for SIR and SEIR models, comparing multiple GPR variants for parameter estimation. 

\section{Model Selection in the Two-stage Framework}
\label{sec:app:model-select}
\begin{figure}[H]
  \centering
  \begin{subfigure}[b]{0.49\textwidth}
    \centering
    \includegraphics[width=\textwidth]{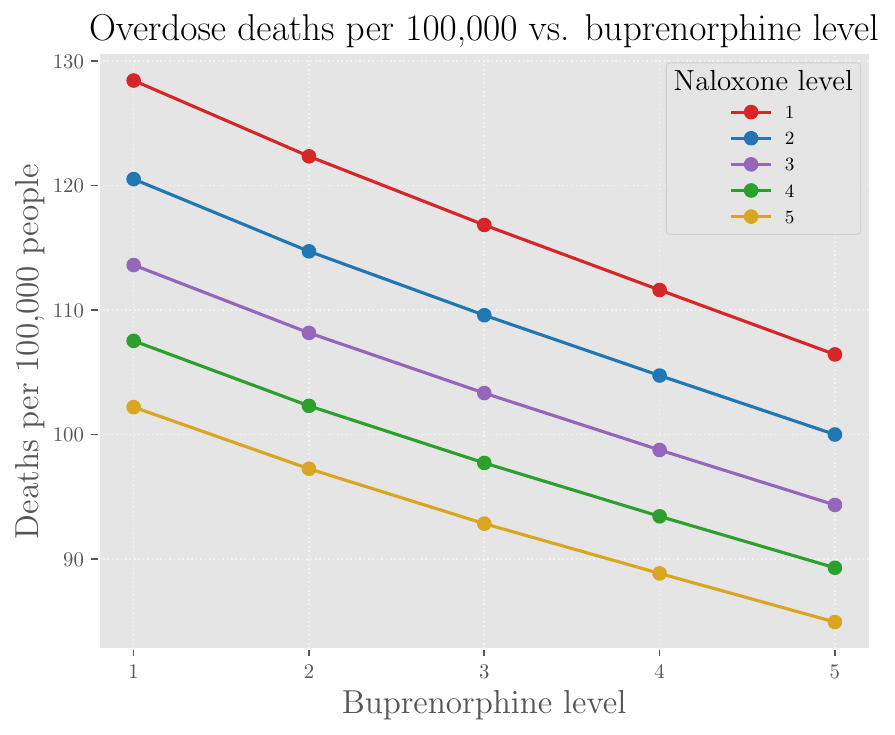}
    \caption{{\small Overdose deaths per 100{,}000 people in Pennsylvania vs.\ buprenorphine level after five years; separate lines show naloxone levels 1–5. Lines are nearly parallel (no crossings), indicating no interaction.}}
  \end{subfigure}
  \hfill
  \begin{subfigure}[b]{0.49\textwidth}
    \centering
    \includegraphics[width=\textwidth]{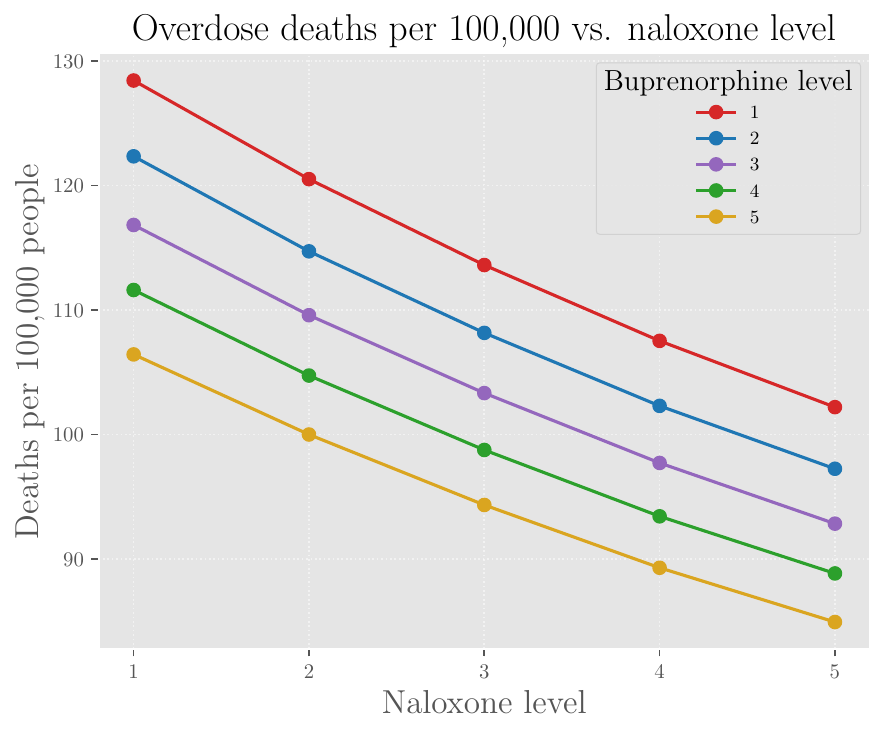}
    \caption{{\small Overdose deaths per 100{,}000 people in Pennsylvania vs.\ naloxone level after five years; separate lines show buprenorphine levels 1–5. Trends are again near-parallel, indicating no interaction. }}
  \end{subfigure}

  \vskip\baselineskip
  \caption[]{\small {\bf Factorial plots over the $5\times5$ intervention grid.} The near-parallel trends in both figures provide no visual evidence of interaction between naloxone and buprenorphine. A simple main-effects model, $z(n,b \mid c) = \mu_0(\mathbf{x}_c) + \mu_n(\mathbf{x}_c) \cdot n + \mu_b(\mathbf{x}_c) \cdot b$, can therefore be sufficient for outcome-stage metamodeling (outcomes shown are cumulative five-year overdose deaths, averaged over 500 simulation replications).}
  \label{fig:model_selection_slices}
  
\end{figure}

\textbf{Response function design.} 
To assess the interaction between naloxone and buprenorphine, we use two factorial plots showing the cumulative deaths per 100,00 people in Pennsylvania (statewide) over the five-year study period (2015-2019) for different treatment levels (Figure~\ref{fig:model_selection_slices}): panel~(a) shows mean overdose deaths per 100{,}000 versus buprenorphine level with separate lines for naloxone levels 1-5; panel~(b) reverses the roles. In both panels, the lines are nearly parallel with no systematic crossings, indicating additive main effects and little evidence of interaction. This supports our choice of the simple main-effects model in Equation~\ref{eq:response_surface2} for our main analysis. This observation is consistent with the small magnitudes of the estimated interaction coefficients $\mu_{nb}$ in our robustness checks (Figure~\ref{fig:interaction-coefficients}). 

\textbf{GPR Kernel Design.}
\label{sec:app:kernel-design}
The kernel function evaluates the covariance structure between any two input feature vectors, denoted by $\mathbf{x}_c$ and $\mathbf{x}_{c'}$ where $\mathbf{x}_c\in\mathbb{R}^d$ represents the $d$-dimensional vector of county-level covariates used by the Gaussian process.
Different kernel functions may be employed depending on the characteristics of the underlying process being modeled.

The radial basis function (RBF) kernel is one of the most widely used covariance functions in Gaussian process modeling and is defined as
\begin{align}
    k_{{RBF}}(\mathbf{x}_c,\mathbf{x}_{c'}) = \exp(-\dfrac{d(\mathbf{x}_c,\mathbf{x}_{c'})^2}{2l^2}),
    \label{eq:RBF}
\end{align}
where $l$ is the length scale parameter and $d(\cdot,\cdot)$ is the Euclidean distance between the input points.
RBF kernels, also known as squared exponential kernels, are widely used in GPR modeling due to their ability to model smooth, nonlinear relationships \shortcite[Chapter~4]{rasmussen2006gaussian}. In prior work, RBF kernels have been used to encode similarity in continuous variables such as time \shortcite{gramacy2020surrogates}, geographic distance \shortcite{banerjee2008gaussian}, or patient-level covariates in healthcare studies \shortcite{brochu2010tutorial}.

In constructing our custom kernel, we incorporate four radial basis function (RBF) kernels. Candidate socio-economic features were first identified based on their potential relevance to explaining variation in overdose mortality across counties, including unemployment, poverty, the primary care physicians rate, rurality index, and racial composition. We then applied a greedy feature selection procedure, sequentially adding features that reduced out-of-sample prediction error and discarding those that did not yield improvement.
The final kernel structure includes four RBF components, each corresponding to a selected contextual feature: the county’s geographic location (via centroid coordinates), median household income, population density, and the percentage of black residents, 
as follows:
\begin{align}
k(\mathbf{x}_c,\mathbf{x}_{c'}) = &k_{RBF}(\mathbf{x}_{c,1:2},\mathbf{x}_{c',1:2}) + k_{RBF}(\mathbf{x}_{c,3},\mathbf{x}_{c',3}) \nonumber  \\& +
k_{RBF}(\mathbf{x}_{c,4},\mathbf{x}_{c',4}) + k_{RBF}(\mathbf{x}_{c,5},\mathbf{x}_{c',5}),
\label{eq:kernelcomposition}
\end{align}
where $\mathbf{x}_{c,1:2}$ corresponds to the county centroid coordinates (latitude and longitude), $\mathbf{x}_{c,3}$ denotes median household income, $\mathbf{x}_{c,4}$ denotes population density, and $\mathbf{x}_{c,5}$ denotes the percentage of Black residents. Each RBF component has a distinct length-scale parameter ($l$) that is optimized to maximize marginal likelihood on data, enabling feature-specific adaptation in smoothness and complexity across contextual covariates.

Table~\ref{tab:kernel_comparison} reports the relative error and outcome SNR for all kernel configurations considered during greedy feature selection, evaluated at approximately 3,000 simulation samples. The outcome SNR is 
computed as the average signal-to-noise ratio ($\sigma/\mu$) of response function predictions drawn from the GP posterior across all counties and treatment conditions at each iteration, and serves as an operationalizable 
diagnostic for evaluating metamodel uncertainty without requiring held-out ground-truth data. Configurations with substantially elevated outcome SNR, such as the five-feature kernels, indicate numerical instability and are 
discarded regardless of their relative error.

\rev{Among the stable configurations, the two-feature kernel RBF($L$)+RBF($D$) and the four-feature kernel RBF($L$)+RBF($D$)+RBF($I$)+RBF($B$) achieve comparable relative error (8.66\% and 8.97\%). We include income and racial composition not because they improve outcome prediction, which is comparable between the two, but because they enrich the similarity structure over which the Gaussian process interpolates, giving the model a finer basis for identifying which counties are  alike. These features are well-documented correlates of county-level overdose  burden and of access to harm-reduction and treatment services \shortcite{friedman2026charting, goedel2020association}, making them informative dimensions of county similarity; we do not claim that population composition alters a community's responsiveness to intervention. The kernel encodes similarity only and has no directionality: the sign and form of the estimated treatment effects reside entirely in the response-function coefficients ($\mu_n$, $\mu_b$), which are predominantly negative, so that each additional level of naloxone or buprenorphine dispensing is associated with a proportional reduction in overdose mortality.}

\begin{table}[h]
\centering
\renewcommand{\arraystretch}{1.2}
\caption{Outcome SNR and relative error after approximately 3,000 simulation 
samples. Feature codes: L~=~location, D~=~population density, 
I~=~income, B~=~Black population percentage, U~=~unemployment rate, 
R~=~rurality index.}
\begin{tabularx}{\linewidth}{@{}X @{\hspace{1cm}} l c c@{}}
\toprule
\textbf{Kernel design} & \textbf{Features} & \textbf{Outcome SNR} & \textbf{Rel.\ Error (\%)} \\ \midrule
RBF($L$)                                                    & L             & 0.0502 & 9.51  \\ \hline
RBF($L$) + RBF($I$)                                         & L, I          & 0.0576 & 9.92  \\ \hline
RBF($L$) + RBF($D$)                                         & L, D          & 0.0472 & 8.66  \\ \hline
Mat\'{e}rn($L$) + RBF($D$)                                  & L, D          & 0.0503 & 9.59  \\ \hline
RBF($L$) + RBF($D$) + RBF($I$) + RBF($B$)                  & L, D, I, B    & 0.0529 & 8.97  \\ \hline
RBF($L$) + RBF($D$) + RBF($I$) + RBF($U$)                  & L, D, I, U    & 0.0502 & 10.06 \\ \hline
RBF($L$) + RBF($D$) + RBF($I$) + RBF($B$) + RBF($U$)       & L, D, I, B, U & 0.6860 & 57.69 \\ \hline
RBF($L$) + RBF($D$) + RBF($I$) + RBF($B$) + RBF($R$)       & L, D, I, B, R & 3.4289 & 43.02 \\
\bottomrule
\end{tabularx}
\label{tab:kernel_comparison}
\end{table}

\begin{figure}[h]
    \centering
    \begin{subfigure}[b]{0.32\textwidth}
        \includegraphics[width=\textwidth]{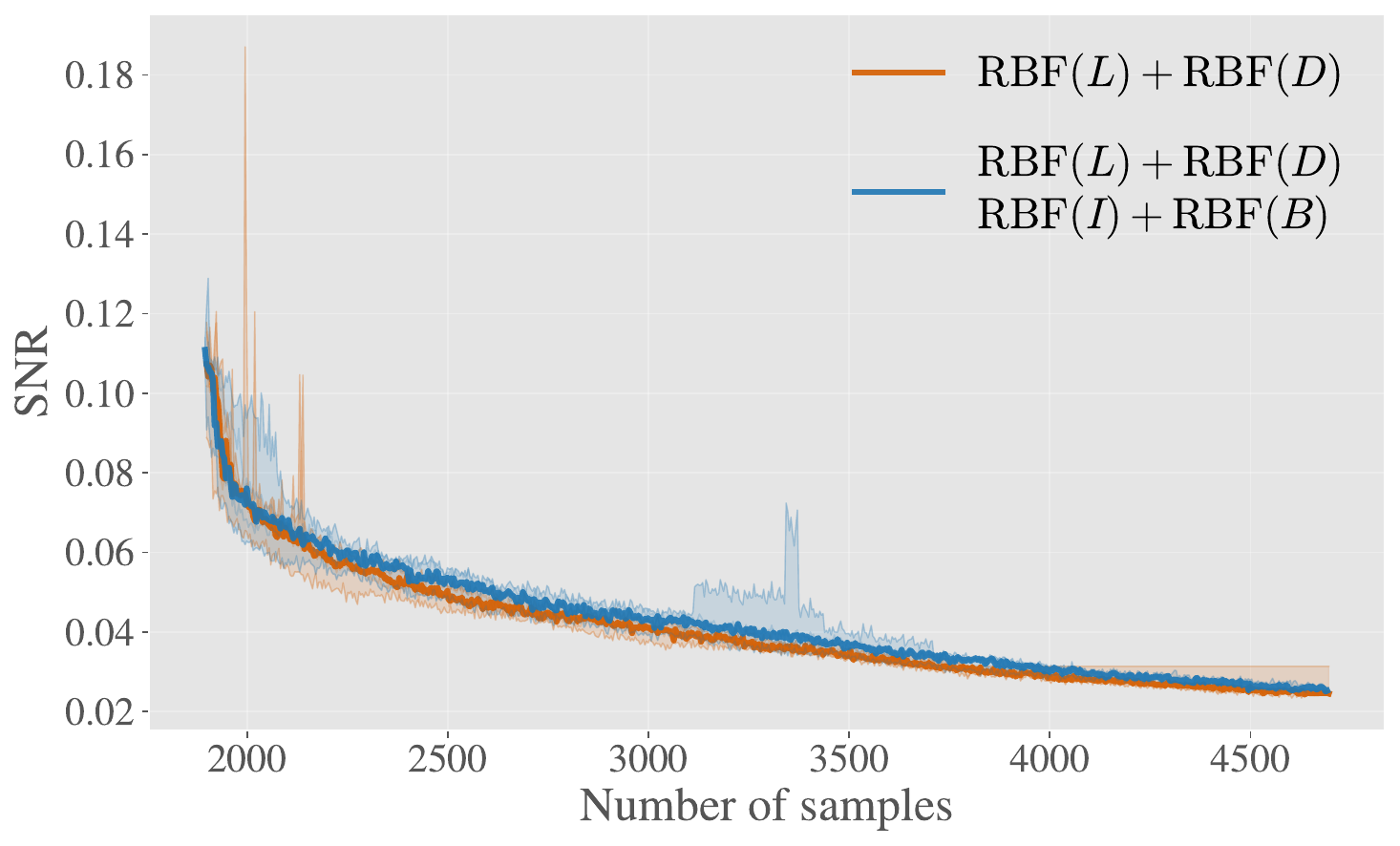}
        \caption{Kernel complexity}
    \end{subfigure}
    \hfill
    \begin{subfigure}[b]{0.32\textwidth}
        \includegraphics[width=\textwidth]{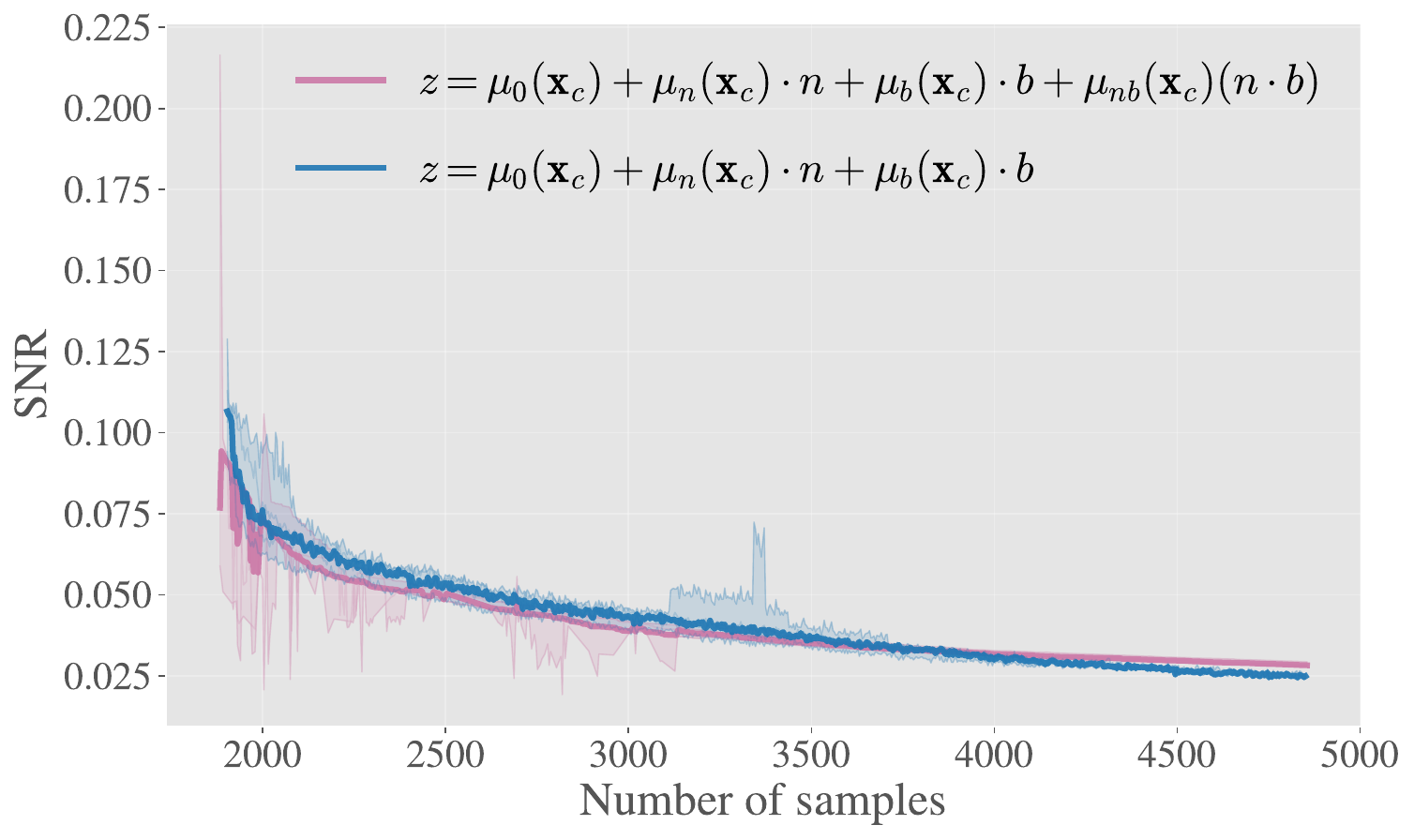}
        \caption{Response function complexity}
    \end{subfigure}
    \hfill
    \begin{subfigure}[b]{0.32\textwidth}
        \includegraphics[width=\textwidth]{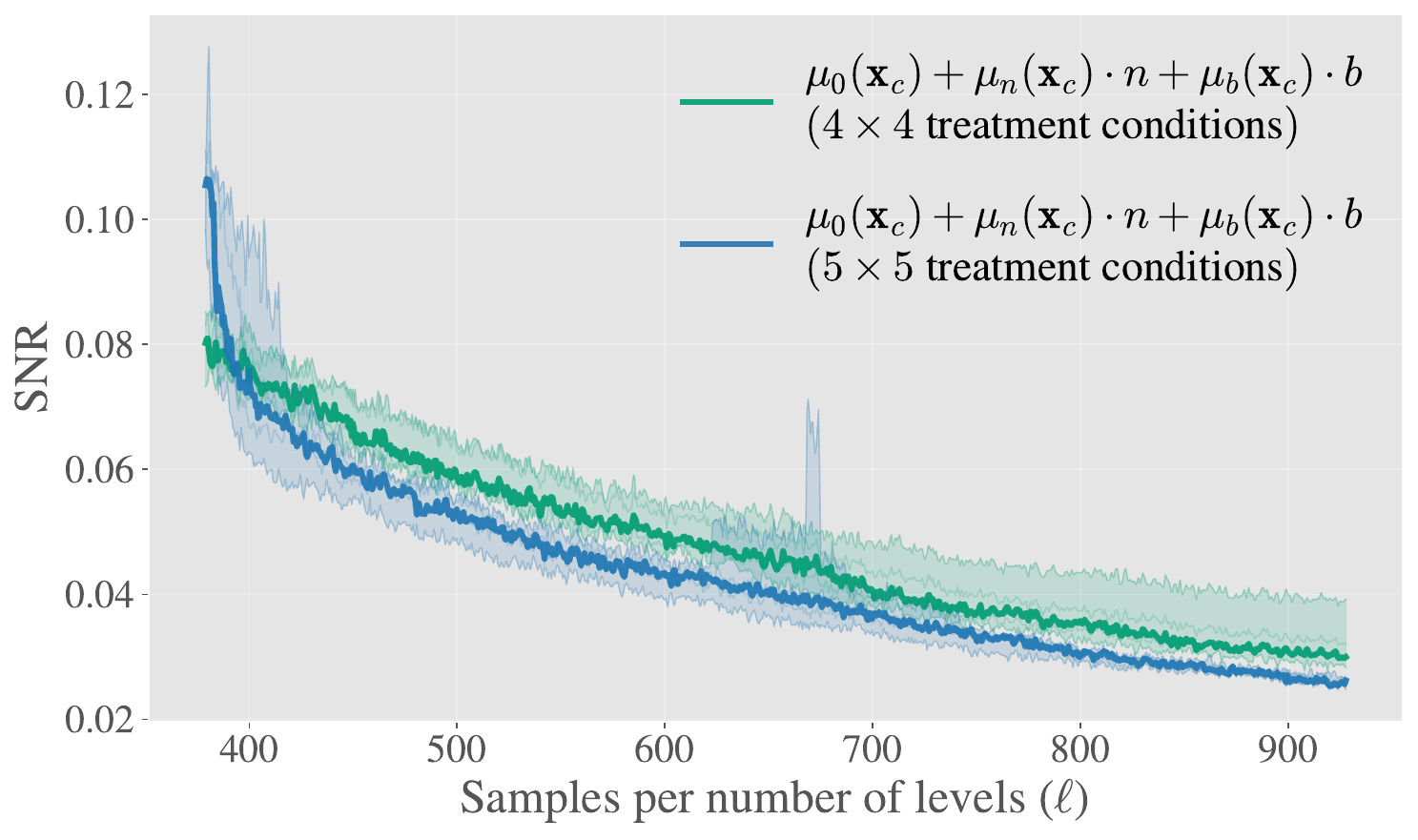}
        \caption{Design space complexity}
    \end{subfigure}
    \caption{Outcome SNR trajectories replicating the three complexity comparisons of Figure~4, kernel complexity (a), response function complexity (b), and design space complexity (c), confirming that outcome SNR yields consistent conclusions with relative error across all three comparisons.}
    \label{fig:snr_complexity}
\end{figure}

\textbf{Outcome SNR as an Alternative Evaluation Metric.}
Figure~\ref{fig:snr_complexity} replicates the three complexity comparisons of Figure~4, kernel complexity, response function complexity, and design
space complexity, using outcome SNR in place of relative error. For kernel complexity, the conclusions are fully consistent: the more complex kernel exhibits higher outcome SNR, mirroring Figure~4(a). For response function complexity, the interaction model exhibits lower outcome SNR than the main-effects model, because the additional interaction term $\mu_{nb}(\mathbf{x}_c)(n \cdot b)$ provides richer characterization of 
the treatment response, reducing relative posterior uncertainty. Similarly, for design space complexity, the $5\times5$ grid exhibits lower outcome SNR than the $4\times4$ grid, as the larger intervention space provides more treatment conditions from which to average posterior uncertainty, yielding a more precise response function estimate. In both cases, the model with lower outcome SNR also achieves lower relative error, so the metric correctly identifies the better-performing model. Outcome SNR therefore remains a reliable operational metric for online model monitoring and stopping decisions from the GP posterior without requiring held-out data, but should be interpreted in context: lower SNR indicates either higher certainty or richer model structure.

\section{FRED Agent-Based Modeling of Opioid Use Disorder}
\label{sec:fred-oud}
The Framework for Reconstructing Epidemiological Dynamics (FRED) is an open-source, agent-based modeling platform developed 
to simulate the spread of infectious diseases \shortcite{grefenstette2013fred}. While originally designed to enhance understanding of epidemic dynamics, FRED has proven effective as a decision-support tool for public health planning and intervention policy development. A key strength of FRED is its use of synthetic populations derived from real U.S. Census data, which enhances the realism and credibility of its simulations \cite{guclu2016agent}. Moreover, FRED builds on the developers’ extensive experience with earlier simulation models, allowing it to overcome many limitations found in previous approaches.

{\textbf{Synthetic Population.}} FRED assigns each individual in the simulation to a specific geographic region, using the U.S. synthetic population database developed by RTI International \cite{rti}, which provides detailed, geographically stratified demographic data. For every agent, FRED generates comprehensive socioeconomic and demographic attributes (e.g., age, education level, income) as well as health-related characteristics (e.g., symptom severity, infection history, vaccination records). Each agent is linked to a specific household and is also assigned to institutions such as schools, workplaces, or prisons. These geographic assignments implicitly encode spatial relationships, including the distance between agents and their assigned locations. Agents in FRED are capable of making individual-level decisions related to health behaviors, such as accepting vaccinations, staying home when ill, or keeping a sick child home from school.

{\textbf{Opioid Use Disorder Model.}}  The rise of Opioid Use Disorder (OUD) has led to a substantial increase in morbidity and mortality in the United States, with over 100,000 overdose deaths (ODDs) reported in 2023 alone. Opioids, including prescription opioids, heroin, and synthetic opioids, are the leading cause of these deaths \shortcite{CDCOpioids}. \citeNP{jalal2018opioiddynamics} analyze the opioid crisis over a span of more than 40 years and found that the current wave of overdose deaths is part of a long-term trend that has persisted across several decades. These observations highlight the importance of developing a robust model to understand the dynamics of OUD. \rev{The OUD model used in this study, including its structure, data sources, 
and calibration, is described in \shortcite{ahmed2026oud}.}

The OUD model is based on a set of health/disorder states, where an individual transitions from one state to another based on specific probabilities. These transition probabilities were estimated using literature and by calibrating the model to actual overdose death rates. At the start of the simulation, agents begin in different health states. The agents may transition to other health states. For example, a non-use may move to a prescribed opioid use state (i.e., receiving a prescription from an accredited physician) or to an opioid misuse state. From either of these states, the agent can then transition into the OUD state, which may lead to ODD, another cause of death, or entry into treatment.

To formally describe agent transitions within the OUD simulation model, we use logistic regression to define three key probabilities that govern agent movement across health states. 
\rev{We present here the three transition equations that depend on the intervention and supply covariates studied in this paper: naloxone dispensing, buprenorphine dispensing, opioid dispensing, and fentanyl seizure rates. The remaining transitions in the OUD model are fixed at values derived from the literature and expert input, as documented in \shortcite{ahmed2026oud}.}
Each probability is expressed in terms of a logistic regression model.

\begin{align}
    \log\left(\frac{p_1}{1 - p_1}\right) &= \beta_0 + \beta_1 \cdot \text{opioid dispensing rate} \label{eq:p1} \\
    \log\left(\frac{p_2}{1 - p_2}\right) &= \beta_2 + \beta_3 \cdot \text{buprenorphine dispensing rate} \label{eq:p2} \\
    \log\left(\frac{p_3}{1 - p_3}\right) &= \beta_4 + \beta_5 \cdot \text{fentanyl seizure rate} - \beta_6 \cdot \text{naloxone dispensing rate} \label{eq:p3}
\end{align}

\noindent
\noindent
Here, $\beta_0,\dots,\beta_6$ are the logistic-regression coefficients internal to the OUD simulation model (with a slight abuse of notions, we have used the same symbols as the metamodel's response-function coefficients $\beta_0,\beta_n,\beta_b$ in Equation~\eqref{eq:regression-estimates} of the main text). Equation~\eqref{eq:p1} models the transition from the nonuser state to prescription opioid use, with the opioid dispensing rate serving as a key predictor. Equation~\eqref{eq:p2} defines the likelihood of an individual in the OUD state entering treatment as a function of buprenorphine availability. Finally, Equation~\eqref{eq:p3} captures the probability of an overdose death, incorporating both fentanyl seizure rate and naloxone dispensing rate to represent the balance between risk and harm reduction. These transition equations serve as the foundation for modeling agent behavior under varying intervention scenarios and contextual risk environments. 

The objective is to evaluate the effect of interventions, primarily involving two medications: Naloxone and Buprenorphine. Naloxone is an antidote used to reverse opioid overdose, while Buprenorphine is a treatment medication used to support recovery and help individuals return to the nonuse state. The availability of these medications defines the intervention level, and the number of ODDs in a given geographic area is treated as the intervention outcome. 

\begin{figure}[h]
    \centering
    \includegraphics[scale=0.5]{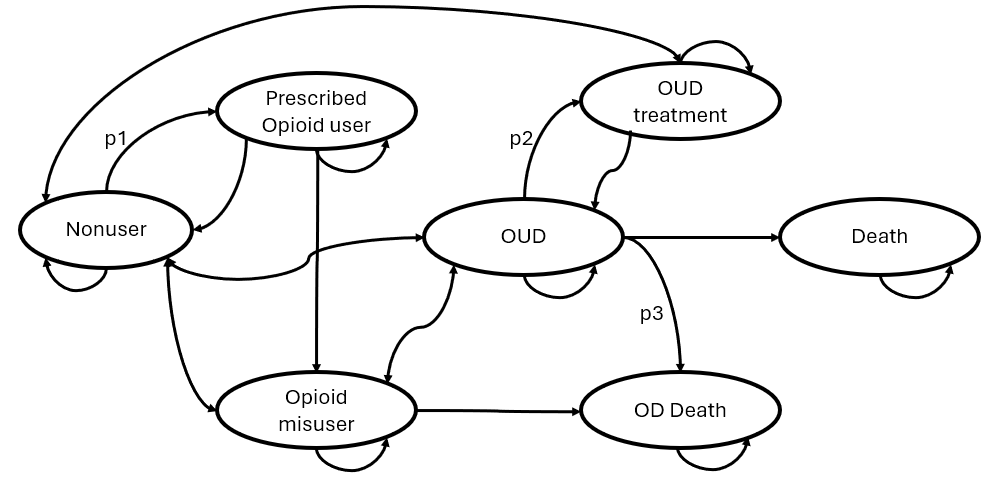}
    \caption{ {\bf State transition diagram for the OUD model.} Transition probabilities 
$p_1$, $p_2$, and $p_3$ are defined by Equations~\eqref{eq:p1}-\eqref{eq:p3}, respectively.}
    \label{fig:oudFlow}
\end{figure}

\section{Calibrating the Opioid Use Disorder Agent-Based Model in Different Counties}
\label{sec:app:calibration}

Model calibration is a critical step in simulation-based studies, particularly when the model includes parameters that are difficult to observe or directly measure. In the context of our OUD simulation model, calibration refers to the process of adjusting unobserved transition probabilities so that the simulated outputs closely match real-world data—most notably, historical overdose death rates. Accurate calibration ensures that simulation results are both credible and reflective of the complex dynamics observed in actual populations.

In this study, we employ Incremental Mixture Importance Sampling (IMIS) \shortcite{imisraftery2010, menzies2017bayesian} as the calibration algorithm. IMIS is a Bayesian technique that combines the strengths of importance sampling and adaptive proposal distributions. It incrementally builds a mixture of proposal distributions that efficiently explore the high-probability regions of the posterior. This makes IMIS particularly well-suited for models with complex, multimodal likelihood surfaces and moderate-dimensional parameter spaces.
The calibration process begins by defining a prior distribution over the uncertain model parameters and specifying a set of target statistics derived from observed data, typically, annual opioid overdose deaths by county and year. The likelihood function measures how well a given parameter configuration reproduces these observed outcomes. IMIS iteratively samples from and updates the proposal distribution to concentrate on regions of the parameter space with high posterior probability.

Because calibration is computationally intensive and requires a large number of simulation runs for each parameter set, we limit the calibration to a representative subset of counties. Specifically, we select six counties, Allegheny, Philadelphia, Erie, Dauphin, Clearfield, and Columbia, based on population size and death trends. These are chosen to represent three types of counties: large (Allegheny and Philadelphia), medium (Erie and Dauphin), and small (Clearfield and Columbia). Each pair of counties is selected to reflect varying geographic and epidemiological characteristics. This decision strikes a balance between computational feasibility and the need for generalizable insights across diverse county profiles. 
To evaluate calibration accuracy, we compared modeled outcomes against observed county-level overdose mortality and examined posterior convergence of the transition-related coefficients (Equations ~\eqref{eq:p1}-\eqref{eq:p3}). Figure~\ref{fig:calib} shows that, across all six calibrated counties, the model closely matches observed mortality patterns and yields well-identified posterior distributions for the calibrated parameters.

We emphasize that calibration is an expert-driven and resource-intensive process. Each calibration attempt involves multiple simulation replications, validation against mortality curves, visual inspection of trajectories, and iterative refinement of assumptions. Due to these demands, full calibration for all counties is infeasible within reasonable time and resource constraints.
Therefore, we select six prototype counties for full calibration, stratified by population size and geography. The calibrated parameters for these counties are then generalized to the remaining counties via a similarity-based assignment procedure. 
Given the effort required, we perform full calibration only for these six counties and use a similarity-based assignment procedure to generalize the calibrated parameters to non-calibrated counties.

\textbf{Generalizing calibrated parameters using county Similarity.} 
For each county, we summarize opioid-related trends over 2015–2019 using a small set of features that capture both overall levels and their temporal changes. These features are used to assign each non-calibrated county to a calibrated county with the closest opioid-related covariates. Specifically, we compute summary measures of overdose mortality and treatment dispensing that reflect average magnitude and the estimated slope of each time series over the study period, rather than year-by-year values. Overdose mortality is represented by two quantities: the average overdose death rate over the study period and the estimated slope of overdose mortality over time. Treatment and supply indicators (opioid dispensing, naloxone, buprenorphine, and fentanyl seizures) are summarized by the magnitude of their estimated time-series slopes. The county population is included as an additional contextual feature.
These features are standardized across counties and combined into a feature vector for each county. \rev{These matching features, like the geographic and 
socio-economic covariates used in the GPR kernel, serve to define proximity 
between counties for parameter transfer and spatial interpolation, rather than 
to predict overdose outcomes directly.} Using this representation, each non-calibrated county is assigned to the most similar calibrated prototype county by minimizing Euclidean distance in the feature space. This matching emphasizes similarity in relative patterns and trends rather than absolute levels.

The resulting mapping enables the transfer of calibrated transition-related coefficients from the six prototype counties to all remaining counties, while preserving each county’s observed dispensing rates and contextual inputs, thereby eliminating the need for county-by-county calibration and substantially reducing computational and manual effort.
Figure~\ref{fig:generalize_calib} summarizes this parameter generalization across Pennsylvania, illustrating how non-calibrated counties inherit transition-related coefficients from their most similar calibrated prototype county.

\begin{figure}[H]
    \centering
    \begin{subfigure}[b]{0.39\textwidth}
        \includegraphics[width=\linewidth]{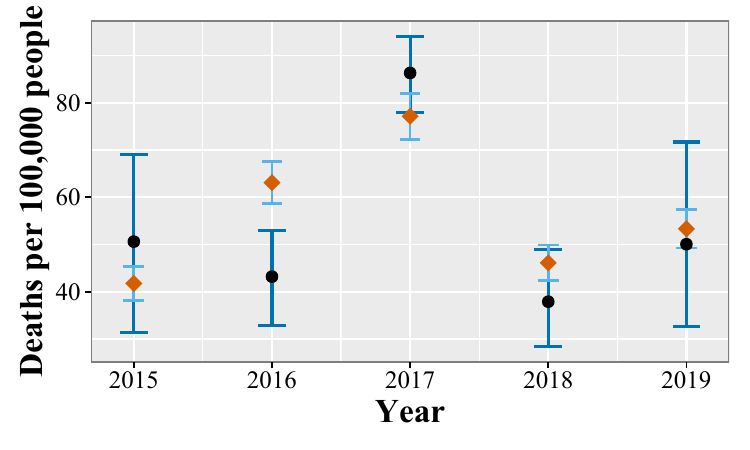}
        \caption{\tiny{Allegheny calibration fit}}
    \end{subfigure}
    \quad
    \begin{subfigure}[b]{0.54\textwidth}
        \includegraphics[width=\linewidth]{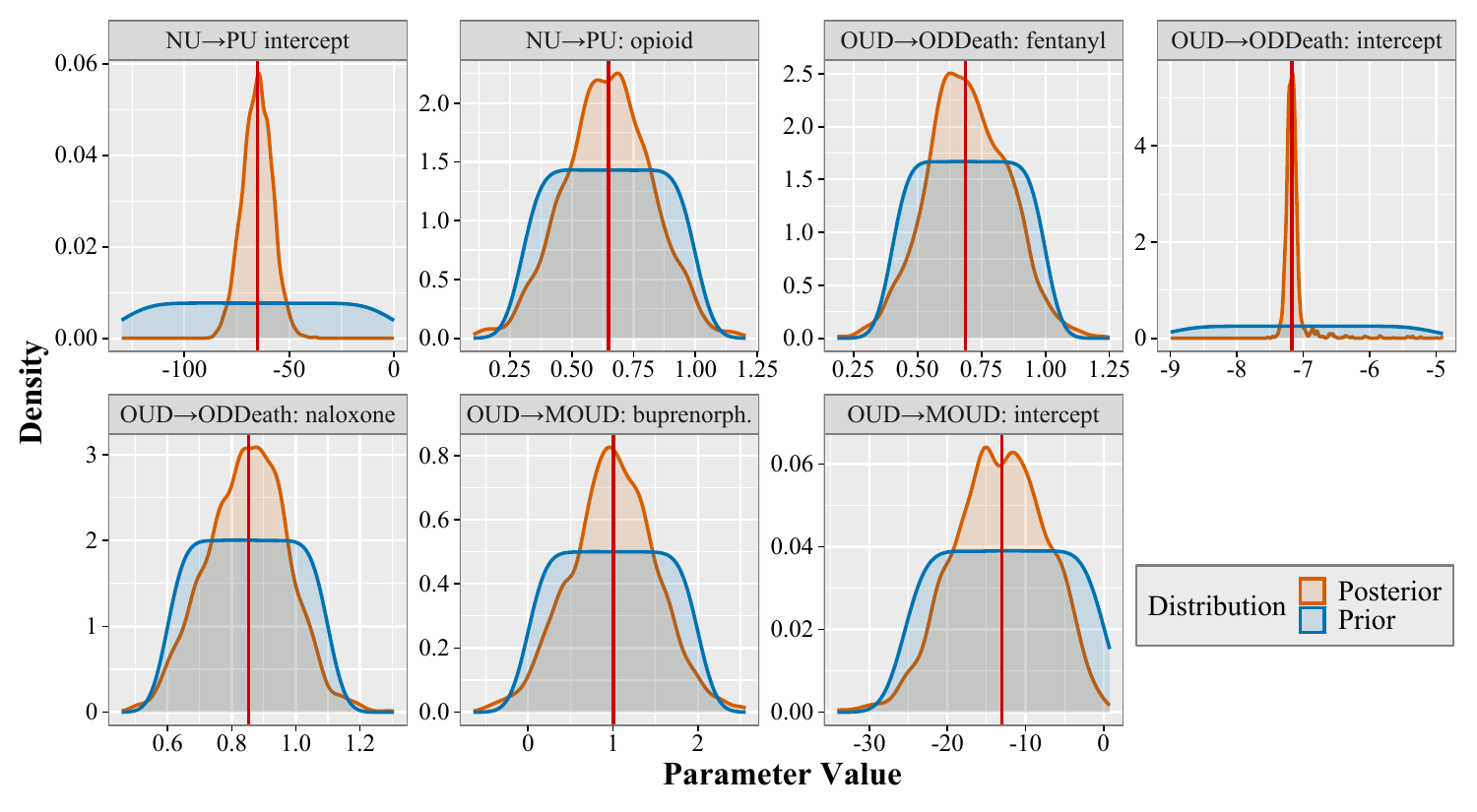}
        \caption{\tiny{Allegheny posterior distributions of calibrated model coefficients}}
    \end{subfigure}
    \vskip\baselineskip
    \begin{subfigure}[b]{0.39\textwidth}
        \includegraphics[width=\linewidth]{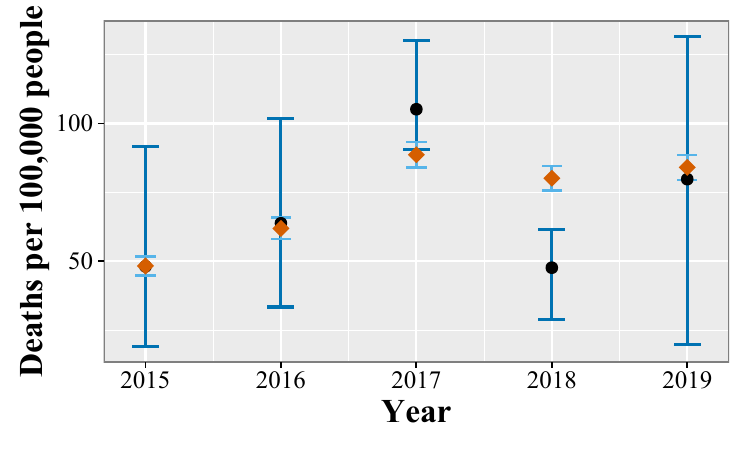}
        \caption{\tiny{Philadelphia calibration fit}}
    \end{subfigure}
    \quad
    \begin{subfigure}[b]{0.54\textwidth}
        \includegraphics[width=\linewidth]{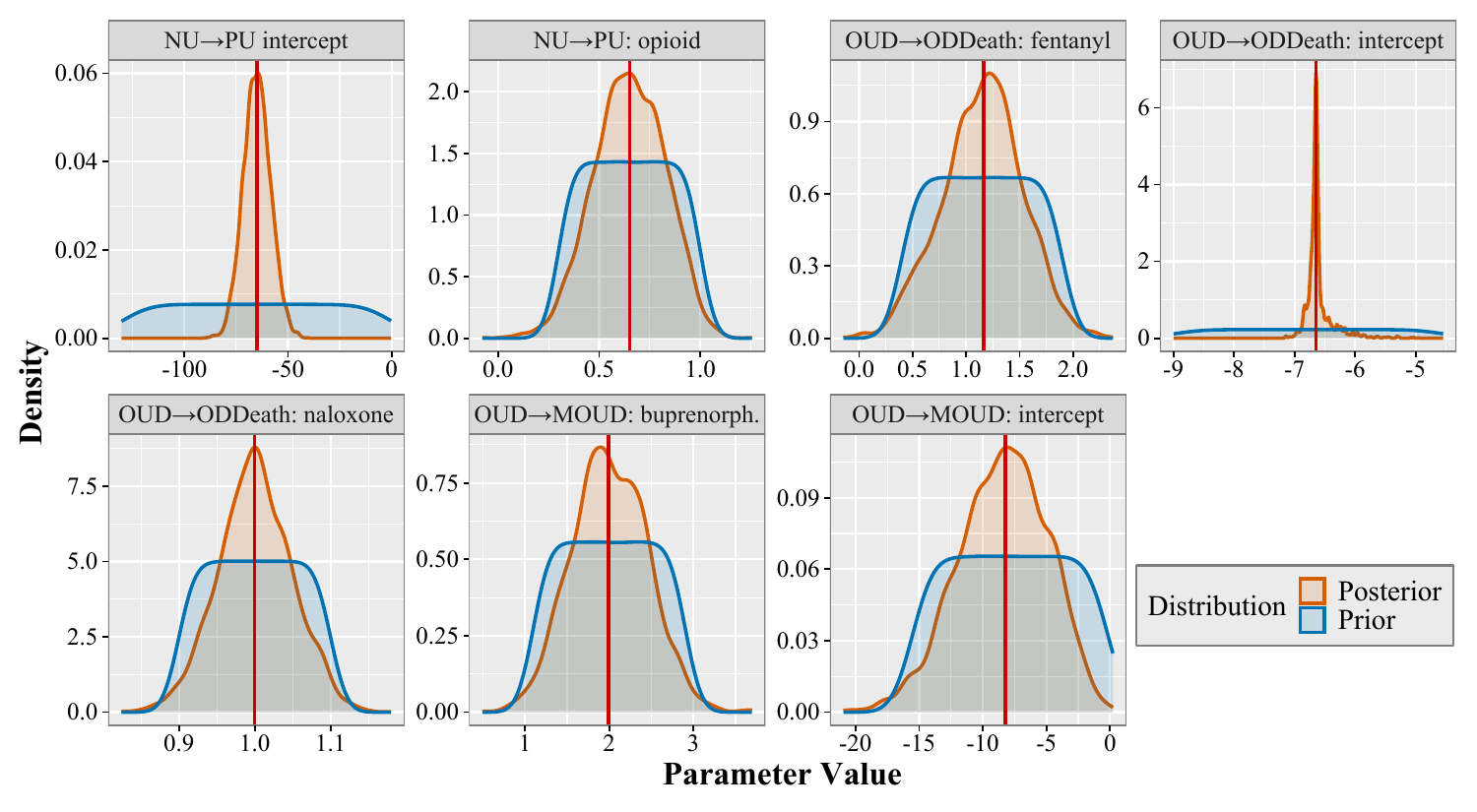}
        \caption{\tiny{Philadelphia  posterior distributions of calibrated model coefficients}}
    \end{subfigure}
    \vskip\baselineskip
    \begin{subfigure}[b]{0.39\textwidth}
        \includegraphics[width=\linewidth]{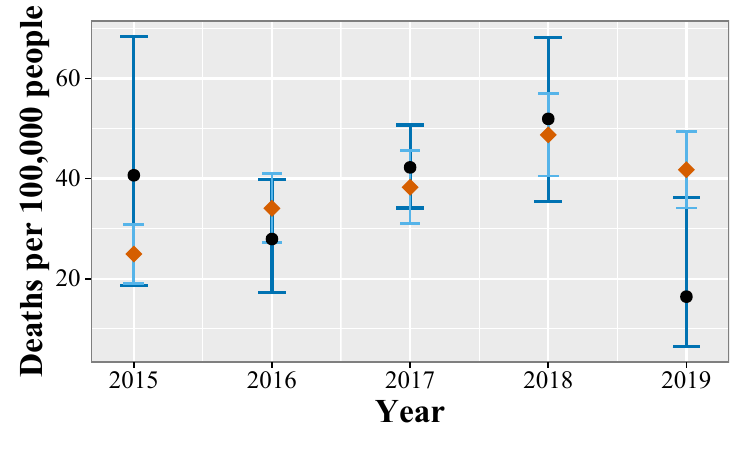}
        \caption{\tiny{Dauphin calibration fit}}
    \end{subfigure}
    \quad
    \begin{subfigure}[b]{0.54\textwidth}
        \includegraphics[width=\linewidth]{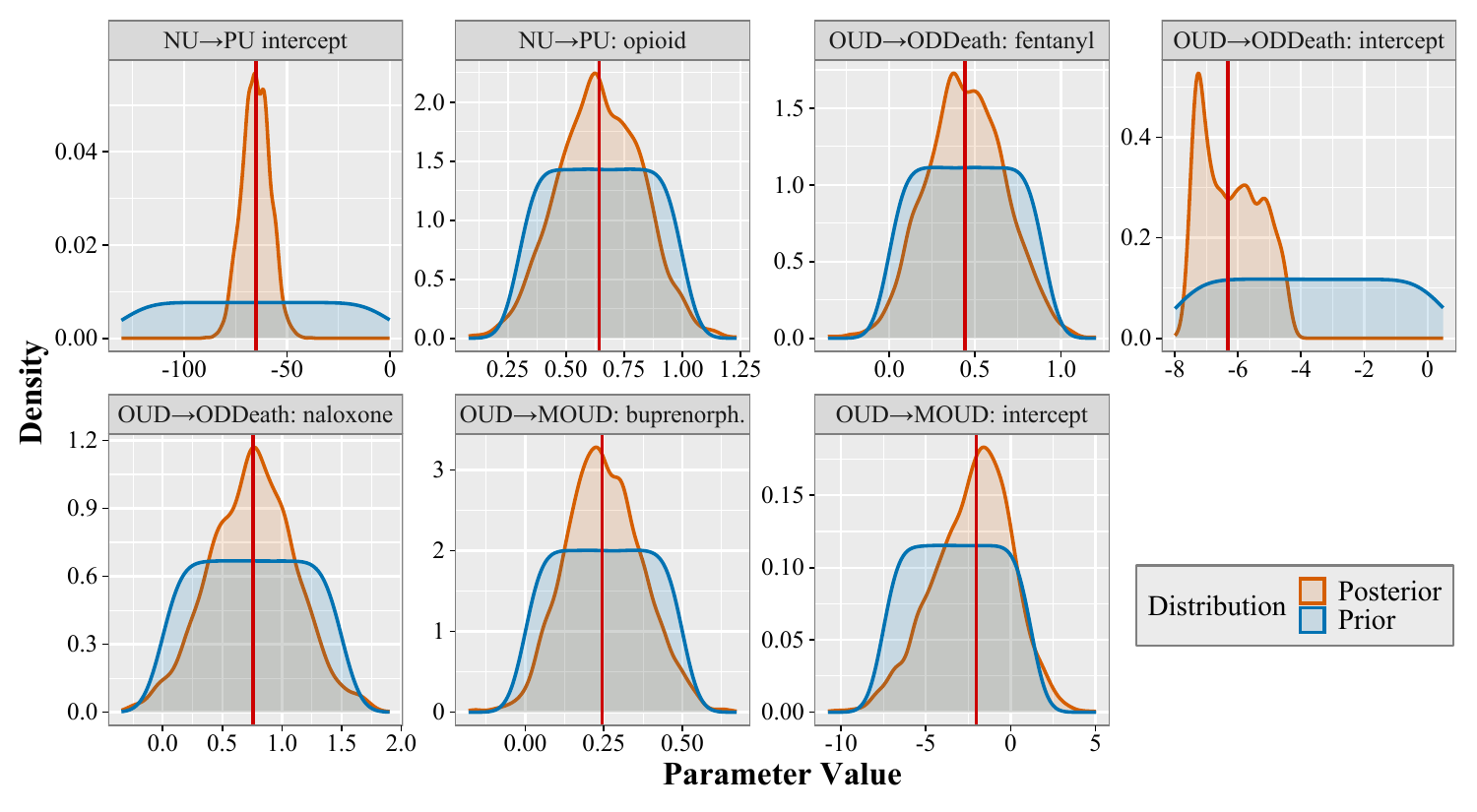}
        \caption{\tiny{Dauphin posterior distributions of calibrated model coefficients}}
    \end{subfigure}
        

    \begin{subfigure}[b]{0.39\textwidth}
        \includegraphics[width=\linewidth]{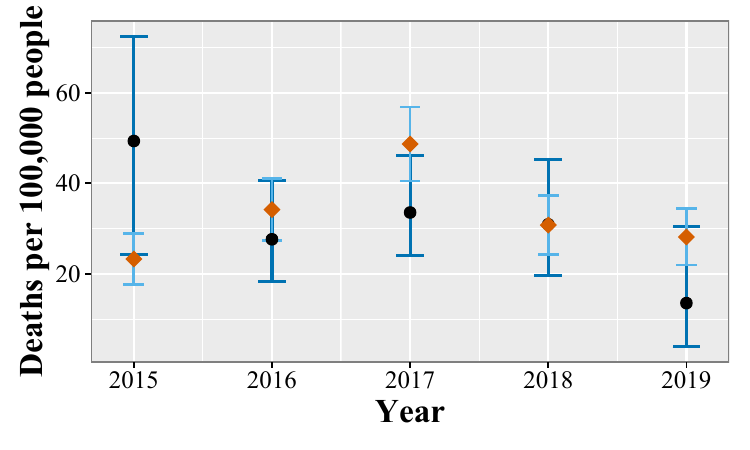}
        \caption{\tiny{Erie calibration fit}}
    \end{subfigure}
    \quad
    \begin{subfigure}[b]{0.54\textwidth}
        \includegraphics[width=\linewidth]{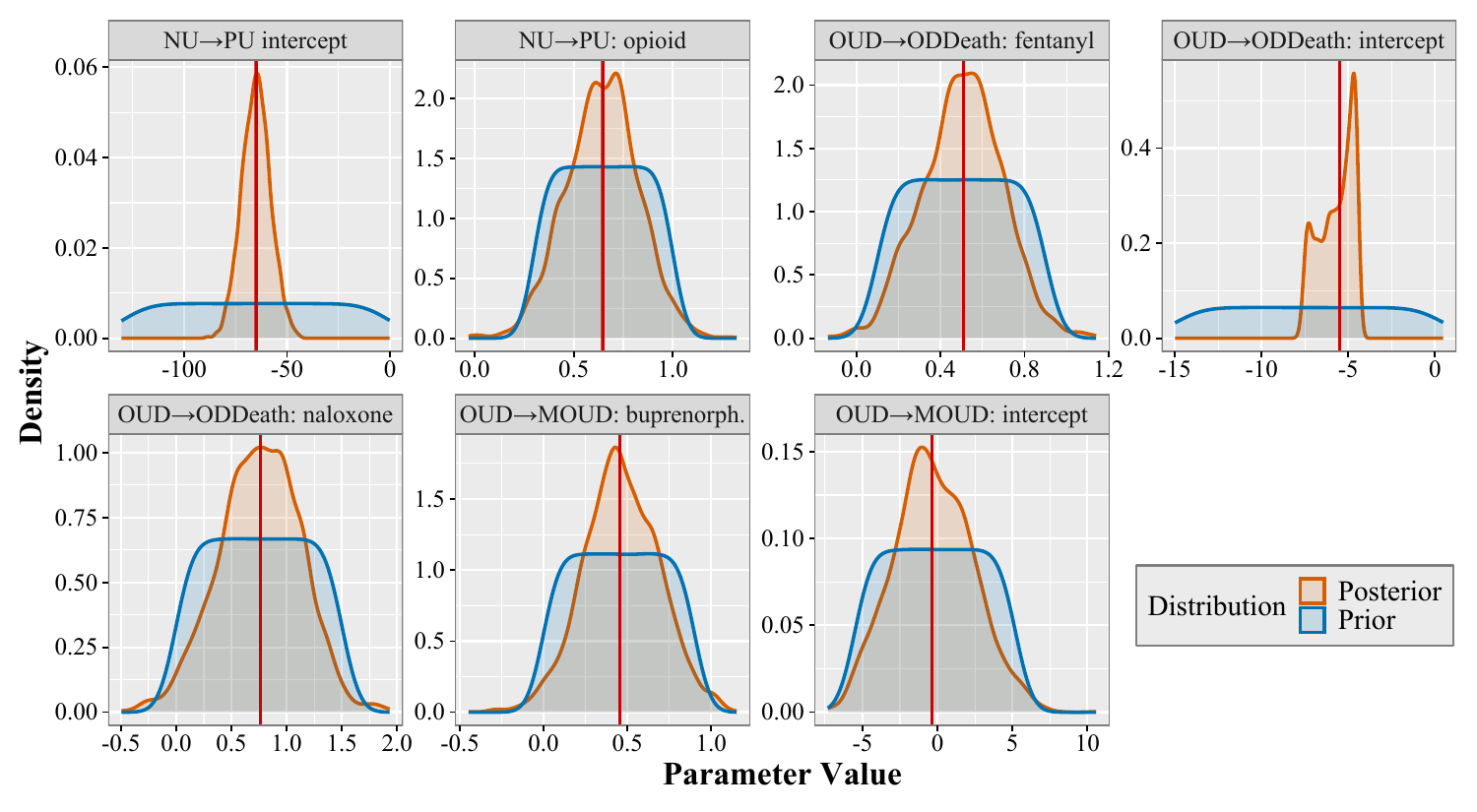}
        \caption{\tiny{Erie posterior distributions of calibrated model coefficients}}
    \end{subfigure}
    \vskip\baselineskip
\end{figure}
\begin{figure}[h]
  \ContinuedFloat 
  \centering
    \begin{subfigure}[b]{0.39\textwidth}
        \includegraphics[width=\linewidth]{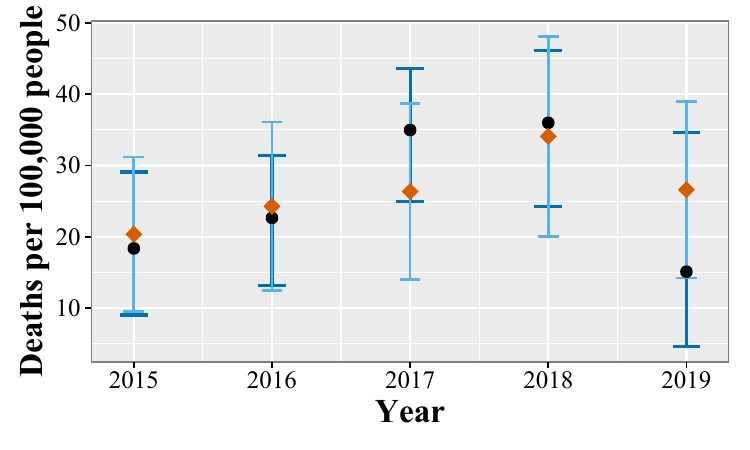}
        \caption{\tiny{Columbia calibration fit}}
    \end{subfigure}
    \quad
    \begin{subfigure}[b]{0.54\textwidth}
        \includegraphics[width=\linewidth]{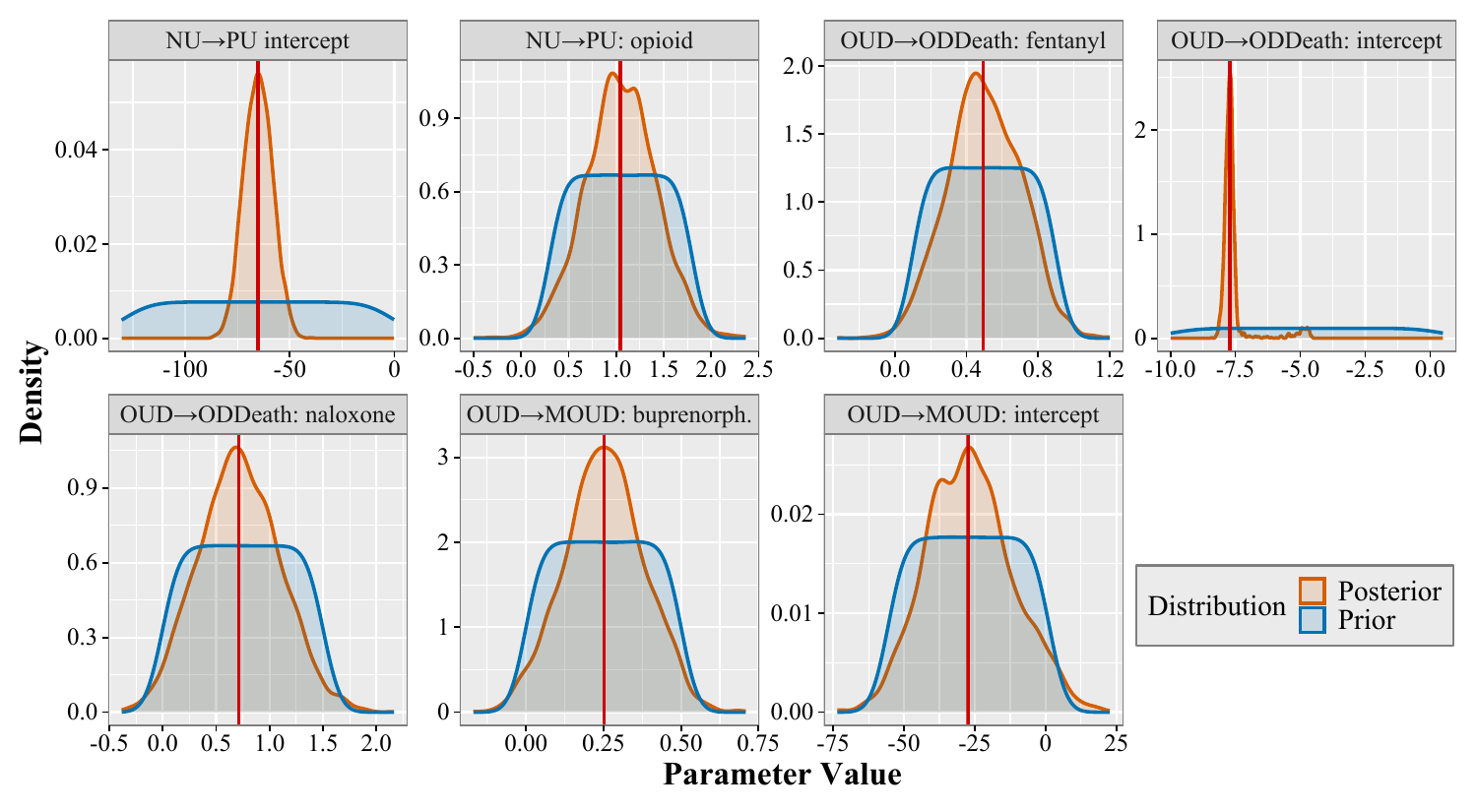}
        \caption{\tiny{Columbia  posterior distributions of calibrated model coefficients}}
    \end{subfigure}
    \vskip\baselineskip
    \begin{subfigure}[b]{0.39\textwidth}
        \includegraphics[width=\linewidth]{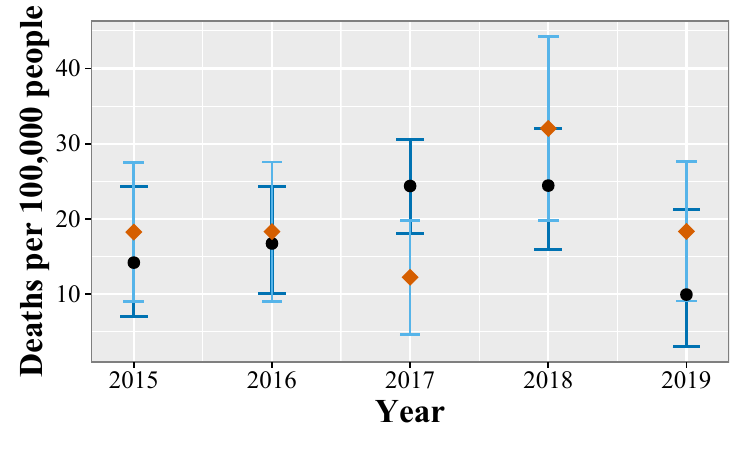}
        \caption{\tiny{Clearfield calibration fit}}
    \end{subfigure}
    \quad
    \begin{subfigure}[b]{0.54\textwidth}
        \includegraphics[width=\linewidth]{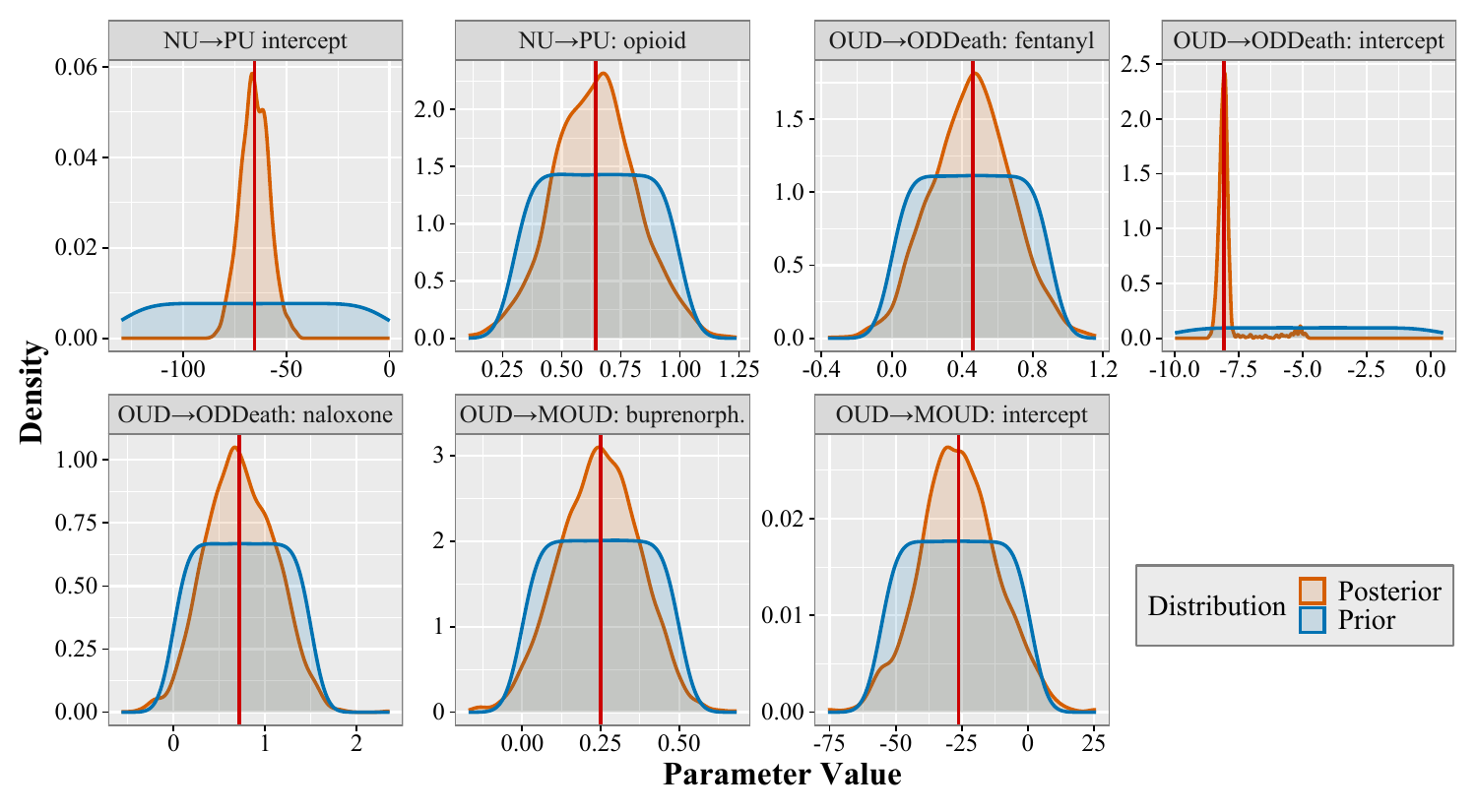}
        \caption{\tiny{Clearfield posterior distributions of calibrated model coefficients}}
    \end{subfigure}  
    \caption{{\bf Calibration performance and parameter identification for the opioid use disorder model.} Left panels show model fit to observed county-level overdose mortality targets, where orange points denote observed mortality estimates with associated confidence intervals, and black points with blue credible intervals represent model outputs. While right panels display posterior distributions of the transition-related coefficients defined in Equations~\eqref{eq:p1}-\eqref{eq:p3}. across the six calibrated counties (Allegheny, Philadelphia, Dauphin, Erie, Columbia, and Clearfield).}
    \label{fig:calib}
\end{figure}

\begin{figure}[H]
    \centering
    \includegraphics[scale=0.45]{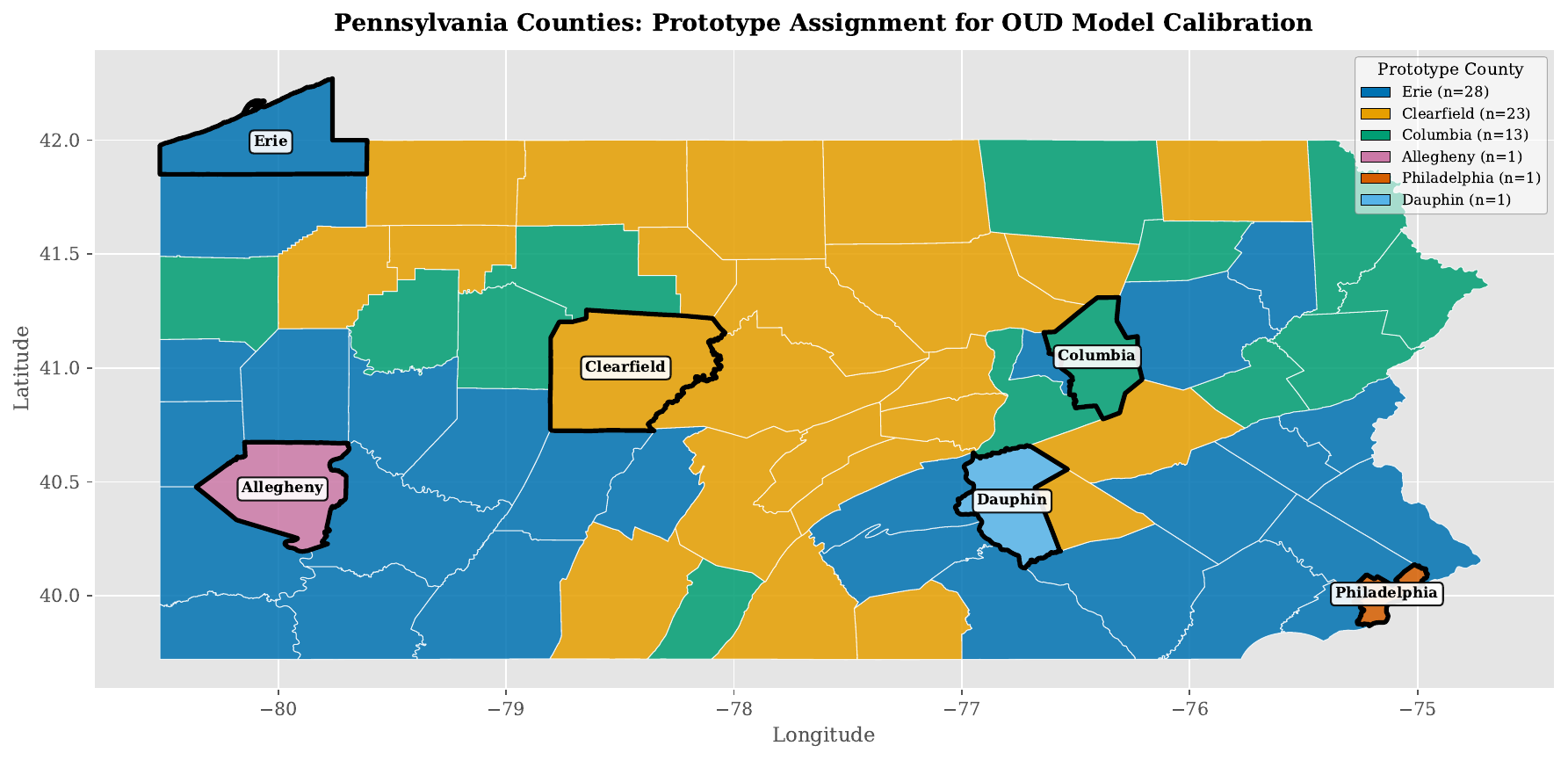}
    \caption{{\bf Generalization of calibrated parameters across Pennsylvania.} Six counties (highlighted with bold borders) were fully calibrated; remaining non-calibrated counties were matched to the nearest calibrated one based on similarity in overdose mortality patterns and treatment dispensing trends (opioid, naloxone, buprenorphine, and fentanyl seizure rates). Legend shows calibrated prototype county and number of assigned counties (n).}
    \label{fig:generalize_calib}
\end{figure}

\section{Implementation Details and Computational Constraints}
\label{sec:implementation}
The simulation experiments are executed on the University of Pittsburgh’s Center for Research Computing (CRC) cluster, using the Simple Linux Utility for Resource Management (SLURM) job arrays for high-throughput scheduling. Our study involves simulating 67 counties and 25 treatment conditions, each with a target of 1000 replications. This yields more than 1.6 million individual simulation runs. The high replication count serves two purposes: (i) it delivers a low-variance “brute-force’’ benchmark against which to judge metamodel accuracy, and (ii) it highlights the prohibitive cost of exhaustive simulation. Each simulation utilizes the FRED platform, which models agent-level interactions across synthetic populations with complex disease transmission and treatment dynamics, making each run both memory- and compute-intensive.

The CRC cluster imposes several job submission constraints that introduce bottlenecks in execution. One such constraint is the wall-clock time penalty: jobs requesting longer wall-clock limits are deprioritized in the queue, which significantly delays execution for long-running tasks. Unfortunately, our simulations, especially for densely populated counties or high-treatment levels, require extended runtimes to complete due to the stochastic nature and agent-level detail of the FRED model. As a result, we must balance the specification between a long enough wall-clock to avoid job termination and a short enough request to avoid queuing delays.
Another challenge lies in memory management. SLURM enforces strict limits on per-job memory usage, and insufficient allocation can lead to job eviction, segmentation faults, or incomplete logs. On the other hand, excessive memory requests increase wait times and reduce overall cluster utilization. To address this, we use heuristics based on population size, treatment intensity, and historical memory profiles to dynamically scale resource requests. However, occasional resubmission and manual intervention are still required to recover from failures and adjust job configurations.
Storage constraints further complicate large-scale simulation. Each simulation generates outputs including agent histories, event logs, and aggregated outcomes. With over a million simulations, the cumulative storage footprint becomes significant. We address this by routinely compressing outputs, writing only summary statistics when feasible, and periodically deleting intermediate files once they are processed into metamodel training datasets.

\end{document}